\documentclass[aps,prx,twocolumn,twoside,superscriptaddress,notitlepage]{revtex4-1}
\usepackage{bm}
\usepackage{mathrsfs}
\usepackage{amssymb}
\usepackage{amsmath}
\usepackage{amsfonts}
\usepackage{xcolor}
\usepackage{natbib}
\usepackage{hyperref}
\usepackage{graphicx}
\usepackage[ruled, noline]{algorithm2e}
\usepackage{amssymb}
\newcommand{\ket}[1]{\left|#1\right\rangle}
\newcommand{\sket}[1]{|#1\rangle}
\newcommand{\bra}[1]{\left\langle#1\right|}
\newcommand{\braket}[2]{\langle#1|#2\rangle}
\newcommand{\llaves}[1]{\lbrace #1\rbrace}
\newcommand{\be}{\begin{equation}}

\newcommand{\ee}{\end{equation}}
\newcommand{\abs}[1]{\left|#1\right|}
\newcommand{\cS}{{\cal S}}
\newcommand{\cA}{{\cal A}}
\newcommand{\cR}{{\cal R}}
\newcommand{\cO}{{\cal O}}
\newcommand{\E}[2]{\mathbb{E}_{#1}\left[#2\right]}
\DeclareMathOperator*{\argmax}{arg\,max}
\newcommand{\Pt}{\mathbf{P}_{t}}
\newcommand{\Rt}{\mathbf{R}_{t}}

\begin{document}

\title{Real-time calibration of coherent-state receivers:\\ learning by trial and error}


\author{M. Bilkis}
\affiliation{F\'{\i}sica Te\`{o}rica: Informaci\'{o} i Fen\`{o}mens Qu\`{a}ntics, %
Departament de F\'{\i}sica, Universitat Aut\`{o}noma de Barcelona, ES-08193 Bellaterra (Barcelona), Spain}

\author{M. Rosati}
\affiliation{F\'{\i}sica Te\`{o}rica: Informaci\'{o} i Fen\`{o}mens Qu\`{a}ntics, %
Departament de F\'{\i}sica, Universitat Aut\`{o}noma de Barcelona, ES-08193 Bellaterra (Barcelona), Spain}

\author{R. Morral Yepes}
\affiliation{F\'{\i}sica Te\`{o}rica: Informaci\'{o} i Fen\`{o}mens Qu\`{a}ntics, %
Departament de F\'{\i}sica, Universitat Aut\`{o}noma de Barcelona, ES-08193 Bellaterra (Barcelona), Spain}

\author{J. Calsamiglia}
\affiliation{F\'{\i}sica Te\`{o}rica: Informaci\'{o} i Fen\`{o}mens Qu\`{a}ntics, %
Departament de F\'{\i}sica, Universitat Aut\`{o}noma de Barcelona, ES-08193 Bellaterra (Barcelona), Spain}

\begin{abstract}
The optimal discrimination of coherent states of light with current technology is a key problem in classical and quantum communication, whose solution would enable the realization of efficient receivers for long-distance communications in free-space and optical fiber channels. In this article, we show that reinforcement learning (RL) protocols allow an agent to learn near-optimal coherent-state receivers made of passive linear optics, photodetectors and classical adaptive control. Each agent is trained and tested in real time over several runs of independent discrimination experiments and has no knowledge about the energy of the states nor the receiver setup nor the quantum-mechanical laws governing the experiments. Based exclusively on the observed photodetector outcomes, the agent adaptively chooses among a set of $\sim3\cdot10^{3}$ possible receiver setups, and obtains a reward at the end of each experiment if its guess is correct. At variance with previous applications of RL in quantum physics, the information gathered in each run is intrinsically stochastic and thus insufficient to evaluate exactly the performance of the chosen receiver. Nevertheless, we present families of agents that: (i) discover a receiver beating the best Gaussian receiver after $\sim 3 \cdot 10^{2}$ experiments; (ii) surpass the cumulative reward of the best Gaussian receiver after $\sim 10^{3}$ experiments; (iii) simultaneously discover a near-optimal receiver and attain its cumulative reward after $\sim 10^{5}$ experiments. Our results show that RL techniques are suitable for on-line control of quantum receivers and can be employed for long-distance communications over potentially unknown channels.
\end{abstract}

\maketitle

\section{Introduction}
Quantum state discrimination (QSD) is the problem of determining the state of a quantum system among a set of possible candidates. It constitutes a fundamental primitive in quantum information processing, with applications ranging from long-distance communication~\cite{helstromBOOK,holevoBOOK,Takeoka14,Waseda10,Waseda11,Guha11,Krovi2014,Rosati16c,Zwolinski2018},  cryptography~\cite{Huttner1995,Dusek2000,Grosshans2002,Gisin2002,Bergou2007,Bennet14,Chabaud2019,Pirandola2019} and, recently, quantum machine learning~\cite{SCHULD2019,Sentis2010,Sentis2014,Guta2010a,Lloyd2018,Fanizza2019,Blank19,Sergioli2019,Benedetti2019}.

In the past few years, the use of machine learning methods to deepen the understanding of fundamental physics has become a standard technique~\cite{Carleo2016,Torlai2016,VanNieuwenburg2017,Carrasquilla2017,Torlai2018,Melnikov2018,Fosel2018,Wallnofer2019,Bukov2017,Niu2019}.
Machine learning can be classified as supervised, unsupervised and reinforcement learning (RL). 
In particular, RL studies the behaviour of an agent interacting with an environment via observations, actions and rewards. The goal is to optimize such interactions in order to maximize a suitable figure of merit, e.g., the expected reward over time.
Combinations of these three machine-learning classes have recently led to out-performing the best human GO player, discovering strategies never played before~\cite{Silver2016}. Recently, RL techniques have also been proved successful in quantum information, e.g., in the design of novel quantum experiments~\cite{Melnikov2018}, quantum error-correction codes~\cite{Fosel2018}, quantum communication protocols~\cite{Wallnofer2019} and optimal control of quantum systems~\cite{Bukov2017,Niu2019}.

In the present work we consider the discrimination of two coherent states with passive linear optics, photodetectors and discrete-time classical adaptive control. This is a prototypical problem in quantum information theory~\cite{helstromBOOK,Osaki1996,RevGauss}, of great technological significance for long-distance communication~\cite{Takeoka2008,Waseda10,Waseda11,Guha11,Krovi2014,Rosati16c}: the optimal measurement to discriminate two coherent states is known~\cite{helstromBOOK,Osaki1996} but its implementation is demanding at the state of the art, i.e., via the so-called Dolinar receiver~\cite{Dolinar1973,Kennedy1973a,Geremia2004,Takeoka2005,Takeoka2006,Cook2007,Takeoka2008,DaSilva2013,Nair2014, Rosati16a,DiMario2018a} that requires asymptotically many control rounds. Moreover, its extension to multiple states is not fully understood~\cite{Becerra2013a,Nair2014,Muller2015}, although it may bring us a step closer to achieving the Holevo communication capacity of real-world channels.

We propose an innovative and experimentally appealing approach to the problem: the search for optimal discrimination strategies is cast as a test-bed for RL, by studying how well an agent can perform in calibrating a receiver by means of \textit{model-free} methods.
The nature of our approach is particularly appealing for scenarios where an accurate description of the system is not possible, e.g., due to intrinsic complexity, experimental constraints or imperfections, untrusted devices or simply lack of knowledge. This is precisely the case for applications of coherent-state discrimination in communication scenarios, where discriminating multiple hypotheses may require tuning long sequences of gate parameters~\cite{Guha2012,Wilde2013,Rosati16b,Rosati2017}, the detectors may be affected by losses and dark-counts~\cite{Geremia2004,Muller2015}, the actual communication channel may add different kinds of noise depending on the physical implementation~\cite{Waseda10,Waseda11,Xiang2017,DiMario2019} and device-independent security may be additionally required~\cite{Gisin2002,Pirandola2019}.

In this article we show that a RL agent can achieve near-optimal control of a coherent-state detector when it has zero prior knowledge of: (i) the energy of the coherent states themselves; (ii) the actual operations that the detector performs; (iii) the underlying quantum-mechanical laws governing the system. By trial and error, the RL agent has to sequentially press buttons and select actions according to previous measurement outcomes and at a final stage guess for one of the possible hypotheses. A non-zero reward is given only if the guess is correct. By repeating the procedure over several episodes (or runs), the agent earns experience and \textit{learns a near-optimal discrimination protocol and guessing rule with the resources at its disposal}.

Our approach differs from previous applications of RL in quantum information~\cite{Melnikov2018,Fosel2018,Wallnofer2019,Bukov2017,Niu2019} at least in three crucial aspects: (i) our agents can simultaneously learn and be tested in a completely model-free setting; (ii) each reward is obtained directly from a single-shot experiment and not indirectly inferred from a known model or from several runs of the experiment as in the case were the reward is, say, a target fidelity or a success probability; (iii) we will not only be concerned about finding near-optimal detectors but also, importantly, on the actual on-line success rate of the agents as measured by the cumulative reward.


We tackle the problem in three stages of increasing complexity: first, in the model-aware setting, where the outcome probability function of the receiver is known, we find the optimal action sequence by solving the Bellman equation via dynamic programming~\cite{Bellman2003,Geremia2004}; second, in the model-free setting, where the receiver is completely unknown, we apply Watkins' Q-Learning~\cite{Watkins1989,Sutton2018}, a standard RL method whose update rule approximates the optimal Bellman equation; third, in the model-free setting we study the trade-off between exploiting potential optimal strategies and exploring new ones, by applying two state-of-the-art methods adapted from bandit theory~\cite{banditbook,Auer2002,Russo2018, qlprovably}, thus enhancing the learning speed or accuracy of our agents. With these methods, in the model-aware setting we are able to compute numerically the optimal success probability and set of actions for several control rounds. Moreover, in the model-free setting we are able to construct agents that surpass the performance of the best Gaussian receiver~\cite{Takeoka2008} after $\sim 3 \cdot 10^{2}$ episodes and its cumulative reward after $\sim 10^{3}$ episodes and then attain near-optimal performance ($>97\%$ optimal) after $\sim 10^{5}$ episodes, searching on a parameter-space of size $\sim 3\cdot10^3$.
Our results provide a flexible and comprehensive ensemble of methods both in the model-aware and model-free settings that enable the on-line optimization of small quantum devices and the benchmarking of their performance. Furthermore, the methods we propose can be enhanced by the use of deep-learning techniques~\cite{Silver2016}, which would allow their application to more complex problems and devices, e.g., multi-state QSD and the study of generalization performance.

The article is organized as follows. In Sec.~\ref{sec:prelim} we introduce our QSD problem, the receiver architecture and the target function for a RL agent controlling the receiver. In Sec.~\ref{sec:sequen} we present the theoretical framework of standard RL methods, introducing the state-action value function, the Bellman equation and Q-Learning. In Sec.~\ref{sec:appliRL} we describe the implementation of these methods and analyze their performance in terms of the cumulative reward.  The {bandit problem} is introduced here as a basic framework to study, quantify and optimize the real-time performance of agents over sequential learning strategies.  We analyse and compare  the performance of standard and bandit-inspired learning strategies in a variety of experimentally relevant settings. We conclude in Sec.~\ref{sec:future} by mentioning possible extensions of our work.

\tableofcontents

\section{Preliminaries}\label{sec:prelim}
We consider the discrimination of two electromagnetic signals with opposite phases, described by two coherent states of the field, $\ket{\pm\alpha}$, whose energy is proportional to $|\alpha|^{2}$. When the energy of the signals approaches zero, i.e., $|\alpha|^{2}\ll1$, quantum effects become evident and it becomes impossible to discriminate between them perfectly.

Any binary discrimination protocol is described compactly by a quantum positive-operator-valued measurement (POVM), $\mathcal{M}=\llaves{M_{0},M_{1}}$ with $M_{1,2}\geq0$ and $M_{1}+M_{2}=\mathbb{I}$. Defining the $k$-th hypothesis as $\alpha^{(k)}=(-1)^{k}\alpha$, with prior probability $p_{k}$, the probability of obtaining outcome $\hat{k}$ given that hypothesis $k$ was true is $p(\hat{k}|\alpha^{(k)})=\bra{\alpha^{(k)}}M_{\hat{k}}\ket{\alpha^{(k)}}$ and the best guess is given by the most likely hypothesis given that outcome. Thus the average success probability over all outcomes is given by
\be\begin{aligned}
P_{s}(\alpha,\mathcal{M})&=\sum_{\hat{k}=0,1}\max_{k=0,1}p(\alpha^{(k)},\hat{k})\\
&=\sum_{\hat{k}=0,1}\max_{k=0,1}p(\hat{k}|\alpha^{(k)})p_{k}.
\end{aligned}\ee
For non-orthogonal quantum states, this quantity is bounded below $1$ by the so-called Helstrom bound~\cite{helstromBOOK}, which in our case reads
\be\label{eq:hel}
\small P_{s}^{(hel)}(\alpha)=\max_{\mathcal{M}}P_{s}(\alpha,\mathcal{M})=\frac12\left(1+\sqrt{1-e^{-4\abs{\alpha}^{2}}}\right),
\ee
where the optimization is carried out over all two-outcome POVMs;
note that the Helstrom probability tends to $1/2$ for $|\alpha|\rightarrow0$, i.e., the states become indistinguishable at very low energies. The optimal Helstrom measurement that attains Eq.~\eqref{eq:hel} is a difficult projection on a superposition of $\ket{\pm\alpha}$, i.e., a Schr\"odinger-cat-like state, which cannot be realized with simple linear-optical operations~\cite{Takeoka2005}. Quite surprisingly, Dolinar~\cite{Dolinar1973} showed that Eq.~\eqref{eq:hel} can be asymptotically attained by continuous-time control of a displacement operator; his receiver has since been extended to the discrete-time scenario by Takeoka et al.~\cite{Takeoka2005}. Nevertheless, the practical implementation of these receivers still proves demanding at present~\cite{Cook2007,Becerra2013a}, due to various experimental limitations. Moreover, in a general communication scenario, the states will be transferred through a noisy channel and could be subject to various kinds of noise, e.g., loss, thermal noise and phase diffusion~\cite{Muller2015,DiMario2019}.

\begin{figure}[t]
    \centering
    \includegraphics[trim={2cm 7cm 5cm 7cm}, scale=.25]{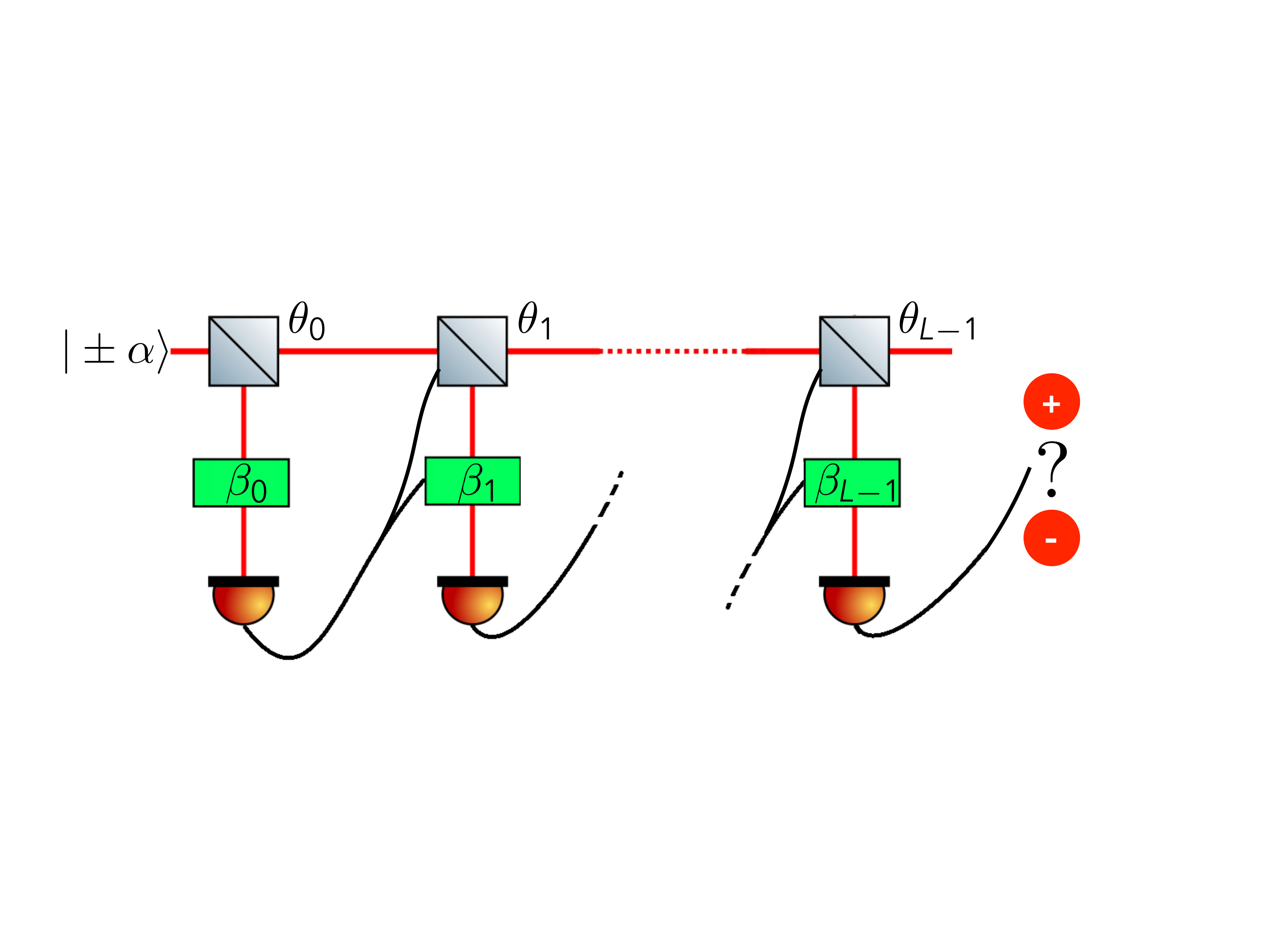}
    \caption{We depict the experimental setup of the receiver considered. For $L \rightarrow \infty$ one gets Dolinar reciever.}
    \label{fig:setup}
\end{figure}
Based on these premises, we aim to construct a model-free RL agent that, without any knowledge of the problem at hand nor of the receiver setup, learns to tune the receiver's parameters in order to maximize its success probability. In this way, when placed in a real-life situation, the agent will be able to train and optimize the receiver for the specific experimental conditions it encounters in real time. The receiver we consider comprises passive linear optics, photodetectors and classical feed-forward, structured into successive processing layers $\ell=0,\cdots,L$, as depicted in Fig.~\ref{fig:setup}; this receiver is known to attain the Helstrom probability in the limit $L\rightarrow\infty$~\cite{Takeoka2005}. For each layer $\ell<L$, the following operations are applied:
\begin{enumerate}
\item\label{ops:start} The input signal $\ket{\alpha}$ is split on a beamsplitter (BS) of transmissivity $\theta$, effectively extracting a fraction $1-\theta$ of the energy for detection. The BS transforms the input signal and vacuum states as
\be
\ket{\alpha}\ket{0}\mapsto\sket{\alpha\sqrt{\theta}}_{\rm tr}\sket{\alpha\sqrt{1-\theta}}_{\rm ref},
\ee
where the added phase of the second mode has been corrected via a proper phase-shift, not shown in the figure.
\item The reflected part of the signal undergoes a displacement operation $D(\beta)$, realizable via interference with a strong coherent signal on a small-reflectivity BS, not shown in the figure. The resulting state is $\sket{\tilde{\alpha}(\beta,\theta)}=\sket{\alpha\sqrt{1-\theta}+\beta}$.
\item\label{ops:end} The displaced signal is measured via a on/off photodetector, which detects no photon, i.e., outcome $o_{\ell+1} = 0$, with conditional probability
\be\label{eq:singLayProb}\small
p(o_{\ell+1}=0|\alpha,(\beta,\theta))=\abs{\braket{0}{\tilde{\alpha}(\beta,\theta)}}^{2}=e^{-\abs{\tilde{\alpha}(\beta,\theta)}^{2}},
\ee
and detects one or more photons, i.e., $o_{\ell+1}=1$, with probability $1-p(o_{\ell+1}=0|\alpha,(\beta,\theta))$.
\item The transmitted part of the signal enters layer $\ell+1$.
\end{enumerate}
Finally, the last processing layer $\ell=L$ consists in elaborating a guess $\hat{k}$ of the true hypothesis $k$, based on previous measurement outcomes and parameter choices.

For an initial coherent state $\ket\alpha$, the input state at the $\ell$-th layer is $\ket{\alpha_{\ell}}=\ket{\alpha\sqrt{\theta_{0}\cdots\theta_{\ell-1}}}$.
Since the experimenter can use all the past history $h_{\ell}=(a_{0},o_{1},\cdots,a_{\ell-1},o_{\ell})$, with $h_{0}=\emptyset$, to decide the next value of $(\beta,\theta)$ and the final guess, the total set of parameters over all possible histories is of exponential size in $L$. We label them compactly as $a_\ell(h_{\ell})=(\beta_{h_{\ell}},\theta_{h_{\ell}})$ and $a_{L}(h_{L})=\hat{k}$, omitting the label $\ell$ or the dependence on $h_{\ell}$ when it is clear from the context.
Hence, the average success probability of this strategy over all possible outcomes' sequences $o_{1:L}=(o_{1},\cdots,o_{L})$ can be written as
\be\label{eq:detProb}
\small P_{s}(\alpha,\{a_\ell\})=\sum_{o_{1:L}} \prod_{\ell=1}^{L}p(o_{\ell}|\alpha^{(k)},a(h_{\ell-1})) \; p_{k}\Big|_{k=a(h_{L})},
\ee
where $\{a_{\ell}\}$ is the total set of actions over all histories and we have written the conditional probability of the sequence of outcomes $o_{1:L}$ factors as a product of single-layer conditional probabilities, \eqref{eq:singLayProb}.

In the model-aware setting, this expression can be optimized using dynamic programming, as we show in Sec.~\ref{ssec:dp}, finding the set of optimal parameters $\llaves{a^{*}_\ell}$ for any given $\alpha$ and $L$:
\begin{equation}
  \llaves{a^{*}_\ell} = \underset{\llaves{a_{\ell}}}{\text{arg max }} P_s(\alpha, \llaves{a_\ell})
\end{equation}
As a shorthand we denote the optimal success probability (over the available actions) as
\begin{equation}\label{eq:OptSuc}
  P_{*}^{(L)}(\alpha) = \underset{\llaves{a_{\ell}}}{\text{ max }} P_s(\alpha, \llaves{a_\ell}),
\end{equation}
and omit the label L when it is clear from the context.

In the model-free setting instead, the agent has no knowledge of Eqs.~(\ref{eq:singLayProb}, \ref{eq:detProb}), so it must resort to exploring the set of possible parameters and sample from the probability of \eqref{eq:detProb} during several runs of the experiment to discover an optimal choice of parameters and guessing rule by trial and error.


\section{Sequential decision-making}\label{sec:sequen}
The framework of RL is based on the interaction between an agent and an environment during several episodes ~\cite{Sutton2018}.
At each time-step $\ell=0,\cdots,L$ of each episode $t=1,\cdots,T$, the agent observes the environment in a \textit{state} $s_{\ell}^{(t)}\in\cS$ and chooses an \textit{action} $a_\ell^{(t)}\in\cA$; as a consequence, the agent enjoys a \textit{reward} $r_{\ell+1}^{(t)}\in\cR$ and observes a new state of the environment, $s_{\ell+1}^{(t)}\in\cS$; where $\cS$, $\cA$ and $\cR$ stand for the sets of states, actions and rewards the agent may experience.


The environment is usually modeled to be Markovian: its dynamics is completely determined by the last time-step via the \textit{transition function} $\tau(s',r|s,a)$, i.e., the conditional probability of ending up in a state $s'$ and conferring a reward $r$, given that the previous state was $s$ and the agent took an action $a$; the next future states accessible from $s$ are thus restricted to $\cS(s) = \{ s' : \tau(s'| s) \neq 0 \} \subseteq \cS$. The agent does not have control of nor access to the transition function, but it will influence the dynamics of the environment by choosing actions according to an interaction \textit{policy} $\pi(a|s)$, i.e., the conditional probability of performing an action $a$ when the observed environment's state is $s$; hence the available actions at a given state may be restricted to a subset $\cA(s) \subseteq \cA$. This setting is usually known as a Markov decision process (MDP).

Informally, the agent's objective is to interact with the environment through an \textit{optimal policy} $\pi^{*}$, such that the total reward acquired during an episode is as high as possible. To achieve this goal, a \textit{value function} is assigned to each state and optimized over all possible policies, 
as further explained below in Sec.~\ref{ssec:valbell}.

The Markov assumption is justified whenever the agent's observations provide a complete description of the state of the environment $s_{\ell}$. However, in general this is not the case, and the agent has only access to \textit{partial observations} $o_{\ell}\in\cO$ at each time-step. Such observations would not allow to determine the dynamics even if $\tau$ was known, and they are generated from the current state and the previous action. In RL literature this is called a partially-observable MDP (POMDP) and developing methods to solve it efficiently constitutes an active area of research~\cite{Singh1994,Mnih2013,Shani2013,Egorov2015,Zhu2018}; usually, the problem is tackled by first reducing it to an effective MDP. The most straightforward approach is to define an effective state that contains all the past history of observations and actions up to a given time-step, i.e., $h_{\ell}=(a_0,o_{1},\cdots,a_{\ell-1},o_\ell)$. In this way, the dynamics observed by the agent can always be described by an effective MDP with transition function $\tau(h',r|h,a)$, which is unknown to the agent and determined by the underlying environmental transition function. Clearly, this approach makes the problem intractable for large time-steps, since the number of states increases exponentially in $L$. In the model-aware setting, one can condense the history in a belief distribution over the states, $b_{o'}(s')=p(s'|o',a,b_{o})$, i.e., the probability that the environment is in state $s'$ given the current observation $o'$, the previous action $a$ and the belief at the previous time-step $b_{o}$. The belief has an initial value $b(s)$ equal to the prior distribution over the initial states and at each time-step it is updated using Bayes' rule.
In the following parts of this Section we will introduce several tools for MDPs, which can be immediately adapted to POMDPs by exchanging the unknown state with the history $h$ or the belief $b_o(s)$.

\subsection{Value functions and the Bellman equation}\label{ssec:valbell}
The agent's objective is to acquire as much reward as possible during an episode. As a matter of fact this strongly depends on the agent's policy. At the end of episode $t$, in which a sequence of $L$ tuples 
$\llaves{(s_\ell, a_\ell, r_{\ell+1})}_{\ell = 0}^{L}$ has been experienced (with $s_{L+1}$ a \textit{terminal state}, and $L$ generally varying among different episodes), the agent's performance after each time-step $\ell$ can be evaluated using the so-called return,
\be \label{eq:returnGT}
G_{\ell}^{(t)}=\sum_{i=0}^{L-\ell}\gamma^i r_{i+\ell+1}^{(t)},
\ee
i.e., the weighted sum of rewards obtained at all future time-steps, with a \textit{discount factor} $\gamma\in(0,1]$, which weighs more the rewards that are closer in the future. Note that for infinite-horizon MDPs, i.e., $L\rightarrow\infty$, it must hold $\gamma<1$ to ensure that $G_\ell$ remains finite.

By introducing the return, it is straightforward to assign a value to a state $s$ for a given interaction policy $\pi$, via the so-called \textit{value function}:
\begin{equation}\label{eq:vFunc}
v_\pi(s)=\E{\pi}{G_\ell | s_\ell=s},
\end{equation}
which is the expected return over all possible trajectories that start from state $s$, take actions according to policy $\pi$ and whose dynamics is governed by $\tau$. In other words, the value function measures how convenient it is to visit state $s$ when policy $\pi$ is being followed. Note that this quantity is completely determined by the future trajectories accessible from $s$ and hence its dependence on the time-step $\ell$ can have at most the effect of restricting the set of states on which $v_{\pi}(s)$ is supported at that time; we keep this dependence implicit unless otherwise stated. By writing explicitly the expected value for the first future time-step in Eq.~\eqref{eq:vFunc} and then applying the definition of $v$ recursively, it is easy to show that the state-value function satisfies, for any policy, the following Bellman equation~\cite{Bellman2003}:
\be\label{eq:vBell}\small
v_\pi(s) = \sum_{a\in\cA,s'\in\cS,r\in\cR}\tau(s',r|s,a)\pi(a|s)\left(r+\gamma v_\pi(s')\right).
\ee
This equation relates the value of a state $s$ with that of its nearest neighbours $s'$, which can be reached with a single action from $s$, and with the corresponding reward obtained by performing such action.

The problem can then be solved by finding an optimal policy $\pi^{*}$, namely one that maximizes the state-value function for each $s$ (also called optimal value function), and thus satisfies the optimal Bellman equation:
\be\begin{aligned}\label{eq:vBellOp}
v^{*}(s)&:=v_{\pi^*}(s)=\max_\pi v_\pi(s)\\
&=\max_{a\in\cA}\sum_{s'\in\cS,r\in\cR}\tau(s',r|s,a)\left(r+\gamma v^{*}(s')\right).
\end{aligned}\ee
Similarly, one can define the state-action value function (or Q-function) as the expected return when starting from state $s$ and performing action $a$:
\be
Q_\pi(s,a) = \E{\pi}{G_\ell | s_\ell=s, a_\ell=a},
\ee
which is related to the state-value function by $v_\pi(s)=\sum_{a\in\cA} \pi(a|s) Q_\pi(s,a)$. Thus, the optimal policy $\pi^*$ can also be obtained by maximizing the Q-function, with a corresponding optimal Bellman equation
{\small\begin{eqnarray}\label{eq:qBellOp}
Q^{*}(s,a)&&:=Q_{\pi^*}(s,a)=\max_\pi Q_\pi(s,a)\\
&&=\sum_{s'\in\cS,r\in\cR}\tau(s',r|s,a)(r+\gamma \max_{a'\in\cA}Q^{*}(s',a')).\nonumber
\end{eqnarray}}%


\subsection{Model-aware learning}\label{sec:dp}
In the model-aware setting, where the transition function is known, an optimal policy can be efficiently found off-line by optimizing the corresponding state-value function. This problem, known as planning~\cite{Sutton2018}, can be solved for finite-horizon MDPs via dynamic programming methods. We follow the method introduced by Bellman~\cite{Bellman2003}, which makes use of the recursive relation of Eq.~\eqref{eq:vBellOp} to find the optimal policy step by step; for this we assume that every episode deterministically ends at a fixed time-step $L$, and denote by $v^{*}_{\ell}(s)$ the optimal value function of state $s$ at time-step $\ell$ (problems in which $L$ is not fixed can be solved, for instance, via value iteration \cite{Sutton2018}).

Since the optimal policy consists in taking the best possible action from any given state, it can be constructed by concatenation of the optimal policies at each time-step: we start by solving Eq.~\eqref{eq:vBellOp} at the last time-step,
\be\label{eq:dpL}
v^{*}_{L}(s)=\max_{a\in\cA(s_{L})}\sum_{r\in\cR}\tau(r|s,a)r,
\ee
where we have omitted the terminal state $s_{L+1}$ and used the fact that $v_{L+1}(s)=0$.
The solution to Eq.~\eqref{eq:dpL} provides the optimal action at step $L-1$ for each $s$ and the optimal value function $v^{*}_{L}(s)$.
Then we plug the latter into the optimal Bellman equation for the previous time-step, which in turn can be solved to obtain the optimal action and value function $v^{*}_{L-1}(s)$. By repeating this procedure iteratively for each time-step $\ell=L,\cdots,0$, we can obtain the optimal sequence of actions and value functions for any state at any time-step.

\subsection{Q-learning}\label{ssec:qlintr}
In the model-free setting, the agent not only has to find an optimal policy by exploiting valuable actions, but also needs to characterize the environment in the first place by exploring possibly advantageous configurations. This is known as the exploration-exploitation trade-off and lies at the core of RL problems ~\cite{Sutton2018}. In this setting, the Q-function is quite helpful since it associates a value to the transitions determined by taking action $a$ from state $s$ and following policy $\pi$ thereafter.

Q-learning was first proposed by Watkins \cite{Watkins1989}, and it is often used as a basis for more advanced RL algorithms~\cite{Mnih2013, ddpg}.
It is based on the observation that any Bellman operator, i.e., the operator describing the evolution of a value function as in Eqs.~(\ref{eq:vBell},\ref{eq:vBellOp},\ref{eq:qBellOp}), is contractive~\cite{algsrl}. This implies that, under repeated applications of a Bellman operator, any value function converges to a fixed point, which by construction satisfies the corresponding Bellman equation. Thus, in order to find $Q^{*}(s,a)$, Q-learning turns the optimal Bellman equation for $Q$, Eq.~\eqref{eq:qBellOp}, into an update rule for $\hat{Q}(s_{\ell},a_{\ell})$, i.e., the $Q$-function's estimate available to the agent at a given time-step $\ell$ of any episode $t=1,\cdots,T$.

After an interaction step $s_{\ell}\rightarrow a_{\ell}\rightarrow r_{\ell+1}\rightarrow s_{\ell+1}$ is experienced, the update rule for the $Q$-estimate is
{\small\be\begin{aligned} \label{eq:QLUPDATERULE}
\hat{Q}(s_{\ell}, a_{\ell}) & \leftarrow (1-\lambda_t(s_{\ell},a_{\ell}))\hat{Q}(s_{\ell}, a_{\ell})\\
&+ \lambda_t(s_{\ell},a_{\ell}) \left(r_{\ell+1}  + \gamma \max_{a'\in\cA(s_{\ell+1})}\hat{Q}(s_{\ell+1}, a')\right),
\end{aligned}\ee}%
where $\lambda_t(s,a)$ is the learning rate, which depends on the number of times the state-action pair $(s_{\ell},a_{\ell})$ has been visited.
Note that in order to do the update at each time-step $\ell$, it is only necessary to enjoy the next immediate reward $r_{\ell+1}$ and observe the next state $s_{\ell+1}$; this method thereby allows an on-line learning of the MDP. A pseudo-code of the algorithm is given in Algorithm 1.

\begin{algorithm}[h]\label{alg:ql}
  \DontPrintSemicolon
  \SetAlgoNoEnd
  \SetKwInOut{Input}{input}\SetKwInOut{Output}{output}
  \Input{$\hat{Q}(s,a)$ \texttt{arbitrarly initialized} $\forall s \in \cS \; \forall a \in \cA (s)$; \texttt{learning rates} $\lambda_t(s_\ell,a_\ell) \in (0, 1]$, $\epsilon > 0$}
  \Output{$\hat{Q}(s,a) \sim Q^{*}(s,a)$}\;
  \For{ $t$ in $1$ ... $T$  }{
  $\; \; \;$ \texttt{initialize} $ s_0  \; \; \;$ \;
  $\; \; \;$ \For{\texttt{step }$\ell$ \texttt{in episode} $t$}{$\; \; \; \; \; \;$\texttt{take action }$a_\ell$ \texttt{according to }$\pi$ (e.g. $\epsilon$\texttt{-greedy})\; $\; \; \; \; \; \;$\texttt{observe reward }$r_{\ell+1}$\texttt{ and next state }$s_{\ell+1}$ \; $\; \; \; \; \; \; $\texttt{update }$\hat{Q}(s_\ell, a_\ell)$ \texttt{according to: }\; $\; \; \; \; \; \; \hat{Q}(s_{\ell}, a_{\ell}) \leftarrow \hat{Q}(s_{\ell}, a_{\ell})  + \lambda (s_\ell, a_{\ell}) [r_{\ell+1} +$\; $\; \; \; \; \; \;  \gamma \max_{a'} \hat{Q}(s_{\ell+1}, a') - \hat{Q}(s_{\ell}, a_{\ell}) ]$\;$\; \; \; \; \; \;$\If{$s_{\ell+1}$\texttt{ is terminal state}}{$\; \; \; \; \; \; \; \; \; \;$ break}$\; \; \; \; \; \;$\Else{}{$\; \; \; \; \; \; \; \; \; \;s_{\ell} \leftarrow s_{\ell+1}$}
  }
}
\caption{Q-learning pseudo-code.}
\end{algorithm}

After a large number $n$ of iterations of the update rule Eq. \eqref{eq:QLUPDATERULE} for all state-action couples, the convergence of the $Q$-estimate to the optimal $Q$-function is guaranteed by two general conditions on the learning rate (also known as Robinson conditions) ~\cite{Watkins1989,Sutton2018}:
{\small\be\begin{aligned}\label{eq:qLConv}
\hat{Q}(s,a)&\underset{k\rightarrow\infty}{\rightarrow}Q^{*}(s,a)\;\,\forall s\in\cS, a\in\cA(s)\\
&\text{ iff }\sum_{t(s,a)}\lambda_{t}(s,a)=\infty,\;\sum_{t(s,a)}\lambda_{t}(s,a)^{2}<\infty,
\end{aligned}\ee}%
where the sums are taken over all interactions at which a given state-action couple is visited.
Once the optimal $Q$-function is obtained, an optimal deterministic policy can be constructed by ``going greedy'' with respect to it, i.e., $\pi^{*}(a|s) = \delta(a,\argmax_{a\in\cA} Q^{*}(s,a))$ for all $s\in\cS$, where $\delta(x,y)$ is a Kronecker delta.

In RL literatue, Q-learning is classified as an off-policy method~\cite{Sutton2018}, meaning that it learns the state-action values of a \textit{target policy} - in this case the optimal policy - by taking actions according to an \textit{interaction policy}, generally differing from the first one. The standard Q-learning method commits to an $\epsilon$-greedy interaction policy, where with probability $\epsilon$ the agent chooses a random action and otherwise it chooses the greedy action that maximizes the current $Q$-estimate. However, as we will see below, more general strategies can be considered.


\section{Model-free reinforcement learning of discrimination strategies}\label{sec:appliRL}
In the following we frame the optimization of the receiver described in Sec. \ref{sec:prelim} into a RL context, in which an agent has to attain optimal reward-per-episode rate (success rate) by departing from a situation of complete ignorance of the experiment.
For simplicity, we assume that the sender and receiver have a shared reference frame, so that we can take the states and displacements to be real, $\alpha,\beta\in\mathbb{R}$, without loss of generality. 

The notation introduced in Sec.~\ref{sec:prelim} is straightforward to translate into the RL notation of Sec.~\ref{sec:sequen}:
\begin{itemize}
\item Each episode $t$ corresponds to an independent discrimination experiment, with a new default state $s_{0}=\alpha^{(k)}$ sampled from $p_{k}$, $k\in\{0,1\}$; we set $\gamma=1$ since the process has finite horizon;
\item Each episode consists of $L+1$ time-step $\ell=0,\cdots,L$, corresponding to the $L$ detection layers followed by the final guessing stage;
\item The possible states of the environment at time-steps $\ell$ are $s_{\ell}=\alpha^{(k)}_{\ell}$, i.e., the transmitted part of $s_{0}$ at that layer; 
\item The agent is not aware of the state $s_{\ell}$, in particular it does not know which hypothesis is true, but it can observe the measurement outcome $o_{\ell}$, $0 < \ell \leq L$;
\item The actions $a_\ell$ available at time-step $0 \leq \ell <L$ are the displacements $\beta_{\ell}$ and BS parameters $\theta_{\ell}$ available at that layer, conditioned on the history of observations $h_{\ell}$, while at the last step they constitute the guess, $a(h_{L})=\hat{k}\in\{0,1\}$;
\item The reward $r\in\{0,1\}$ is non-zero only at the end of the episode and provided that the guess is correct, hence the transition function for the environment is
\begin{align}
\label{eq:sTr}&\tau(\alpha^{(k')}_{\ell+1}|\alpha^{(k)}_\ell,a_{\ell})=\delta(k',k)\;\;\forall\ell\leq L,\\
&\tau(r_{L+1}|\alpha^{(k)}_L,a_L)=\delta(r_{L+1},1)\delta(a_L,k),
\end{align}
were we omitted the trivial reward for $\ell\leq L$.
\end{itemize}

\subsection{Benchmarking the success probability via dynamic programming}\label{ssec:dp}
In order to benchmark the performance of our RL agent, we start by considering a model-aware POMDP where the agent knows the amplitude $|\alpha|$ of the optical signals and the transition probabilities; its task is to optimize the success probability of Eq. \eqref{eq:detProb}.
In this case, we define $b_{\ell}(k) = p(\alpha^{(k)}_{\ell}|o_{\ell},a_{\ell-1},b_{\ell-1})$ to be the belief distribution over the states $\alpha^{(k)}$, after performing action $a_{\ell-1}$ and observing $o_{\ell}$, with prior belief $b_{\ell-1}(k)$. The initial value of the prior is $b_0(k)=p_k$ and its update rule follows Bayes' theorem:
\begin{eqnarray}\label{eq:belUp}
b_{\ell}(k) = \frac{p(\alpha^{(k)}_{\ell}|o_{\ell}, a_{\ell-1}) \;b_{\ell-1}(k)}{\sum_{k} p(\alpha^{(k)}_{\ell}|o_{\ell}, a_{\ell-1}) \; b_{\ell-1}(k)}.
\end{eqnarray}
%

The optimal Bellman equation, Eq.~\eqref{eq:dpL}, for the state-value function of this POMDP at step $L$ reads
\be\label{eq:opBellLQSD}
v^{*}_{L}(b_{L})= \max_{k} b_{L}(k),
\ee
which means that at the last step, if the final belief distribution over the states is known, the best guess is the hypothesis with maximum likelihood.
The optimal Bellman equation at step $\ell<L$ instead reads
{\small\be\label{eq:opBellEllQSD}
v^{*}_{\ell}(b_{\ell})
=\max_{a\in\cA(s_{\ell})}\sum_{o_{\ell+1}\in\cO}\sum_k p(o_{\ell+1}|\alpha^{(k)}_{\ell},a_{\ell})b_{\ell}(k)v^{*}_{\ell+1}(b_{\ell+1}).
\ee}%
\begin{figure}[t!]
    \centering
    \includegraphics[width=.45\textwidth,trim={1cm 1cm 5cm 5cm},clip]{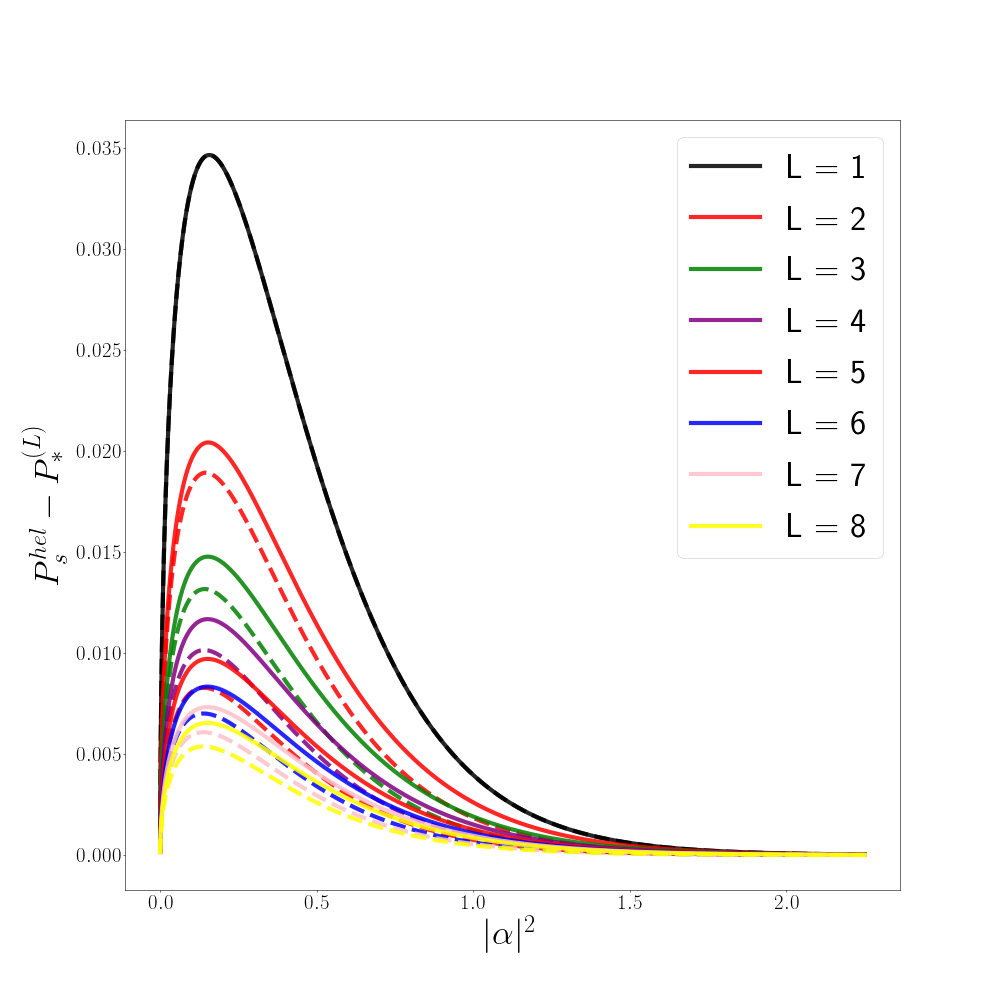}
    \caption{We show the difference between the best probability of success attainable for a fix $L$ and the optimal probability of success in discriminating two coherent states. The results were obtained by dynamic programming, exaplained in Sec.\ref{sec:dp}. Solid lines correspond to fixed attenuations $\theta_{\ell}$ such that the input state of each layer has equal amplitude $\alpha_{\ell}^{(k)}=\frac{\alpha^{(k)}}{\sqrt{L-1}}$ for all $\ell$, whereas dashed lines correspond to the probability of success optimized also on conditional attenuations. } 
    \label{fig:dp}
\end{figure}%
These equations can be solved iteratively by inserting the solution $v^{*}_{\ell+1}(b_{\ell+1})$ into the equation for $v^{*}_{\ell}(b_{\ell})$, starting with $\ell=L-1$ and $v^{*}_{L}(b_{L})$ found in Eq.~\eqref{eq:opBellLQSD}. Note that, since $v^{*}_{\ell+1}(b_{\ell+1})$ is computed for a discrete set of values of the belief distribution, these cannot always coincide with the values, determined by Eq.~\eqref{eq:belUp}, needed to solve Eq.~\eqref{eq:opBellEllQSD} and hence we use interpolation methods to obtain them.

The maximum success probability attainable with the receiver is equal to the optimal value function at step $\ell=0$, since the latter corresponds to the expected reward starting from the initial belief distribution:
\be\label{eq:maxLDP}
v^{*}_{0}(b)=\mathbb{E}_{\pi^{*}}[r_{L+1}|b_{0}(k)]=P^{(L)}_*(\alpha)
\ee
as can be seen by repeated applications of Eqs.~(\ref{eq:belUp},\ref{eq:opBellEllQSD}) and detailed in Appendix~\ref{app:optVal}

In Fig.~\ref{fig:dp} we show the optimal success probability obtained with this method as a function of $|\alpha|^{2}$ and for up to $L=8$ layers. We also show the results at fixed $\theta_{\ell}$ such that the input state of each layer has equal amplitude $\alpha_{\ell}^{(k)}=\frac{\alpha^{(k)}}{\sqrt{L-1}}$ for all $\ell$ (dashed lines). We observe that for all $L\geq2$ there is an energy threshold above which allowing adaptive optimization of the attenuations gives a better success probability than adding one layer with fixed attenuations.

\subsection{Learning a near-optimal receiver via Q-learning}\label{ssec:qLRes}
In this Subsection we present the results obtained by a RL agent based on Q-Learning with $\epsilon$-greedy interaction policy. The experiment is modelled as a POMDP, which can be reduced to an effective MDP for the history of observations and actions $h_{\ell}$, as explained in Sec.~\ref{sec:sequen}. The update rule for the $Q$-function is given by Eq.~\eqref{eq:QLUPDATERULE} with $s\rightarrow h$ and learning rates $\lambda_t(h,a)=N_{t}(h,a)^{-1}$, the inverse of the number of times a state-action pair has been visited. This choice guarantees convergence as per Eq.~\eqref{eq:qLConv}.
As for dynamic programming, the optimal value of the success probability of Eq.~\eqref{eq:detProb} is obtained by maximizing the optimal $Q$-function at time-step $\ell=0$:
\be\small\label{eq:opBell0}
\max_{a_{0}}Q^{*}(a_{0})=\max_{a_{0}}\mathbb{E}_{\pi^{*}}[r_{L+1}|a_{0}]= P_*^{(L)}(\alpha),
\ee
where we have omitted the default history state $h_{0}=\emptyset$; this is detailed in Appendix~\ref{app:optVal}. At variance with the model-aware case, where the guessing rule was obtained straightforwardly from the Bellman equation at the last time-step, the optimization of Eq.~\eqref{eq:opBell0} includes a non-trivial search for the optimal guessing rule, determined by the optimal $Q$-function.
\begin{figure}[t!]
    \centering
    \includegraphics[width=.48\textwidth,trim={6cm 0 6cm 5cm},clip]{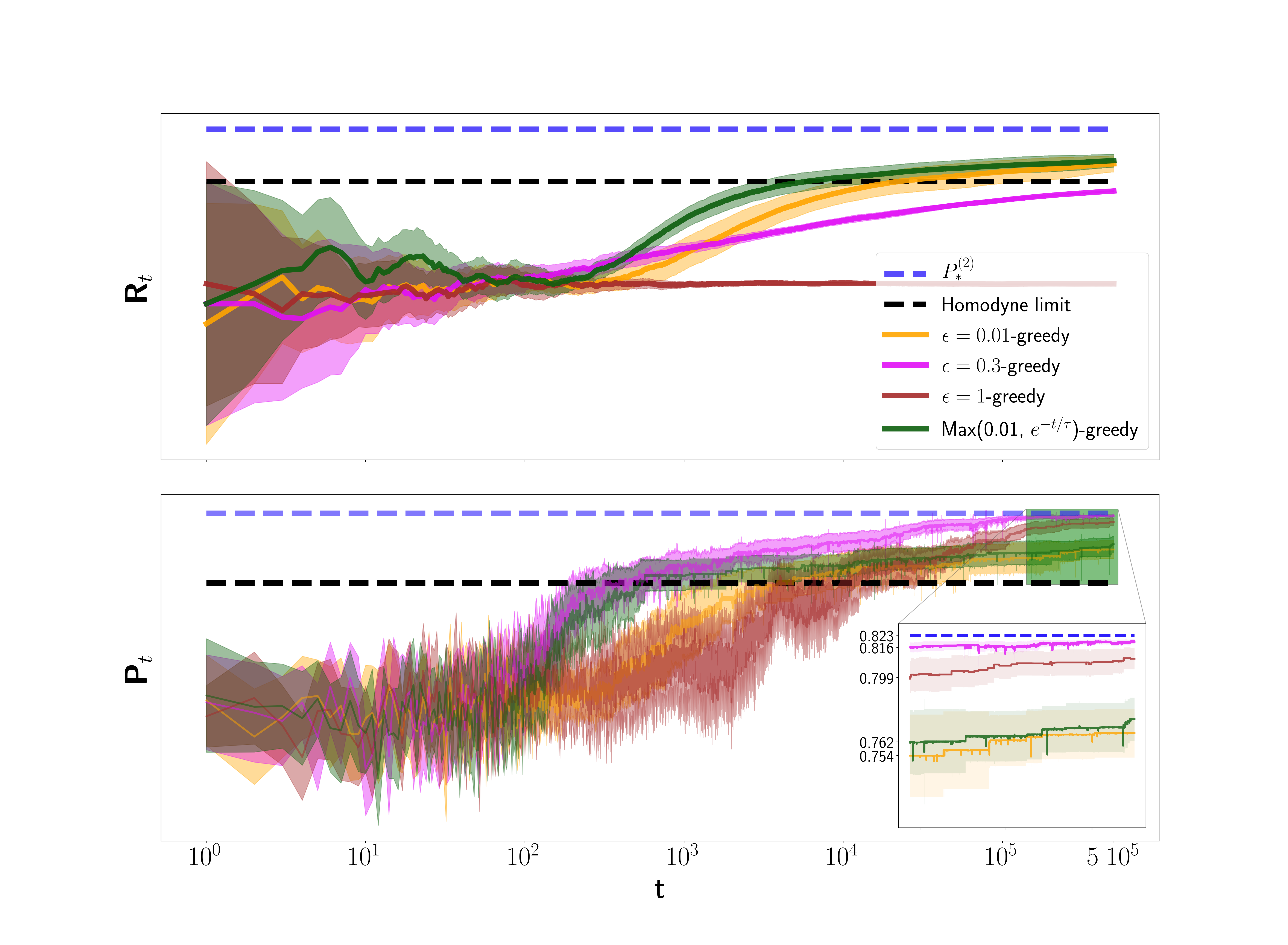}
    \caption{We benchmark traditional Q-learning with different \textit{schedules} on $\epsilon$ as the episode number increases. The figures of merit are averaged over $A=48$ agents and show the corresponding uncertainty region. }
    \label{fig:compEps}
\end{figure}

We evaluate the performance of the agent using two figures of merit as a function of the number of episodes elapsed so far, $t$: (i) the cumulative return per episode (also called average reward per episode)
\be
\mathbf{R}_{t}=\frac1t\sum_{i=1}^{t}G^{(i)}_{0}=\frac1t\sum_{i=1}^{t}r_{L+1}^{(i)},
\ee
where $r_{L+1}^{(i)}=\{1,0\}$ stands for the correctness of the guess made at episode $i$, and (ii) the success probability of the best actions according to the agent, at the current episode,
\be
\mathbf{P}_{t}=P_{s}(\alpha,\{a_{\ell}^{(t)*}\}),
\ee
where the best actions $\{ a_{\ell}^{(t)*} \}$ at episode $t$ are obtained by going greedy with respect to the current $Q$-estimate, i.e.,
{\small
\begin{eqnarray}
  a_{0}^{(t)*}(h_{0})&=&\argmax_{a \in \cA(h_0)} \hat{Q}(h_0,a) \rightarrow h_1^{*} = (o_1, a_0^{(t)*})  \\  \nonumber
  a_{1}^{(t)*}(h^{*}_{1})&=&\argmax_{a \in \cA(h^{*}_1)} \hat{Q}(h^{*}_1,a) \rightarrow h_2^{*} = \big( a_0^{(t)*}, o_1, a_1^{(t)*}(h_1^{*}), o_2 \big) \\ \nonumber
  &...& \\ \nonumber
  a_{L}^{(t)*}(h_L^{*}) &=&\argmax_{a \in \cA(h^{*}_L)} \hat{Q}(h^{*}_L,a).  \nonumber
\end{eqnarray}
}%
The first figure of merit, $\Rt$, is usually employed to describe the learning process in RL and it evaluates the success rate obtained by the agent so far.
On the other hand, the second figure of merit, $\Pt$, is standard in QSD and in our context it evaluates the best strategy discovered by the agent so far.

As $t\rightarrow\infty$, for a \textit{good} learner it is expected that $\Rt\rightarrow\Pt$, i.e. with enough learning time the average reward should tend to the success probability for the best actions found by the agent, which in turn should converge to the optimal success probability $P_{*}^{(L)}$. Therefore, the learner is not only expected to find a good discrimination strategy, but to also follow it: the interaction policy should tend to the optimal policy. This feature is captured by the evolution of $\Rt$ over different episodes: a good learner is asked to obtain as much reward as possible \textit{during} the learning process.

Here and in the rest of the article, we restrict to $L=2$ interaction layers and fix the attenuation coefficients to give equal amplitude at each layer, since the difference in success probability is small compared with the additional number of episodes one would need to learn it, as shown in Sec.~\ref{ssec:dp}. We choose a resolution of $21$ points for each displacement, each one ranging from -1 to 1 with step 0.1, leading to a fairly large state-action space: the agent has $3528$ possible Q-values to learn from (including the last guess). We note that each discretized displacement is an independent action or ``button'' in the eyes of the agent ---the agent is dispossessed of any notion of closeness between buttons corresponding to similar values of $\beta$.
 As the behaviour of the RL agent strongly depends on the actions chosen at early episodes, we averaged the learning curves over 24 agents. Our results are compared with: (i) the maximum success probability attainable with this number of layers and discretization of displacements, Eq.~\eqref{eq:OptSuc}, and (ii) the success probability attainable via a standard homodyne measurement, which is optimal among Gaussian receivers~\cite{Takeoka2008}.

In Fig.~\ref{fig:compEps} we plot these two figures of merit for $Q$-learning agents with three different $\epsilon$-greedy interaction policies: (i) a completely random one, i.e., $\epsilon=1$, (ii) a 0.3-greedy one, i.e., $\epsilon=0.3$, and (iii) a dynamic one (exp-greedy) that becomes exponentially greedier as time passes, i.e., $\epsilon (t)=\max\{e^{- \frac{t}{\tau}},0.01\}$; this standard choice assures that at initial episodes the agent favours exploration, whereas at $t = \tau \log \frac{1}{\epsilon_0}$ the agent's behaviour collapses to an $\epsilon_0$-greedy policy.

In the first place we note in Fig.~\ref{fig:compEps} that a fully random search over the action space ($1$-greedy policy) leads to the extremely poor cumulative reward per episode of $\Rt\approx 1/2$, even for long times, which is expected because a random guess (last action) leads to $P_s(\alpha, \llaves{a_\ell})=1/2$. Instead, since all the actions will be sampled enough times for the agent to learn the optimal policy, $\Pt$  will converge to optimal value at long enough episode number. Nevertheless, if the action space is large, the fully random strategy will require a large number of episodes to explore each action a significant number of times, and for moderate times a $\epsilon$-greedy strategy might reach a better strategy. Indeed, Fig.~\ref{fig:compEps} shows that the $0.3$-greedy policy has at all episodes a higher $\Pt$ than the 1-greedy one, being $99\% $ the optimal success probability $P_*^{(L=2)}$ at episode $t=10^{5}$. Of course, for $0.3$-greedy policy the agent collects many more rewards (actual correct guesses) than for the $1$-greedy  but it is still limited to $\Rt\approx 0.7 P_{*}^{(L=2)}$. In order to reach a better exploration-exploitation trade-off, it is customary to consider an episode-dependent $\epsilon$, e.g. $\epsilon (t)=\max\{e^{-\frac{t}{\tau}},\epsilon_0\}$.
Fig.~\ref{fig:compEps} shows the results for this tunable interaction policy with $\tau = 2 \cdot  10^{2}$ and $\epsilon_0 = 0.01$. This allows the agent's $\Rt$ to surpass the homodyne limit at about episode $\sim 5 \cdot 10^{3}$ (which is comparable with the size of the action space), while at later times the performance converges to that of the $0.01$-greedy policy. Notice finally that 0.3-greedy discovers a strategy whose $\Pt$ surpasses the homodyne limit at episode $\sim 3 \cdot 10^{2}$.

Our numerical results show that standard Q-learning successfully trains agents that surpass the homodyne limit of optical detection and discover strategies whose error rate is comparable with that of the optimal receiver. This is remarkable, especially taking into consideration that the agents are not initially trained for this task, and run in a model-free setting entirely based on the feedback they get (correct/incorrect) on their guess.
As mentioned above, although many RL schemes focus on extracting the optimal policy from the agent (as measured e.g. by $\Pt$), our central figure of merit, $\Rt$, captures the real performance of the agent, and can actually be assessed by the agent itself. It is hence important to design strategies that not only aim at finding the optimal policy within an episode, but also maximize the cumulative reward per episode, reaching $\Rt \rightarrow P_*^{(L)}$ as fast as possible.  We have seen above an example of such strategy (the exponential greedy) and in the next sections we will study more advanced ones. For this purpose we will first study a simplified setting, called the multi-armed bandit problem, where the intra-episode dynamics is trivial, and the main focus is drawn on how to optimize the inter-episode learning strategy.
In passing, we introduce some theoretical tools in order to study the bandits' learning curves, which are a cornerstone to tackle more challenging situations such as learning optimal policies over a MDP.

\subsection{The multi-armed bandit problem}\label{ssec:bandits}
Multi-armed bandit problems are MDPs with a single default state at which the agent faces a fixed set of actions $a \in \cA$, each one leading to a reward $r \in \cR$ with an unknown probability $\tau(r|a)$. After action $a$ is performed, the reward $r$ is enjoyed and the episode finishes: the return defined in Eq.~\eqref{eq:returnGT} becomes $r$. The situation models a gambler at a row of slot machines that has to decide which arms to pull, how many times to pull each one and in which order, with the aim of maximizing the earned rewards.


The bandit problem is an ideal framework to highlight the aforementioned crucial difference between learning strategies that accomplish the main goal of identifying the optimal policy after a given number of episodes, and more refined strategies that also procure high \emph{de facto} cumulative returns during the learning process.
This is why bandit problems are very relevant in real-life applications where the final success probability is not the only figure of merit, as for example in clinical trials \cite{Thompson1933} where one needs to find the right compromise between advancing in the search of the best treatment while effectively treating current patients.

The general traits of the cumulative return per episode $\Rt$ in Fig.~\ref{fig:compEps} could inspire several ways of quantifying the performance of the learning agent, e.g., (a) the onset episode at which $\Rt$ starts exceeding the random policy; (b) the transient episode at which $\Rt$ reaches a given fraction of the optimal success probability;  (c) the learning speed as quantified by the slope of $\Rt$ after the onset episode; (d) the learning speed at which
$\Rt$ converges to the optimal success probability. Unfortunately, very little is known about these or alternative ways to characterize the learning curves. Nonetheless, bandit theory provides us with a framework were some of these notions can be defined and rigorously studied. In particular, bandit theory defines the so-called \textit{cumulative regret}:
\begin{equation}
  \mathcal{L}_t =  \mathbb{E} \sum_{k=1}^{t} \Big( Q(a^{*}) - Q(a^{(k)}) \Big) = t \;\big(Q(a^{*}) - \mathbb{E}\, \Rt \big),
  \label{eq:CumRegegret}
\end{equation}
where $ \mathbb{E}$ indicates the expected value with respect to different agents following the same strategy, and $a^{(t)}$ is the action actually taken by an agent at episode $t$. The cumulative regret is closely related to the (expected) cumulative return per episode $\Rt$, and quantifies the price to pay, or loss, for taking actions different from the optimal one ($a^{*}$). In other words, it quantifies the difference in earnings of the agent with respect to those of a model-aware super-agent.
One of the most fundamental results in bandit theory is the
Lai-Robbins bound \cite{Lai1985} for the asymptotic expected cumulative regret:
\begin{equation}\label{eq:RLBOUND}
 \mathcal{L}_t \underset{t \gg 1}{\gtrsim} \log t \Big( \sum_{a \in \cA\backslash \{a^{*}\}} \frac{\Delta_a}{\text{KL}(a||*)} + o(1) \Big):=C_{\mathrm{LR}} \log t
\end{equation}
with $\Delta_a = Q(a^{*}) - Q(a)$ and $\text{KL}(a||*)$ the Kullback-Leibler divergence between the reward distributions $\tau(r|a) $ and $ \tau(r|a^{*})$.
Recalling the definition in Eq.~\eqref{eq:CumRegegret} we note that the Lai-Robbins bound characterizes
 the learning curve in the asymptotic regime, indeed the average return over agents, $\mathbb{E}\, \Rt$ can approach the optimal value not faster than  \mbox{$\mathbb{E}\, \Rt \lesssim Q(a^{*})-C_{\mathrm{LR}} \frac{\log t}{t}$}. Let us now briefly present some possible bandit strategies in light of this ultimate performance bound.

The most straightforward policy to use is the $\epsilon$-greedy (already introduced in Sec.\ref{ssec:qlintr}), as described in Algorithm 2.
\begin{algorithm}[H]\label{alg:ql}
  \DontPrintSemicolon
  \SetAlgoNoEnd
  \SetKwInOut{Input}{input}\SetKwInOut{Output}{output}
  \Input{$\hat{Q}(a)$ arbitrarily initialized and learning rates $\lambda_{t}(a) \in (0, 1]$ $\forall a \in \cA\;$, $\epsilon \in (0,1]$}
  \For{ $t$ in $1$ ... $T$  }{
  $\; \; \; \; \; \;$\texttt{generate a random number j}\;
  $\; \; \; \; \; \;$\If{\texttt{j} $\leq \epsilon$}{$\; \; \; \; \; \; \; \; \; \;$\texttt{choose} $a$ \texttt{at random}}$\; \; \; \; \; \;$\Else{}{$\; \; \; \; \; \; \; \; \; \;$\texttt{choose} $a =\argmax_{a \in \cA} \hat{Q}(a)$}\;
  $\; \; \; \; \; \;$\texttt{observe} $r$\;
  $\; \; \; \; \; \;$\texttt{update} $\hat{Q}$: \;$\; \; \; \;  \; \; \; \; \; \;\; \; \hat{Q}(a) \leftarrow \hat{Q}(a) + \lambda_{t}(a) [r - \hat{Q}(a)]$
  }
\caption{$\epsilon$-greedy for bandit problems.}
\end{algorithm}
Note that by choosing the learning rates $\lambda_t(a)$ to be the inverse of the number of times action $a$ was visited up to time $t$, then
\begin{equation}
   \hat{Q}(a)  \underset{t \rightarrow \infty}{\rightarrow} \sum_{r \in \cR} r \; \tau(r|a) \; = Q(a) \; \; \forall a \in \cA.
   \label{eq:Qband}
\end{equation}
As stressed in Sec.\ref{ssec:qLRes}, the $\epsilon$-greedy policy never attains the optimal success rate,  $\mathbb{E} \Rt<Q(a^{*})$ for all $t$, since at every episode there is a finite probability $\epsilon$ that the agent performs a suboptimal action. For this reason the expected cumulative regret grows linearly with $t$. It is then clear that there is room for improvement before reaching the
logarithmic scaling of the ultimate limit of Eq.~\eqref{eq:RLBOUND}. We will present two strategies, one based on Upper Confidence Bounds (UCB) ~\cite{Lai1985,Agrawal1995,Auer2002} and the other called Thompson sampling (TS)~\cite{Thompson1933,Thompson1935,Scott2010,Russo2018}, which substantially improve the performance of $\epsilon$-greedy and may even attain the Lai-Robbins ultimate bound \cite{TSoptimal, Auer2002}.
\begin{figure}
    \centering
    \includegraphics[width=0.5\textwidth]{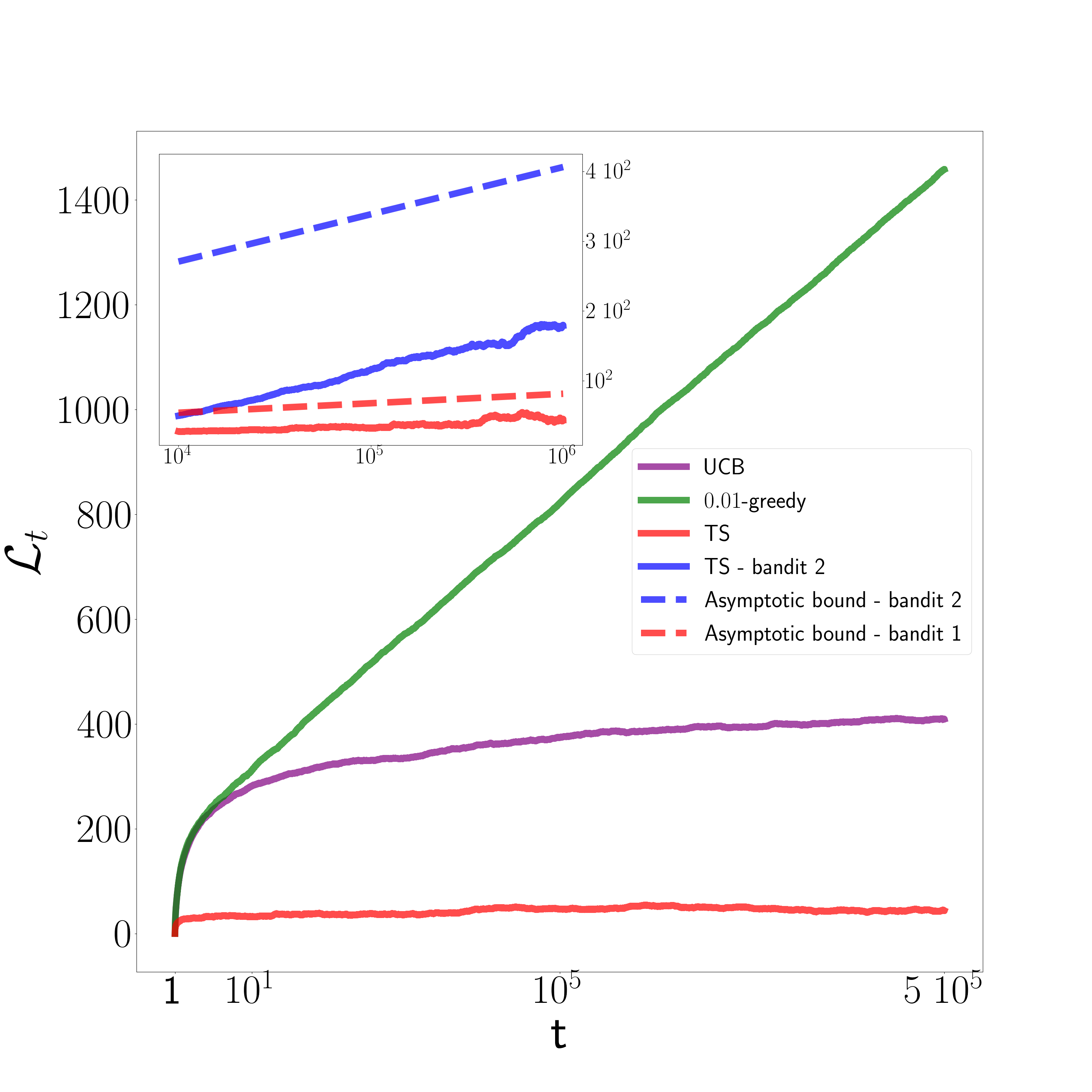}
    \caption{We show the evolution of the cumulative regret for three different strategies. In this case, \textit{bandit problem 1}, the displacements considered are $\beta \in \{0, -\alpha, \beta^{*}\}$, for $\alpha = 0.4$ and $\beta^{*} = -0.74$. All curves are averaged  over $10^{3} $ agents. Furthermore, we compare the asymptotic behaviour of TS, studying \textit{bandit problem 2}, with $\beta \in \{-\alpha, \beta^{*}, -1.5 \alpha\}$} 
    \label{fig:bandfig}
\end{figure}

In UCB, the agent keeps a record of the number of times each action $a$ was selected up to episode $t$, which we denote as $N_{t}(a)$. Hoeffding's inequality bounds the probability that the $\hat{Q}(a)$ underestimates the true value of ${Q}(a)$  by more than $\varepsilon(t)>0$, as
\begin{equation}\label{eq:ucbeq}\small
{\rm Pr}[ \hat{Q}(a)<Q(a)-{\varepsilon}(t) ] \leq e^{- 2 N_{t}(a) {\varepsilon}(t)^{2}} =: \mathcal{P}(t),
\end{equation}
where here and in the rest of this section we assume that $r\in [0,1]$.
Then, for  $N_{t}(a)>0$, the upper confidence bound, defined as
\be
 \label{eq:ucbeq2}
{\rm ucb}_{t}(a):=\hat{Q}(a)+\varepsilon(t) = \hat{Q}(a)+\sqrt{\frac{-\log \mathcal{P}(t)}{2 N_{t}(a)}} ,
\ee
represents an upper bound to the true value ${Q}(a)$ with (high) probability $1-\mathcal{P}(t)$.
This value is used to compare and choose among the different actions, i.e.
$a = \argmax_{a \in \cA} \mathrm{ucb}_t(a)$, and responds to the motto ``optimism under the face of uncertainty'': actions that have not been visited enough are assigned an ``optimistic'' value of the return and hence more chances of being picked; in addition, actions whose Q-estimate is accurate but sub-optimal will have little chances to be picked again. The functional form of $\mathcal{P}(t)$ can be tuned to balance exploration and exploitation. In particular, for the standard choice $\mathcal{P}(t) = t^{-4}$ (here called UCB-1) it can be easily seen that the expected cumulative regret follows the logarithmic scaling ~\cite{Auer2002, banditbook}. In Appendix~\ref{app:ucbTS}  we also discuss a different choice, which is known to saturate the Lai-Robbins bound, but that performs worst than UCB-1 for our setting and  moderately small number of episodes we consider.

\begin{algorithm}[h]\label{alg:ucbandit}
  \DontPrintSemicolon
  \SetAlgoNoEnd
  \SetKwInOut{Input}{input}\SetKwInOut{Output}{output}
  \Input{$\mathcal{P}(t)$, initialize $\hat{Q}(a), N(a)$ to zero $\forall a \in \cA\;$. }

  \For{$t$ in $1, ...,T$ }{
  $\; \; \; \; \; \;$ \If{ $t \leq \big| \cA \big|$}{
  $\; \; \; \; \; \; \; \; \; \; \; \;$\texttt{(choose each action once)} $a = t$}
  $\; \; \; \; \; \;$ \Else{
  $\; \; \; \; \; \; \; \; \; \; \; \;$
  \texttt{choose} $a = \argmax_{a \in \cA} ucb_t(a)$, (Eq.~\eqref{eq:ucbeq2})
  }
  $\; \; \; \; \; \;$ \texttt{observe reward} $r$ \;
  $\; \; \; \; \; \;$ \texttt{record visit: }\;
  $\; \; \; \; \; \; \; \; \; \; \; \; N_t(a)) \leftarrow N_t(a) +1$ \;
  $\; \; \; \; \; \;$ \texttt{update Q-value: } \;
  $\; \; \; \; \; \; \; \; \; \; \; \;  \hat{Q}(a) \leftarrow \hat{Q}(a) + \frac{[r - \hat{Q}(a)]}{N_t(a)}$\;

  }
\caption{UCB for bandit problems.}
\end{algorithm}



Thompson sampling (TS) departs from the standard Q-learning paradigm, which is based on keeping track and be updating the Q-table. 
Instead, TS follows a Bayesian approach and at every episode assigns a full prior distribution (not just an expectation value) for the \emph{expected reward}  $\bar{r}$ of every arm $a$, $f_{t}(\bar{r}|a) \; \forall a \in \cA$. This distribution characterizes the knowledge the agent has about the expected earnings of each arm, $Q(a)$, and at the first episode can be taken to be flat over the whole interval $[0,1]$. The policy then consists in sampling an expected reward $\bar{r}\sim f_{t-1}(\bar{r}|a)$ 
\emph{for each} possible action $a$ and choosing the action with the largest sample $\bar{r}$:
$a=\argmax_{a\in\cA}\{\bar{r}\sim f_{t}(\bar{r}|a)\})$. Finally, the distribution for the chosen action is updated according to the true reward $r$ obtained, using Bayes' theorem.


In order to avoid computationally-expensive Bayesian updates, families of distributions that are closed under the update rule are used. In the case of Bernoulli bandits, beta-distributions are employed since those are precisely their conjugate priors. That is, given
\be\label{eq:betaDistro}
f_{t}(\bar{r}|a)=\text{Beta}(\mu_t(a), \nu_t(a))\propto \bar{r}^{\mu_{t}(a)-1}(1-\bar{r})^{1-\nu_{t}(a)},
\ee
upon obtaining a reward $r$ the prior is updated to a beta distribution with parameters  $\mu_{t+1}(a)=\mu_{t}(a) + r$, $\nu_{t+1}(a)=\nu_{t}(a) + 1-r$, where at the first episode it is $\mu_{1}(a)=\nu_{1}(a)=1 \;\; \forall a \in \cA$  (flat prior). By mimicking the underlying distributions, TS gauges exploration according to the information acquired so far: if a certain action has not been sampled enough at episode $t$, its reward distribution will still be broad and, when sampled, can easily return a higher value of $\bar r$ than that obtained from other (more peaked) distributions; thereby TS will favour to explore such action. At the same time, if a sub-optimal action has been sampled enough episodes, it will be very unlikely that it is sampled again, since the corresponding prior will be highly peaked at low values. The pseudo-code of TS for Bernoulli bandits is described in Algorithm 4.

%

\begin{algorithm}[h]\label{alg:tsberbandit}
  \DontPrintSemicolon
  \SetAlgoNoEnd
  \SetKwInOut{Input}{input}\SetKwInOut{Output}{output}
  \Input{$\mu_1(a), \nu_1(a)$ initialized to one $\forall a \in \cA\;$}
\;
  \For{$t$ in $1, ...,T$ }{
  \;

  $\; \; \; \; \; \;$ \For{ $a$ \texttt{in} $\cA$}{
  $\; \; \; \; \; \; \; \; \; \; \; \;\; \; $\texttt{draw} $\bar{r}_a \texttt{ according to } \text{Beta}(\mu_t(a), \nu_t(a))$} \;
  $\; \; \; \; \; \;$ \texttt{choose} $a = \underset{a}{\argmax} \; \bar{r}_a$\;
  $\; \; \; \; \; \;$ \texttt{observe reward} $r$ \;
  $\; \; \; \; \; \;$ \texttt{update Beta distribution: }\;
  $\; \; \; \; \; \; \; \; \; \; \; \;\; \; \mu_{t+1}(a)=\mu_{t}(a) + r$ \;
  $\; \; \; \; \; \; \; \; \; \; \; \;\; \; \nu_{t+1}(a)=\nu_{t}(a) + 1-r$\;
  }
\caption{TS for Bernoulli bandit problems.}
\end{algorithm}

 Figure \ref{fig:bandfig} shows the performance of different strategies in a 3-armed bandit problem. For this purpose we have considered a simple optical receiver as described in Sec.\ref{sec:prelim}  with a single layer $L=1$. 
 Since there is only a single detector with binary outcome, we have assumed a fixed decision rule. With this, each possible displacement $\beta$ constitutes an action $a_0$ (recalling the notation used in last section) of a bandit problem. 
 The figure shows that the cumulative regret scales linearly with time for the $\epsilon$-greedy strategy, while it has a logarithmic scaling for the UCB and TS strategies. The inset shows the cumulative regret as a function of $\log t$ together with the ultimate bound  given by Lai-Robbins bound.
The achievability of this bound is hard to observe in simulations because the convergence to the asymptotic results is very slow  \cite{workTSFOLK}, i.e. sub-leading constants and terms of order $\log(\log t)$ might be important.
 Nevertheless, in the setting of Fig. \ref{fig:bandfig} we see that the leading term captured by the slope of the curve is consistent with the ultimate bound.


Let us conclude this overview of bandit theory by introducing the \textit{simple regret}, another widely used figure of merit that, as $\Pt$, quantifies how well has the agent learned so far, regardless of his actual performance:
\begin{equation}
\Lambda_t = \mathbb{E} \big(Q(a^{*}) - Q(a^{(t)*}) \big),
\end{equation}
where $a^{(t)*}$ is the agent's \emph{recommendation} of which the optimal action is at episode $t$, which, e.g., in Q-learning would be given by $a^{(t)*}=\argmax_{a}\hat{Q}(a)$. Strategies designed to minimize $\Lambda_t$ will prioritize to learn what the optimal arm to pull is, and the probability that the agent confuses the optimal arm by a sub-optimal one will be exponentially small, hence $\Lambda_t $ will converge to zero exponentially fast. Recent results  \cite{simpleRegretMunoz} show that the exploitation-exploration trade-off manifests itself  in the asymptotic scaling of the simple and cumulative regret in the sense that one imposes lower and upper-bounds on the other, and therefore optimizing one usually affects the performance of the other.

\subsection{Enhancing the agent via UCB and TS} \label{ssec:enh}
In this Subsection we consider two enhanced RL strategies, inspired by the advanced bandit methods introduced in Sec.~\ref{ssec:bandits}, and adapted to our MDP problem.

The first strategy employs the standard Q-learning update rule for the estimate $\hat{Q}$, as described in Eq.~\eqref{eq:QLUPDATERULE}, but it employs UCB to determine the interaction policy at each time-step of each episode, as described in Sec.~\ref{ssec:bandits}, with $\mathcal{P}(t)=t^{-4}$ (see Appendix~\ref{app:ucbTS} for a comparison of different choices on $\mathcal{P}(t)$). This is implemented by keeping count of the number of visits of each history-action couple up to the current episode $t$, i.e., $N_{t}(h_{\ell},a_{\ell})$, which is then used to compute an upper confidence bound, ${\rm ucb}_{t}(h_{\ell},a_{\ell})$ as in Eq.~\eqref{eq:ucbeq2}, for each action $a_{\ell}$ and history $h_{\ell}$. Finally, at time-step $\ell$ the agent chooses the greedy action w.r.t. the UCB, i.e., $a_{\ell}^{(t)}=\argmax_a{\rm ucb}_{t}(h_{\ell},a)$.

The second strategy is instead based entirely on TS, considering each action conditioned on the past history as a bandit problem and rewarding each sequence of actions that led to a successful experiment. In detail, the agent keeps a beta-distribution, Eq.~\eqref{eq:betaDistro}, of the mean reward obtainable at each time-step $\ell$ from each action $a_{\ell}$ given each possible history $h_{\ell}$, i.e., $f_{t}(\bar{r}|h_{\ell},a_{\ell})$. In order to choose a new action at time-step $\ell$ given history $h_{\ell}$, the agent samples an expected reward $\bar{r} \sim f_{t}(\bar{r}|h_{\ell},a_{\ell})$ for each $a_{\ell}$ and selects the action with the largest sample $\bar{r}$. At the end of the episode a reward is obtained as usual, and $f_{t}(\bar{r}|h_{\ell},a_{\ell})$ is updated in a Bayesian way for all the history-action couples visited at the episode. In this case, when computing $\Pt$, the best actions are chosen by going greedy w.r.t. their mean reward distribution $f_{t}(\bar{r}|h_{\ell},a_{\ell})$ ~\cite{Russo2018}.

\begin{figure}
    \centering
    \includegraphics[width=.48\textwidth,trim={6cm 0 6cm 5cm},clip]{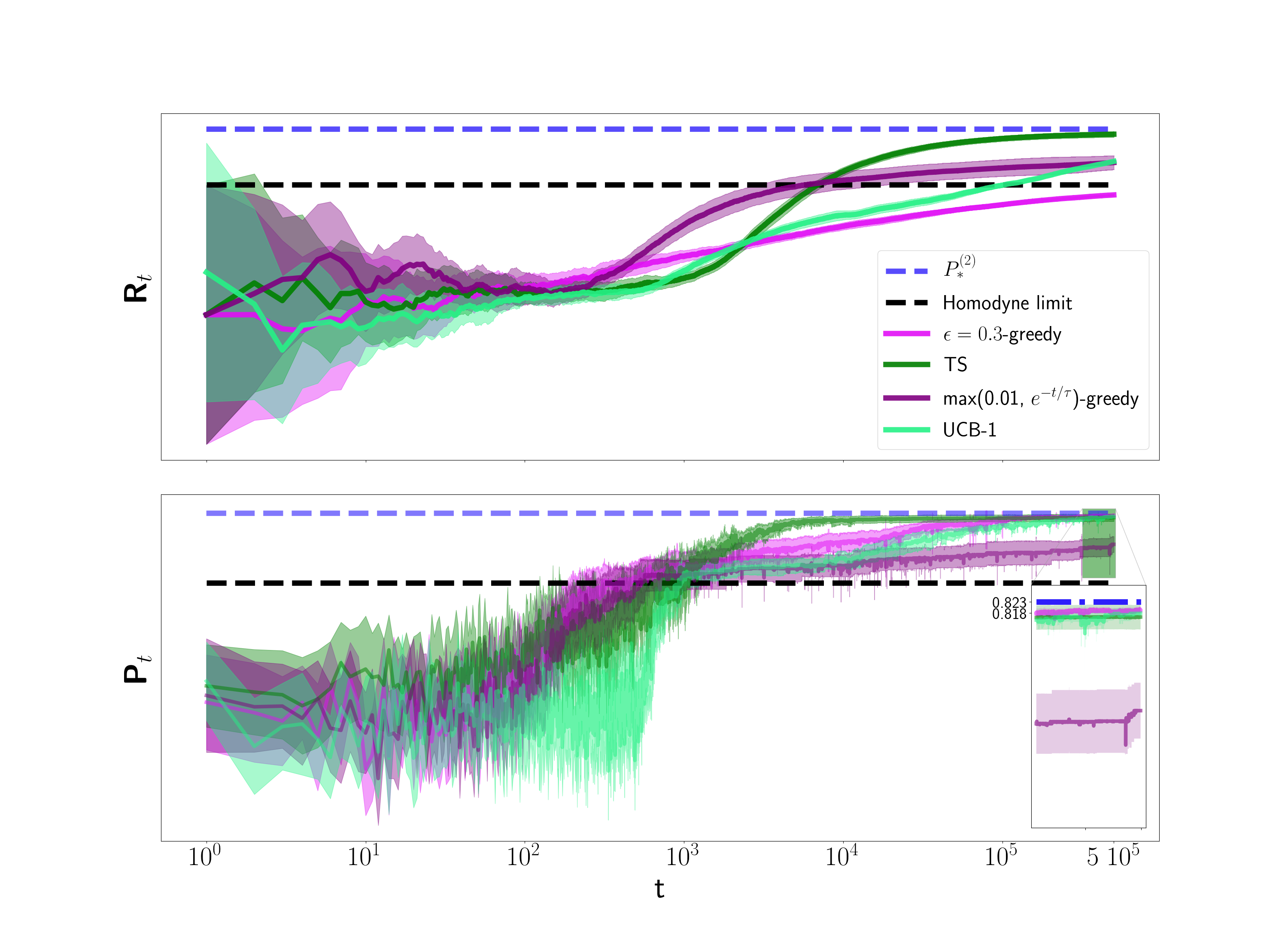}
    \caption{We show the learning curves for the enhanced Q-learning agents via bandit methods. On the upper plot we despict $\Rt$, the agent's success rate per episode, whereas on the bottom plot we despict $\Pt$, the success probability of the agent's recommended actions at episode $t$, $\llaves{a_{\ell}^{(t)*}}$. Each of the learning curves is averaged over 24 agents; the amplitude was fixed to $\alpha = 0.4$.}
    \label{fig:threemethods}
\end{figure}

In Fig.~\ref{fig:threemethods} we plot the two figures of merit $\Rt$, $\Pt$ for agents trained using these two enhanced strategies, as well as for those based on the exp-greedy and $0.3$-greedy strategies, considered in Sec.~\ref{ssec:qLRes}, which had respectively the largest final $\Rt$ and $\Pt$ out of all the analyzed strategies. We observe that UCB performs a thorough exploration of the action space and indeed it is able to attain a value of $\Pt$ close to that of $0.3$-greedy. This result comes at the price of a small $\Rt$ value, which nevertheless shows that UCB has better exploitation properties than $0.3$-greedy;  in particular it has a strikingly larger slope than the latter at long times. As for TS, we observe that this strategy attains the best $\Rt$ values, surpassing exp-greedy at intermediate times. Moreover, TS also radically improves the values of $\Pt$ w.r.t. exp-greedy and it is even able to attain the performance of the other two strategies that favour exploration. Overall, it appears that for our problem TS provides the most profitable balance of exploration and exploitation; we expect this strategy to perform worst in scenarios in which the underlying distribution probability does not belong to the family of distributions used by TS.

\begin{figure}
    \centering
    \includegraphics[width=0.23\textwidth]{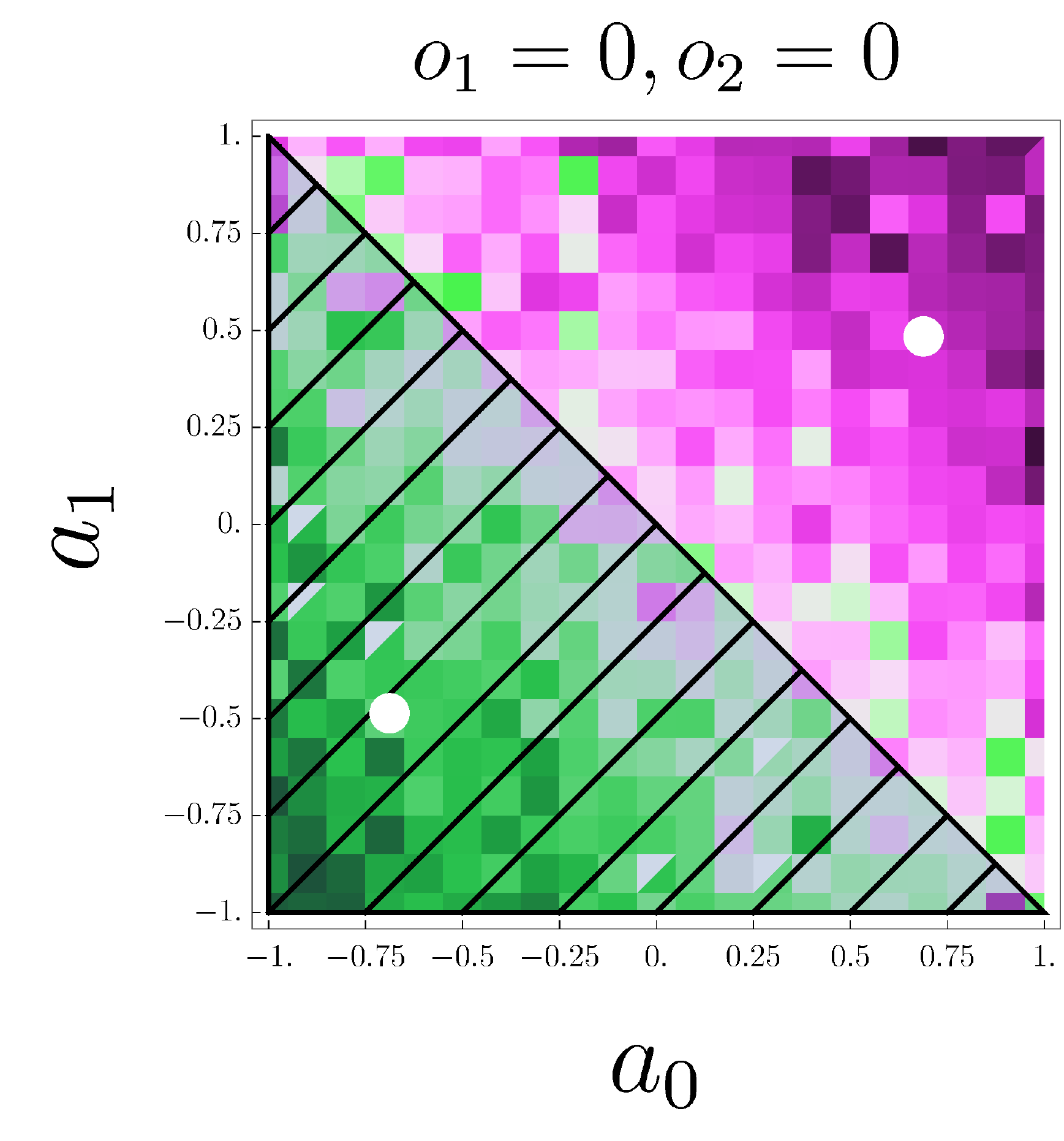}\,\includegraphics[width=0.23\textwidth]{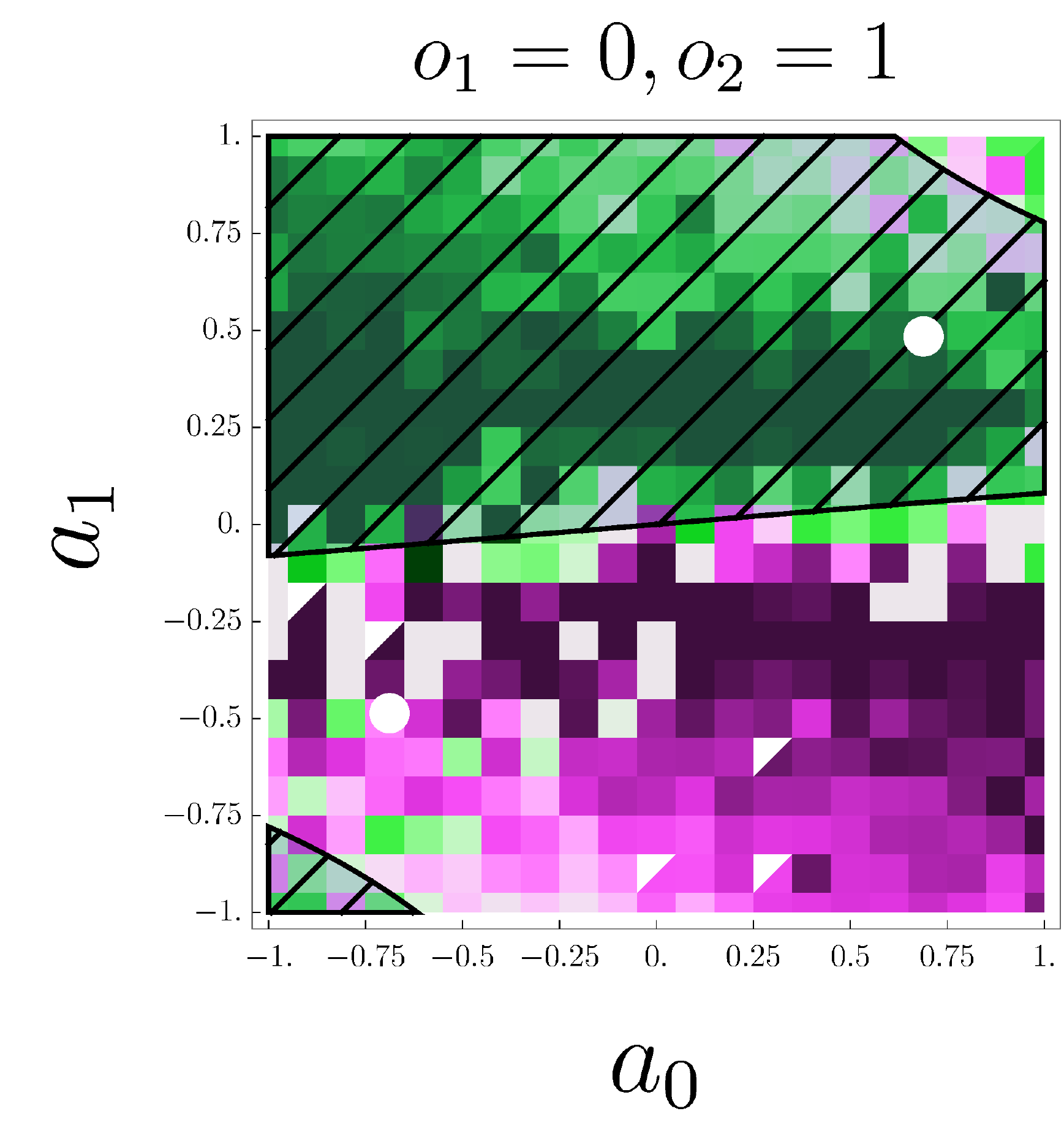}\\
        \includegraphics[width=0.23\textwidth]{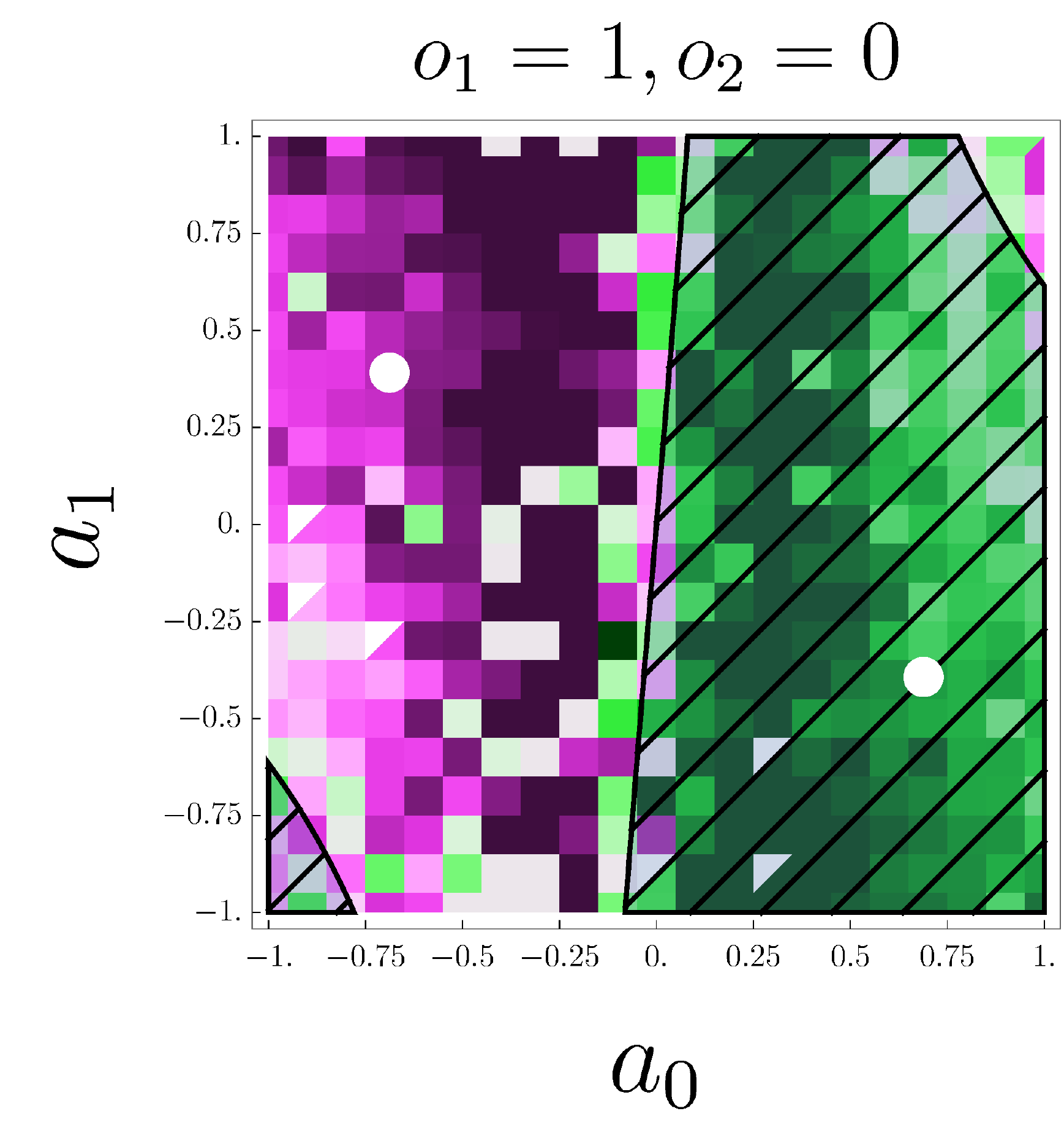}\,\includegraphics[width=0.23\textwidth]{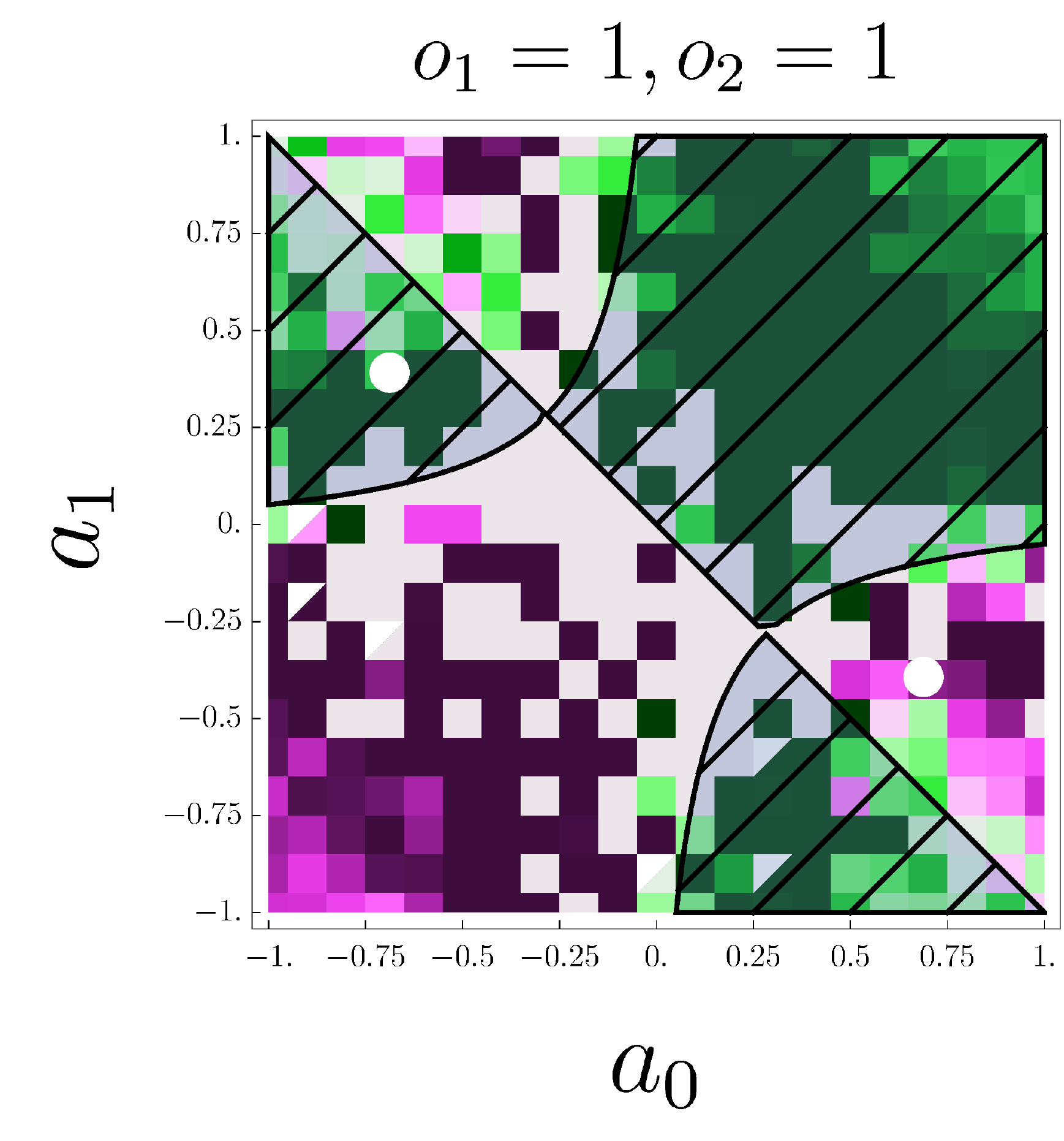}\\
       \includegraphics[width=0.46\textwidth]{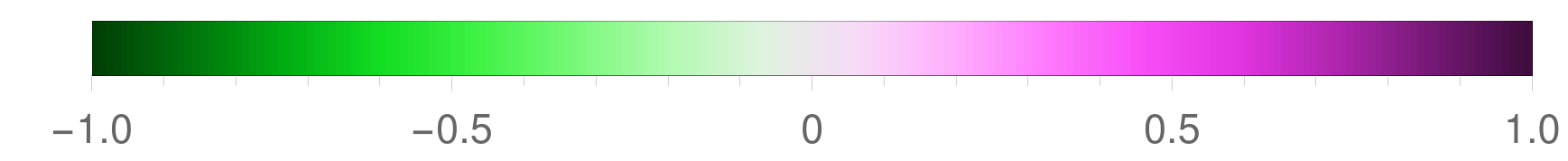}
    \caption{Density plot of the difference between the estimated $Q$-values for guessing ``plus'' and ``minus'' as a function of the displacements at the first and second layer, for each possible sequence of outcomes, with $\alpha=0.4$. The shaded areas correspond to the regions where the optimal guess, taken according to maximum-likelihood, is ``plus''. The white dots corresponds to the optimal values of the displacements for the proper discretization).
    }
    \label{fig:guess}
\end{figure}

In Fig.~\ref{fig:guess} we study the guessing rule discovered by the UCB agent at episode $t= 5 \cdot 10^{5}$. For each sequence of outcomes $o_{1}$, $o_{2}$, we plot the difference between the $Q$-values of guessing for $\ket{-\alpha}$, i.e., $a_{L}=1$, and $\ket{+\alpha}$, i.e., $a_{L}=0$, as a function of the past actions:
\be\label{eq:diff}
\hat{Q} \big( (a_0,o_{1},a_1(h_1),o_{2}),1 \big)-\hat{Q} \big( (a_0,o_{1},a_1(h_1),o_{2}),0 \big).
\ee
Note that the sign of Eq.~\eqref{eq:diff} corresponds to the agent's best guess for the true hypothesis, since the latter is obtained by going greedy towards $\hat{Q}(h_L,a_L)$, as explained in  Appendix~\ref{app:optVal}. We compare these results with the optimal guessing rule in the model-aware setting, plotting a shaded region when the maximum-likelihood guess is $\ket{\pm\alpha}$. The plot shows that UCB agents perfectly learn the guessing rule at the given resolution. Moreover, the difference between the two $Q$-values is more pronounced in the surroundings of the optimal $\beta$ values, meaning that the agents are more confident about their guess in these regions.

\begin{figure}
    \centering
    \includegraphics[width=0.5\textwidth]{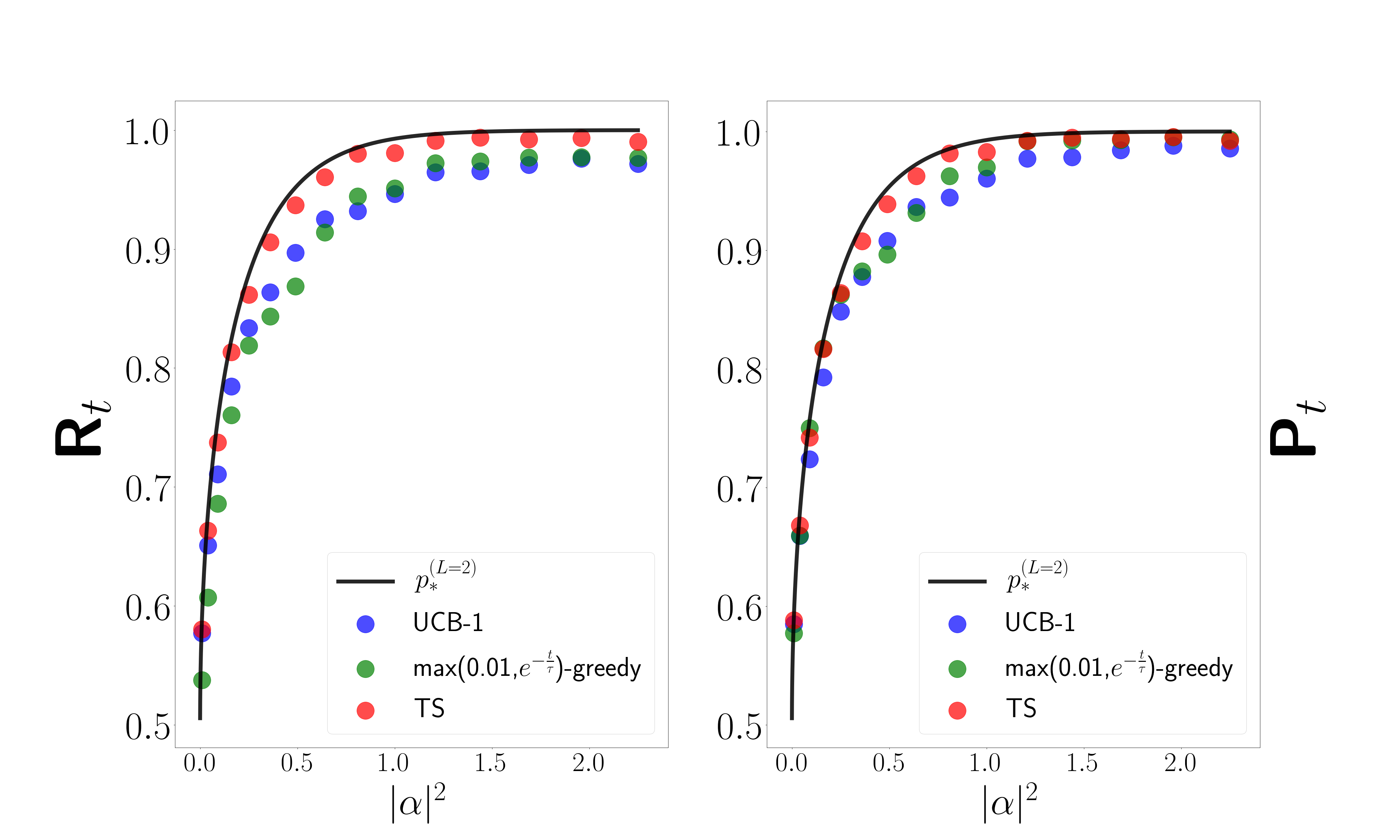}
   \caption{The performance at episode $t= 5\cdot 10^{5}$ of three different RL agents is evaluated as the energy of the coherent states increase. All data points are averaged over 24 agents.}
    \label{fig:energies}
\end{figure}

Finally, we show that RL agent's performance is independent of the coherent states' energy. For this we evaluate, for a range of different amplitudes $|\alpha|$, the values of $\Rt$ and $\Pt$ attained by different the agents at episode $t = 5 \cdot 10^{5}$, comparing them with the optimal success probability $P_*^{(L=2)}$, as can be seen in Fig.~\ref{fig:energies}.

In the following we turn to test model-free methods in realistic experimental scenarios, where the ultimate success probability is affected by the presence of noise.

\subsection{Noise robustness}
In the previous subsections we have shown that our RL agents are able to learn near-optimal discrimination strategies and --- most importantly --- exploit them in real time, employing exclusively the detectors' outcomes and the rewards at the end of each episode. Here we show that these results do not sensibly change in the presence of noise, i.e., that the same agents are able to attain near-optimal performance even when unknown errors affect the experiment and hence the learning process.

Firstly, we consider a common experimental imperfection known as dark counts: due to the presence of background noise, each photodetector of the receiver has a non-zero probability $p_{\rm dc}$ of detecting a photon even when it receives a vacuum signal. Accordingly, the conditional probability of obtaining an outcome $0$ given an input state $\ket\alpha$, Eq.~\eqref{eq:singLayProb}, is modified by a multiplicative factor $(1-p_{\rm dc})$.

\begin{figure}
    \centering
    \includegraphics[width=0.5\textwidth]{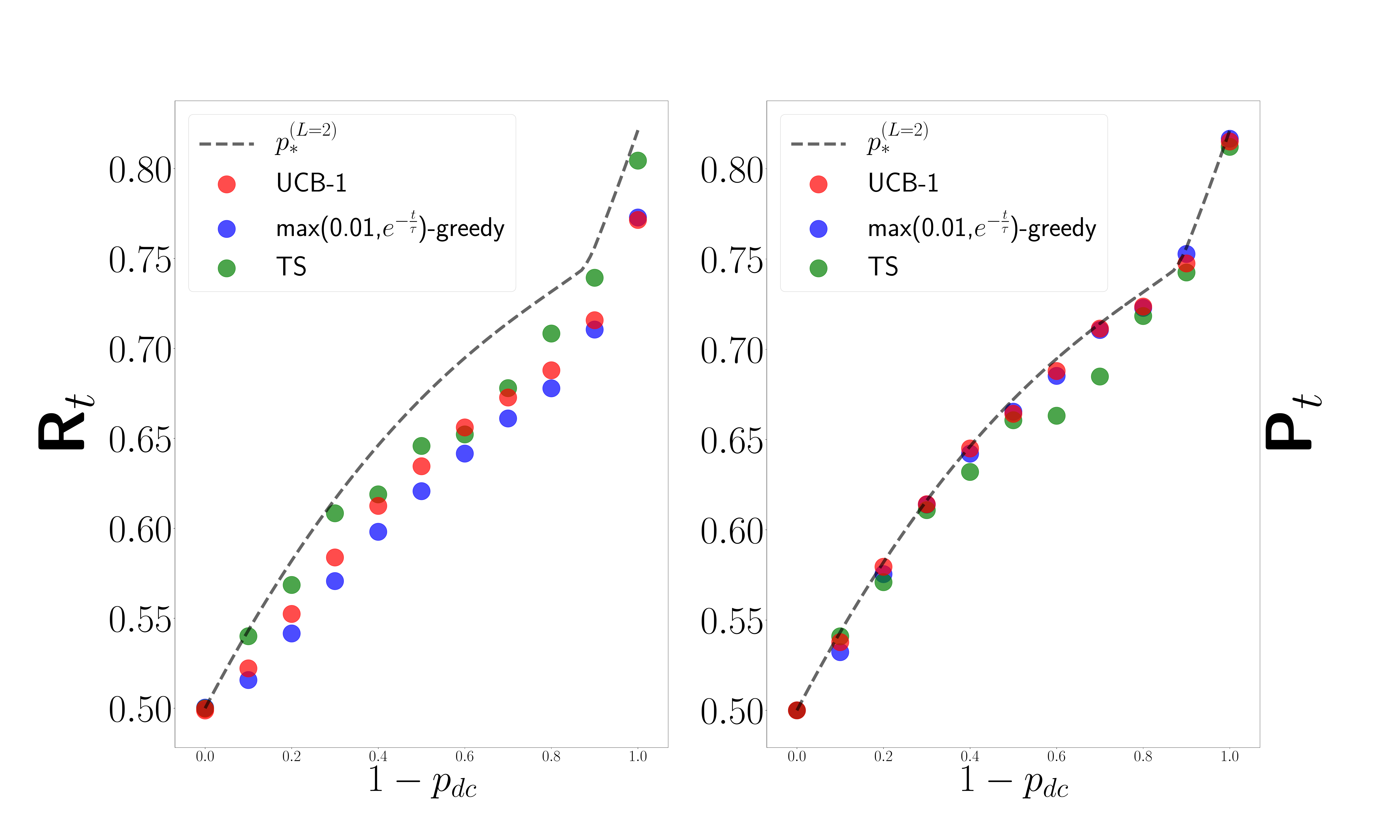}
   \caption{The performance at episode $t= 5\cdot 10^{5}$ of three different RL agents (the same considered in \ref{fig:threemethods}) is evaluated as a function of photo-detection noise. The amplitude of the coherent states is fixed to $\alpha = 0.4$; all data points are averaged over 24 agents.}
    \label{fig:dkResults}
\end{figure}

In Fig.~\ref{fig:dkResults} we plot $\Rt$ and $\Pt$ at time $t=5\cdot10^{5}$ for several RL strategies as a function of $p_{\rm dc}\in[0,1]$, along with the maximum success probability attainable by the corresponding receiver. We see that the final values of $\Pt$ are near-optimal for all values of $p_{\rm dc}$, while $\Rt$ seems to be slightly affected in an intermediate region of values of $p{\rm _{dc}}$. Since the agents operate on a completely model-free basis and the reward system has been chosen to ensure convergence of the value function to the true success probability, it can be expected that they are still be able to learn in the long term, as shown by the high values of $\Pt$ attained. However, since a dark count effectively increases the chance of (not) obtaining a reward for a (correct) wrong action, the time it takes to learn a near-optimal strategy and to start exploiting it might increase, as shown by the behaviour of $\Rt$. Note that for $p_{dc}\sim0.5$ the best guess is the random one and thus easier to learn.

Next, we consider the case where the phase of the incoming signal is flipped before arriving to the receiver, with probability $p_{f}$. In this scenario, if the agent guesses for the correct received phase, the corresponding reward will be zero since the phase initially sent was opposite than the received one. In particular, the probability of observing a string of outcomes $p(o_{1:L}|\alpha,\{a(h_{L-1})\})$ in Eq.~\eqref{eq:singLayProb} is modified such that
\begin{equation}\begin{aligned}
p(o_{1:L}|\alpha,\{a(h_{L-1})\}) &\rightarrow (1-p_f) p(o_{1:L}|\alpha,\{a(h_{L-1})\}) \\
&+ p_f p(o_{1:L}|-\alpha,\{a(h_{L-1})\})
\end{aligned}\end{equation}

\begin{figure}
    \centering
    \includegraphics[width=0.5\textwidth]{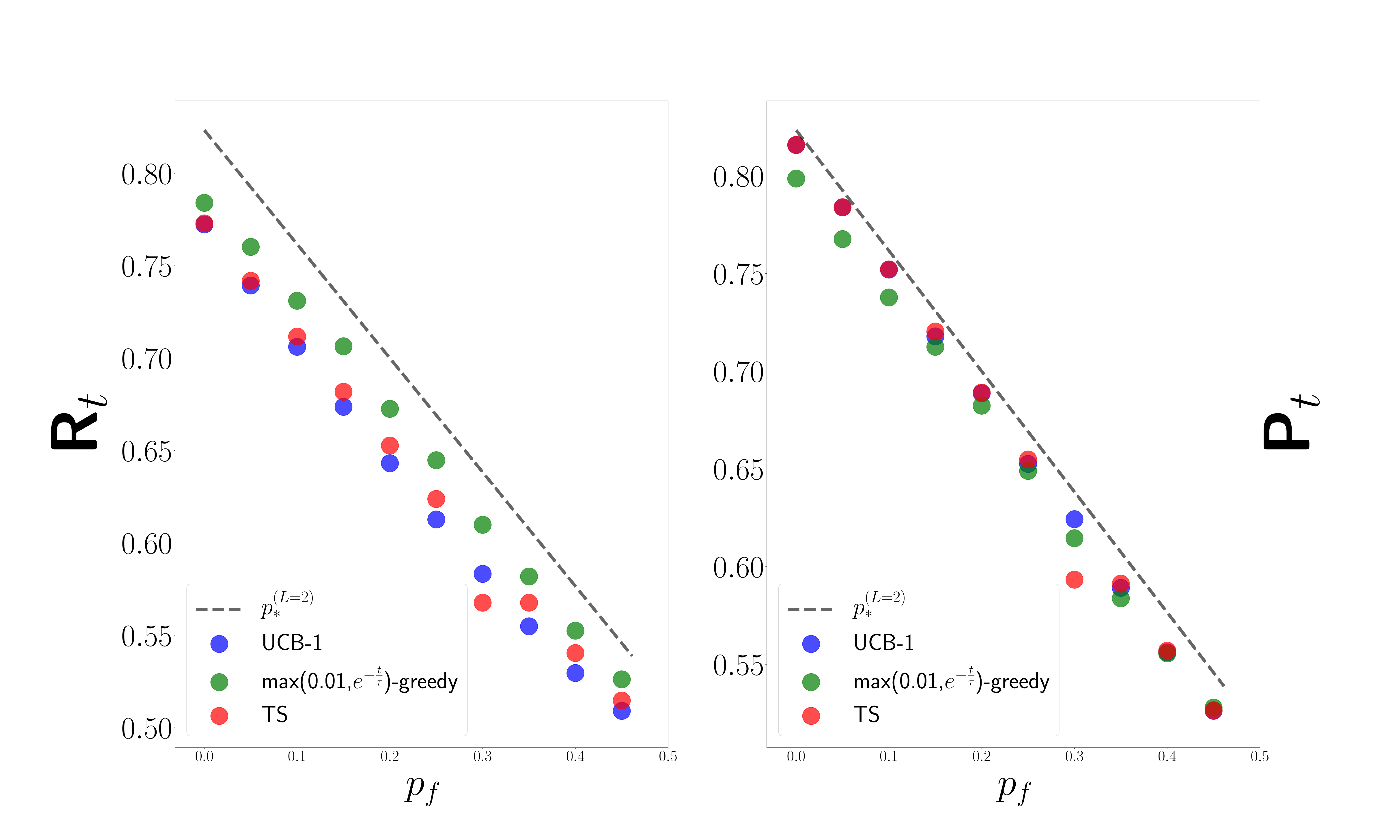}
    \caption{The performance at episode $t= 5\cdot 10^{5}$ of three different RL agents (the same considered in \ref{fig:threemethods}) is evaluated as a function of phase flipping probability before the signal arrives to the receiver. The amplitude of the coherent states is fixed to $\alpha = 0.4$; all data points are averaged over 24 agents.}
    \label{fig:dfResults}
\end{figure}

In Fig.~\ref{fig:dfResults} we despict the values of $\Rt$ and $\Pt$ attained by several agents at episode $t=5\cdot10^{5}$, as a function of $p_{f}\in[0.5,1]$, along with the maximum success probability attainable by the corresponding receiver. As in the case of dark counts, we see that for all values of $p_{f}$, the agents are able to converge to near-optimal $\Pt$ values and they exhibit very small variations in $\Rt$ as $p_{f}$ increases.

\section{Discussion and conclusions}\label{sec:future}

In this article we provided an in-depth study of RL methods for the on-line optimization of coherent-state receivers based on current technology. Such receivers are crucial for the deployment of high-data-rate long-distance communications in free space or optical fiber and they are based on the interplay of several simple quantum gates and measurements, combined to create a complex structure. The RL methods that we analyzed enable to optimize such structure based on the actual experimental conditions and limitations of the communication channel and of the receiver. Thus, they possess a high potential for increasing the flexibility and effectiveness of current receivers and provide a useful addition to the current experimental toolbox. This is even more so the case if we consider that the methods we studied are relatively simple and rely only on ``shallow'' RL techniques, i.e., they are not based on the use of neural networks, which would allow to consider larger state-action spaces, possibly at the cost of a longer training time. We expect that the use of such ``deep'' RL methods could allow to control receivers of multiple and/or multi-mode coherent states, whose best performance is still to be determined at present.

On the other hand, the characterization of the cumulative or simple regret for intermediate times, through general non-asymptotic upper and lower bounds, as well as its extension from bandits to more general MDPs is still an open and active field of research. Quantum technologies can benefit from this progress, and non-trivial quantum features might appear in more general quantum learning scenarios.

Finally, we would like to stress that the RL problem induced by real-time state discrimination is characterized by intrinsically noisy and stochastic rewards. As such, it stands out from other instances where RL has been applied to quantum physics. In particular, even when performing a good set of actions and guessing rule, an agent might still not be rewarded. This is due to two crucial factors: (i) quantum states are intrinsically indistinguishable, i.e., even the best receiver has a non-zero probability of discrimination error; (ii) our methods can be applied in real time to the experiment, hence the binary reward received for a given set of actions is not sufficient to estimate the success probability of the corresponding receiver. Still, the best among our agents are able to reach good configurations and start exploiting them in a number of experiments which is roughly sufficient to try each set of actions only once. This is a key signature of the agents' intelligent behaviour, showing that they make the most out of each reward rather than blindly trying actions at random. Hence we believe that the discrimination problem provides an interesting, rich and flexible sandbox for testing RL in quantum-physics-inspired scenarios and will constitute an interesting line of work at the intersection between these two fields.
\section{Code}
The code developed to obtain the numerical results of this research can be found at \texttt{github.com/matibilkis/marek.git}. Any suggestions, comments and even collaborations are welcome.
\section{Acknowledgments}
This project has received funding from the European Union's Horizon 2020 research and innovation programme under the Marie Sk\l odowska-Curie grant agreement No 845255. We acknowledge support by the Catalan Government for the project QuantumCAT  001-P-001644  (RIS3CAT comunitats) co-financed by the European Regional Development Fund (FEDER), the Spanish MINECO, project FIS2016-80681-P and BES-2017-081836 with the support of AEI/FEDER funds; the Generalitat de Catalunya, project CIRIT 2017-SGR-1127.

\appendix
\section{Optimal state-action values}\label{app:optVal}
In this section we verify that --- by construction --- the optimal policy leads to the maximum success probability $P^{(L)}_*(\alpha)$. 
It is assumed that $Q$ is always associated with the optimal policy $\pi^{*}$; we simplify notation by $Q = Q_{\pi^{*}}$.

At step $L$, given any history $h_{L}$, the actions available to the agent are $\hat{k} = a_L$, i.e. guessing for one of the possible phases of the coherent state. The Q-values at this time-step read as
\begin{equation}
  \begin{aligned}
  Q(h_{L}, a_L) = \mathbb{E}[G_L | h_L, a_L] &= \sum_{r_{L+1}} r_{L+1} \; p(r_{L+1}| h_L, a_L) \\
    &=  p( \alpha ^{(a_L)} | o_{1:L}; a_{0:(L-1)}) ,
  \end{aligned}
\end{equation}
with $o_{1:\ell} = \{o_1, o_2, ..., o_{\ell}\}$ the observations obtained up to the $(\ell)^{\text{\underline{th}}}$ photodetector, and $a_{0:\ell} = \{a_0, a_1 , ..., a_\ell \}$ the actions done up to step $\ell$. Recalling that the optimal action, given $h_{\ell}$, is obtained from Q as $\pi^{*}(h_{\ell}) = \argmax_{a_{\ell}} Q(h_\ell, a_\ell)$, the optimal guess $a_L^{*}$ is the one of maximum-likelihood:
\begin{equation*}
    a^{*}_L = \underset{a_L}{\argmax} \; p( \alpha ^{(k)} | o_{1:L}; a_{0:(L-1)}) \Big|_{k=a_{L}}.
\end{equation*}

By definition of the optimal policy and because optimal Bellman equation Eq.~\eqref{eq:qBellOp} holds, the optimal action to take given history $h_{L-1}$ at step $L-1$ is
{\small
\begin{equation}\begin{aligned}
  &a^{*}_{L-1} = \argmax_{a_{L-1}} \; Q(h_{L-1}, a_{L-1}) \\
  &= \argmax_{a_{L-1}} \sum_{o_L} p(o_{L}|o_{1:(L-1)}; a_{0:(L-1)}) \; \max_{a_{L}}  \; Q(h_L, a_L) \\
  &= \argmax_{a_{L-1}} \sum_{o_L} p(o_{L}|o_{1:(L-1)}; a_{0:(L-1)}) \; \max_{k} \; p( \alpha ^{(k)} | o_{1:L}; a_{0:(L-1)}) \\
  &= \argmax_{a_{L-1}} \; \sum_{o_L}  \; \frac{\max_{k} \; p(o_{1:L} | \alpha ^{(k)}; a_{0:(L-1)}) p_k}{p(o_{1:(L-1)}; a_{0:(L-1)}) }, \\
\end{aligned}\end{equation}
}%
where in the last line we have used Bayes theorem. Following this line of reasoning, we can obtain the optimal actions $a^{*}_{\ell}$ at any time-step. In particular, for $\ell=0$, by recursively applying the optimal Bellman equation (Eq.~\eqref{eq:qBellOp}) we have
{\small
\begin{eqnarray}\label{eq:OptQLL}
Q(h_0, a_0)&=& \sum_{o_1} p(o_1; a_0) \; \underset{a_1}{\max } \; Q(h_1, a_1)  \\ \nonumber
&=& \sum_{o_1} p(o_1; a_0) \; \underset{a_1}{\max } \; \sum_{o_{2}} p(o_{2}|o_{1}; a_1) \; \underset{a_2}{\max } \; Q(h_2, a_2) \\ \nonumber
& = &  \sum_{o_1} p(o_1; a_0) \; \underset{a_1}{\max } \sum_{o_{2}} p(o_{2}|o_{1}; a_1) \; \underset{a_2}{\max } \sum_{o_3} \; (...) \; \Big( \\ \nonumber
& (...) & \; \sum_{o_{L}}  p(o_L|o_{1:(L-1)}; a_{1:(L-1)}) \; \underset{a_L}{\max}\; Q(h_L, a_L)\Big) \\  \nonumber
&=& \sum_{o_1} \; \underset{a_1}{\max } \sum_{o_{2}} \; \underset{a_2}{\max } \sum_{o_3} \; (...) \; \Big( \\ \nonumber
& (...) & \; \sum_{o_{L}}  \; \underset{k}{\max}\; p(o_{1:L}|\alpha^{(k)}; a_{0:(L-1)}) \; p_k \Big). \\  \nonumber
\end{eqnarray}
}
Therefore, by taking the optimal action $a^{*}_0 = \argmax_{a_0} Q(h_0,a_0)$, we obtain
\begin{equation}
  \underset{a_0}{\text{max }} Q(h_0,a_0) = p_*^{(L)}.
\end{equation}
As pointed out in the main text, the value and action-value functions are related via $v_\pi(s) = \sum_{a}\pi(a|s)Q_{\pi}(s,a)$. Therefore, the optimal value function for the initial state is the optimal success probability:
\begin{equation}
  v_*(h_0) = \sum_{a} \delta \big(a, \argmax_a Q(h_0, a)) = Q(h_0, a^{*}_0 \big) = P_*^{L}(\alpha).
\end{equation}

\begin{figure}[t]
    \centering
    \includegraphics[width=0.5\textwidth]{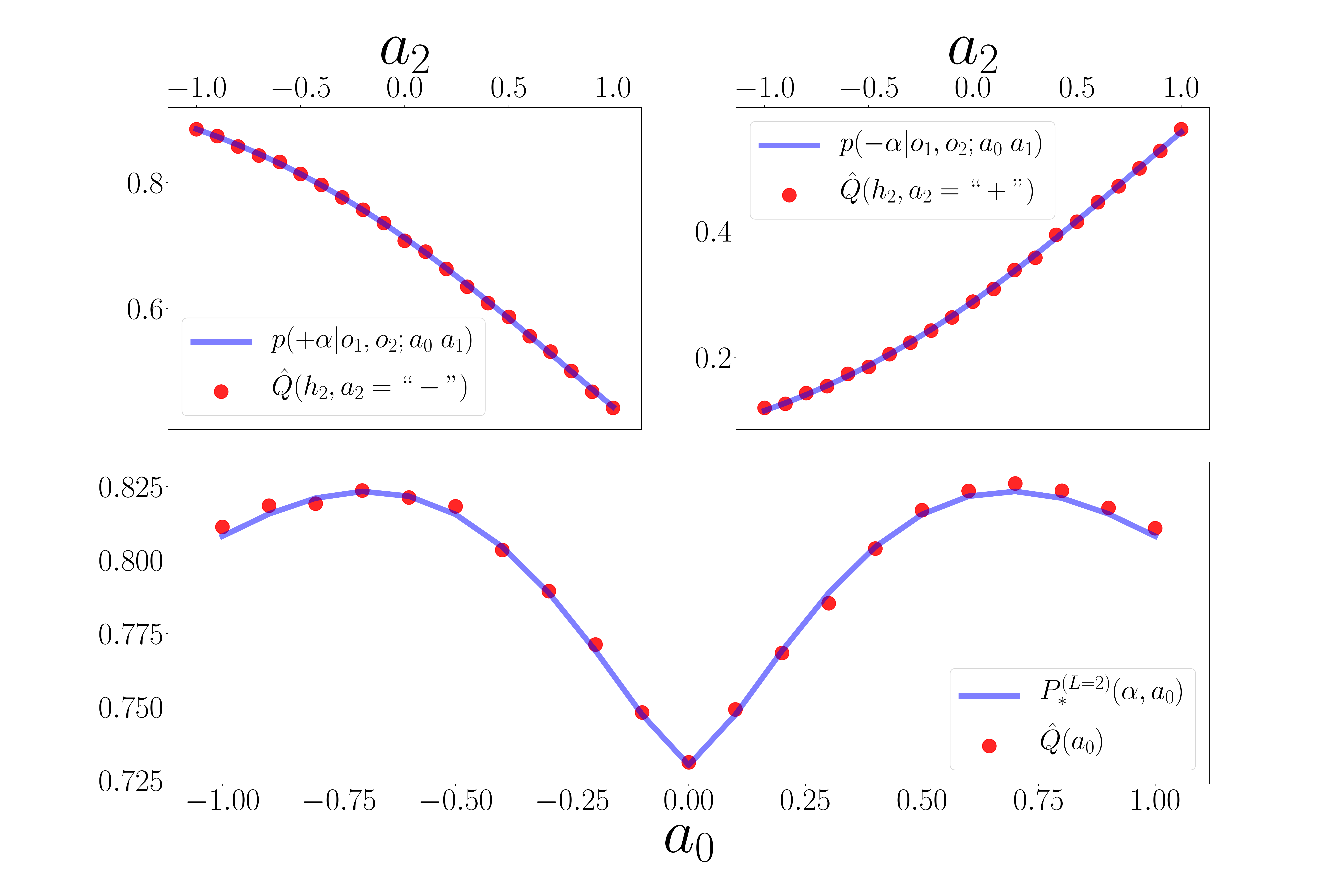}
    \caption{We plot different values of the Q-estimates, after $10^{8}$ episodes of random exploration ($\epsilon = 1$), updating the Q-estimates according to Q-learning (see Algorithm 1). The random exploration is used in order to ensure that, at \textit{finite} number of episodes, all state-action pairs were equally visited on average. The lower plot corresponds to the estimates $\hat{Q}(a_0)$, and it is compared with the optimal success probability as a function of $a_0$, i.e. $P_*^{(L=2)}(\alpha, a_0) =  \sum_{o1}\;p(o_1; a_0) \; \underset{a_1}{\text{max}} \sum_{o_2} p(o_2|h_1, a_1) \; \underset{a_2 = \pm}{\text{max }} p(\pm \alpha | h_2) pr(\pm \alpha)$.}
    \label{fig:profiles}
\end{figure}

In Fig. \ref{fig:profiles} we show several sections of the estimats $\hat{Q}$, using 1-greedy as the interaction policy and each update made according to Algorithm 1, at episode $t = 10^{8}$,

%

\section{Comparison of different UCB strategies}\label{app:ucbTS}
\begin{figure}[t]
    \centering
    \includegraphics[width=0.5\textwidth]{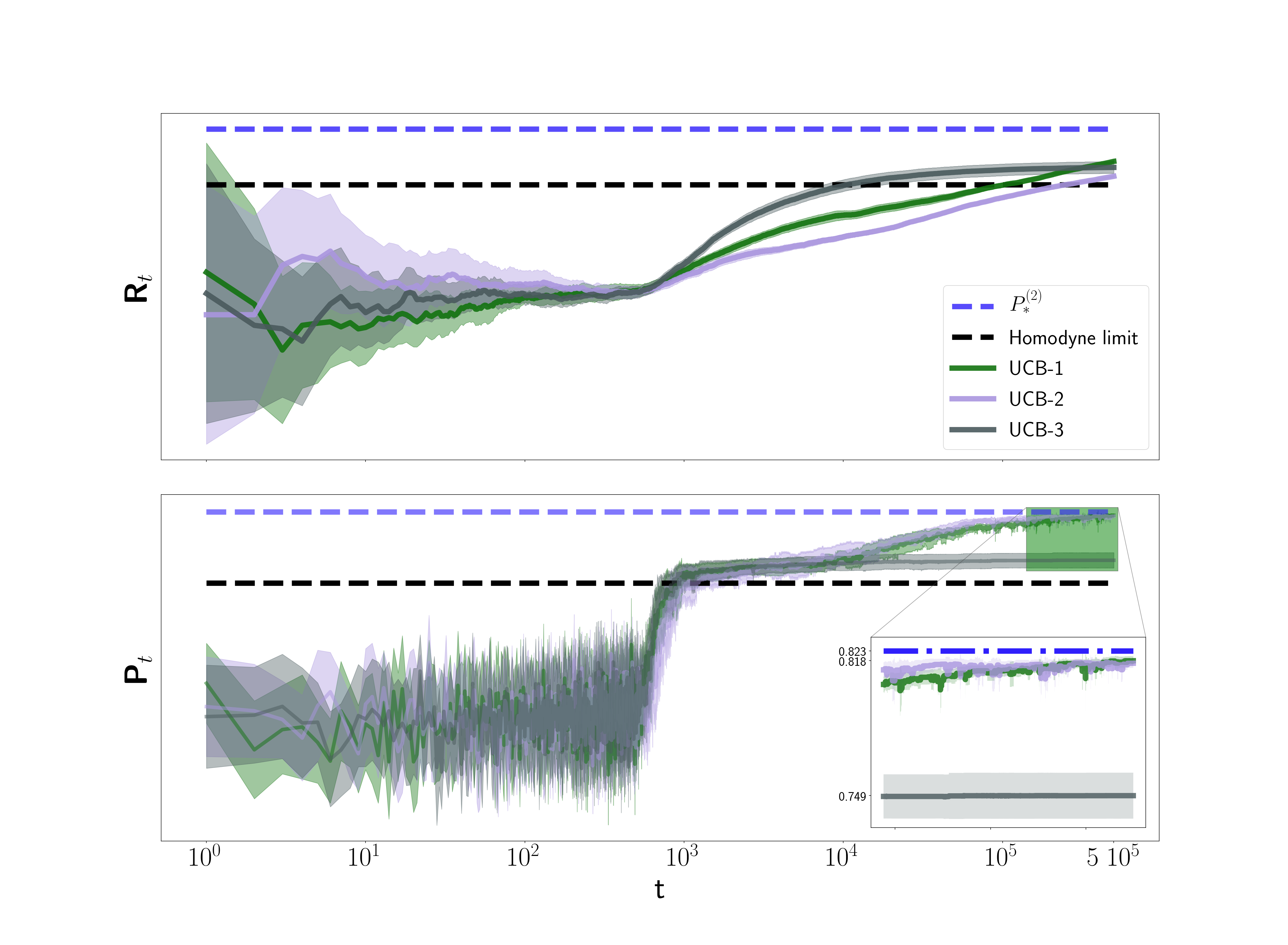}
    \caption{We compare two different variants of UCB showing that the exploration-exploitation trade-off is an intrinsic feature of our problem.}
    \label{fig:ucbs}
\end{figure}

In this section we show numerical studies on how different choices of $\mathcal{P}(t)$ for the UCB strategy can lead to policies whose learning curves for the case $L=2$ exhibit different results. 
As explained in Sec.~\ref{ssec:bandits}, the probability to overestimate the state-action value can be bounded by Hoeffding's inequality. This probability can be forced to depend on the epispode. In Fig.~\ref{fig:ucbs} we show the performance of three different choices of $\mathcal{P}(t)$ for the same receiver considered in Sec.\ref{ssec:qLRes}. First we consider UCB-1, which is the \textit{standard} choice of $\mathcal{P}(t) = t^{-4}$,  which for a bandit problem with  $K=|\cA|$ arms can be easilly proven to have an asymptotic cumulative regret upperbounded by \cite{banditbook}
\be
  \mathcal{L}_t  \leq 8 \sum_{a \in \cA\backslash \{a^{*}\}} \frac{\log t}{\Delta_{k}}+\frac{K \pi^{2}}{3}
\ee
which together with the Lai-Robbins bound of Eq.~\eqref{eq:RLBOUND} implies that $\mathcal{L}_t = O(\log(t))$. Secondly, we consider UCB-2, with a choice of $\mathcal{P}(t)$ proved to be asymptotically optimal in bandit problems \cite{banditbook}. Lastly, an heuristic and instance dependent variation of $\mathcal{P}(t)$, UCB-3, leads to better $\Rt$ only in the short-term, as exploration is damped too fast (which is also reflected in sub-optimal $\Pt$ even in the long term).

\begin{table}
  \begin{tabular}{|c| c |c |c|}
    \hline \\
     Algorithm's name: & UCB-1 & UCB-2 & UCB-3  \\
     \hline \\
     $\mathcal{P}(t)$ & $t^{-4}$ &$\frac{1}{1+t\log^{2}{t}}$  & $t^{\frac{1}{N_t[s,a]}}$ \\
     \hline
  \end{tabular}
\end{table}

%
%

 \bibliographystyle{apsrev4-1}
   \bibliography{library}

\begin{thebibliography}{80}%
\makeatletter
\providecommand \@ifxundefined [1]{%
 \@ifx{#1\undefined}
}%
\providecommand \@ifnum [1]{%
 \ifnum #1\expandafter \@firstoftwo
 \else \expandafter \@secondoftwo
 \fi
}%
\providecommand \@ifx [1]{%
 \ifx #1\expandafter \@firstoftwo
 \else \expandafter \@secondoftwo
 \fi
}%
\providecommand \natexlab [1]{#1}%
\providecommand \enquote  [1]{``#1''}%
\providecommand \bibnamefont  [1]{#1}%
\providecommand \bibfnamefont [1]{#1}%
\providecommand \citenamefont [1]{#1}%
\providecommand \href@noop [0]{\@secondoftwo}%
\providecommand \href [0]{\begingroup \@sanitize@url \@href}%
\providecommand \@href[1]{\@@startlink{#1}\@@href}%
\providecommand \@@href[1]{\endgroup#1\@@endlink}%
\providecommand \@sanitize@url [0]{\catcode `\\12\catcode `\$12\catcode
  `\&12\catcode `\#12\catcode `\^12\catcode `\_12\catcode `\%12\relax}%
\providecommand \@@startlink[1]{}%
\providecommand \@@endlink[0]{}%
\providecommand \url  [0]{\begingroup\@sanitize@url \@url }%
\providecommand \@url [1]{\endgroup\@href {#1}{\urlprefix }}%
\providecommand \urlprefix  [0]{URL }%
\providecommand \Eprint [0]{\href }%
\providecommand \doibase [0]{http://dx.doi.org/}%
\providecommand \selectlanguage [0]{\@gobble}%
\providecommand \bibinfo  [0]{\@secondoftwo}%
\providecommand \bibfield  [0]{\@secondoftwo}%
\providecommand \translation [1]{[#1]}%
\providecommand \BibitemOpen [0]{}%
\providecommand \bibitemStop [0]{}%
\providecommand \bibitemNoStop [0]{.\EOS\space}%
\providecommand \EOS [0]{\spacefactor3000\relax}%
\providecommand \BibitemShut  [1]{\csname bibitem#1\endcsname}%
\let\auto@bib@innerbib\@empty
\bibitem [{\citenamefont {Helstrom}(1976)}]{helstromBOOK}%
  \BibitemOpen
  \bibfield  {author} {\bibinfo {author} {\bibfnamefont {C.~W.}\ \bibnamefont
  {Helstrom}},\ }\href@noop {} {\enquote {\bibinfo {title} {{Quantum Detection
  and Estimation Theory}},}\ } (\bibinfo {year} {1976})\BibitemShut {NoStop}%
\bibitem [{\citenamefont {Holevo}(2012)}]{holevoBOOK}%
  \BibitemOpen
  \bibfield  {author} {\bibinfo {author} {\bibfnamefont {A.~S.}\ \bibnamefont
  {Holevo}},\ }\href {\doibase 10.1515/9783110273403} {\emph {\bibinfo {title}
  {{Quantum Systems, Channels, Information}}}}\ (\bibinfo  {publisher} {De
  Gruyter},\ \bibinfo {address} {Berlin, Boston},\ \bibinfo {year}
  {2012})\BibitemShut {NoStop}%
\bibitem [{\citenamefont {Takeoka}\ and\ \citenamefont
  {Guha}(2014)}]{Takeoka14}%
  \BibitemOpen
  \bibfield  {author} {\bibinfo {author} {\bibfnamefont {M.}~\bibnamefont
  {Takeoka}}\ and\ \bibinfo {author} {\bibfnamefont {S.}~\bibnamefont {Guha}},\
  }\href {\doibase 10.1109/ISIT.2014.6875344} {\bibfield  {journal} {\bibinfo
  {journal} {IEEE Int. Symp. Inf. Theory - Proc.}\ }\textbf {\bibinfo {volume}
  {042309}},\ \bibinfo {pages} {2799} (\bibinfo {year} {2014})},\ \Eprint
  {http://arxiv.org/abs/1401.5132} {arXiv:1401.5132} \BibitemShut {NoStop}%
\bibitem [{\citenamefont {Waseda}\ \emph {et~al.}(2010)\citenamefont {Waseda},
  \citenamefont {Takeoka}, \citenamefont {Sasaki}, \citenamefont {Fujiwara},\
  and\ \citenamefont {Tanaka}}]{Waseda10}%
  \BibitemOpen
  \bibfield  {author} {\bibinfo {author} {\bibfnamefont {A.}~\bibnamefont
  {Waseda}}, \bibinfo {author} {\bibfnamefont {M.}~\bibnamefont {Takeoka}},
  \bibinfo {author} {\bibfnamefont {M.}~\bibnamefont {Sasaki}}, \bibinfo
  {author} {\bibfnamefont {M.}~\bibnamefont {Fujiwara}}, \ and\ \bibinfo
  {author} {\bibfnamefont {H.}~\bibnamefont {Tanaka}},\ }\href {\doibase
  10.1364/JOSAB.27.000259} {\bibfield  {journal} {\bibinfo  {journal} {J. Opt.
  Soc. Am. B}\ }\textbf {\bibinfo {volume} {27}},\ \bibinfo {pages} {259}
  (\bibinfo {year} {2010})}\BibitemShut {NoStop}%
\bibitem [{\citenamefont {Waseda}\ \emph {et~al.}(2011)\citenamefont {Waseda},
  \citenamefont {Sasaki}, \citenamefont {Takeoka}, \citenamefont {Fujiwara},
  \citenamefont {Toyoshima},\ and\ \citenamefont {Assalini}}]{Waseda11}%
  \BibitemOpen
  \bibfield  {author} {\bibinfo {author} {\bibfnamefont {A.}~\bibnamefont
  {Waseda}}, \bibinfo {author} {\bibfnamefont {M.}~\bibnamefont {Sasaki}},
  \bibinfo {author} {\bibfnamefont {M.}~\bibnamefont {Takeoka}}, \bibinfo
  {author} {\bibfnamefont {M.}~\bibnamefont {Fujiwara}}, \bibinfo {author}
  {\bibfnamefont {M.}~\bibnamefont {Toyoshima}}, \ and\ \bibinfo {author}
  {\bibfnamefont {A.}~\bibnamefont {Assalini}},\ }\href {\doibase
  10.1364/JOCN.3.000514} {\bibfield  {journal} {\bibinfo  {journal} {J. Opt.
  Commun. Netw.}\ }\textbf {\bibinfo {volume} {3}},\ \bibinfo {pages} {514}
  (\bibinfo {year} {2011})}\BibitemShut {NoStop}%
\bibitem [{\citenamefont {Guha}(2011)}]{Guha11}%
  \BibitemOpen
  \bibfield  {author} {\bibinfo {author} {\bibfnamefont {S.}~\bibnamefont
  {Guha}},\ }\href {\doibase 10.1103/PhysRevLett.106.240502} {\bibfield
  {journal} {\bibinfo  {journal} {Phys. Rev. Lett.}\ }\textbf {\bibinfo
  {volume} {106}},\ \bibinfo {pages} {1} (\bibinfo {year} {2011})},\ \Eprint
  {http://arxiv.org/abs/1101.1550} {arXiv:1101.1550} \BibitemShut {NoStop}%
\bibitem [{\citenamefont {Krovi}\ \emph {et~al.}(2014)\citenamefont {Krovi},
  \citenamefont {Guha}, \citenamefont {Dutton},\ and\ \citenamefont {{Da
  Silva}}}]{Krovi2014}%
  \BibitemOpen
  \bibfield  {author} {\bibinfo {author} {\bibfnamefont {H.}~\bibnamefont
  {Krovi}}, \bibinfo {author} {\bibfnamefont {S.}~\bibnamefont {Guha}},
  \bibinfo {author} {\bibfnamefont {Z.}~\bibnamefont {Dutton}}, \ and\ \bibinfo
  {author} {\bibfnamefont {M.~P.}\ \bibnamefont {{Da Silva}}},\ }\href
  {\doibase 10.1109/ISIT.2014.6874850} {\bibfield  {journal} {\bibinfo
  {journal} {IEEE Int. Symp. Inf. Theory - Proc.}\ }\textbf {\bibinfo {volume}
  {062333}},\ \bibinfo {pages} {336} (\bibinfo {year} {2014})},\ \Eprint
  {http://arxiv.org/abs/1507.04737} {arXiv:1507.04737} \BibitemShut {NoStop}%
\bibitem [{\citenamefont {Rosati}\ \emph
  {et~al.}(2016{\natexlab{a}})\citenamefont {Rosati}, \citenamefont {Mari},\
  and\ \citenamefont {Giovannetti}}]{Rosati16c}%
  \BibitemOpen
  \bibfield  {author} {\bibinfo {author} {\bibfnamefont {M.}~\bibnamefont
  {Rosati}}, \bibinfo {author} {\bibfnamefont {A.}~\bibnamefont {Mari}}, \ and\
  \bibinfo {author} {\bibfnamefont {V.}~\bibnamefont {Giovannetti}},\ }\href
  {\doibase 10.1103/PhysRevA.94.062325} {\bibfield  {journal} {\bibinfo
  {journal} {Phys. Rev. A}\ }\textbf {\bibinfo {volume} {94}},\ \bibinfo
  {pages} {062325} (\bibinfo {year} {2016}{\natexlab{a}})}\BibitemShut
  {NoStop}%
\bibitem [{\citenamefont {Zwoli{\'{n}}ski}\ \emph {et~al.}(2018)\citenamefont
  {Zwoli{\'{n}}ski}, \citenamefont {Jarzyna},\ and\ \citenamefont
  {Banaszek}}]{Zwolinski2018}%
  \BibitemOpen
  \bibfield  {author} {\bibinfo {author} {\bibfnamefont {W.}~\bibnamefont
  {Zwoli{\'{n}}ski}}, \bibinfo {author} {\bibfnamefont {M.}~\bibnamefont
  {Jarzyna}}, \ and\ \bibinfo {author} {\bibfnamefont {K.}~\bibnamefont
  {Banaszek}},\ }\href {\doibase 10.1364/oe.26.025827} {\bibfield  {journal}
  {\bibinfo  {journal} {Opt. Express}\ }\textbf {\bibinfo {volume} {26}},\
  \bibinfo {pages} {25827} (\bibinfo {year} {2018})},\ \Eprint
  {http://arxiv.org/abs/1806.08401} {arXiv:1806.08401} \BibitemShut {NoStop}%
\bibitem [{\citenamefont {Huttner}\ \emph {et~al.}(1995)\citenamefont
  {Huttner}, \citenamefont {Imoto}, \citenamefont {Gisin},\ and\ \citenamefont
  {Mor}}]{Huttner1995}%
  \BibitemOpen
  \bibfield  {author} {\bibinfo {author} {\bibfnamefont {B.}~\bibnamefont
  {Huttner}}, \bibinfo {author} {\bibfnamefont {N.}~\bibnamefont {Imoto}},
  \bibinfo {author} {\bibfnamefont {N.}~\bibnamefont {Gisin}}, \ and\ \bibinfo
  {author} {\bibfnamefont {T.}~\bibnamefont {Mor}},\ }\href {\doibase
  10.1103/PhysRevA.51.1863} {\bibfield  {journal} {\bibinfo  {journal} {Phys.
  Rev. A}\ }\textbf {\bibinfo {volume} {51}},\ \bibinfo {pages} {1863}
  (\bibinfo {year} {1995})},\ \Eprint {http://arxiv.org/abs/9502020}
  {arXiv:9502020 [quant-ph]} \BibitemShut {NoStop}%
\bibitem [{\citenamefont {Du{\v{s}}ek}\ \emph {et~al.}(2000)\citenamefont
  {Du{\v{s}}ek}, \citenamefont {Jahma},\ and\ \citenamefont
  {L{\"{u}}tkenhaus}}]{Dusek2000}%
  \BibitemOpen
  \bibfield  {author} {\bibinfo {author} {\bibfnamefont {M.}~\bibnamefont
  {Du{\v{s}}ek}}, \bibinfo {author} {\bibfnamefont {M.}~\bibnamefont {Jahma}},
  \ and\ \bibinfo {author} {\bibfnamefont {N.}~\bibnamefont
  {L{\"{u}}tkenhaus}},\ }\href {\doibase 10.1103/PhysRevA.62.022306} {\bibfield
   {journal} {\bibinfo  {journal} {Phys. Rev. A - At. Mol. Opt. Phys.}\
  }\textbf {\bibinfo {volume} {62}},\ \bibinfo {pages} {9} (\bibinfo {year}
  {2000})}\BibitemShut {NoStop}%
\bibitem [{\citenamefont {Grosshans}\ and\ \citenamefont
  {Grangier}(2002)}]{Grosshans2002}%
  \BibitemOpen
  \bibfield  {author} {\bibinfo {author} {\bibfnamefont {F.}~\bibnamefont
  {Grosshans}}\ and\ \bibinfo {author} {\bibfnamefont {P.}~\bibnamefont
  {Grangier}},\ }\href {\doibase 10.1103/PhysRevLett.88.057902} {\bibfield
  {journal} {\bibinfo  {journal} {Phys. Rev. Lett.}\ }\textbf {\bibinfo
  {volume} {88}},\ \bibinfo {pages} {4} (\bibinfo {year} {2002})},\ \Eprint
  {http://arxiv.org/abs/0109084} {arXiv:0109084 [quant-ph]} \BibitemShut
  {NoStop}%
\bibitem [{\citenamefont {Gisin}\ \emph {et~al.}(2002)\citenamefont {Gisin},
  \citenamefont {Ribordy}, \citenamefont {Tittel},\ and\ \citenamefont
  {Zbinden}}]{Gisin2002}%
  \BibitemOpen
  \bibfield  {author} {\bibinfo {author} {\bibfnamefont {N.}~\bibnamefont
  {Gisin}}, \bibinfo {author} {\bibfnamefont {G.}~\bibnamefont {Ribordy}},
  \bibinfo {author} {\bibfnamefont {W.}~\bibnamefont {Tittel}}, \ and\ \bibinfo
  {author} {\bibfnamefont {H.}~\bibnamefont {Zbinden}},\ }\href {\doibase
  10.1103/RevModPhys.74.145} {\bibfield  {journal} {\bibinfo  {journal} {Rev.
  Mod. Phys.}\ }\textbf {\bibinfo {volume} {74}},\ \bibinfo {pages} {145}
  (\bibinfo {year} {2002})},\ \Eprint {http://arxiv.org/abs/0101098}
  {arXiv:0101098 [quant-ph]} \BibitemShut {NoStop}%
\bibitem [{\citenamefont {Bergou}(2007)}]{Bergou2007}%
  \BibitemOpen
  \bibfield  {author} {\bibinfo {author} {\bibfnamefont {J.~A.}\ \bibnamefont
  {Bergou}},\ }\href {\doibase 10.1088/1742-6596/84/1/012001} {\bibfield
  {journal} {\bibinfo  {journal} {J. Phys. Conf. Ser.}\ }\textbf {\bibinfo
  {volume} {84}} (\bibinfo {year} {2007}),\
  10.1088/1742-6596/84/1/012001}\BibitemShut {NoStop}%
\bibitem [{\citenamefont {Bennett}\ and\ \citenamefont
  {Brassard}(2014)}]{Bennet14}%
  \BibitemOpen
  \bibfield  {author} {\bibinfo {author} {\bibfnamefont {C.~H.}\ \bibnamefont
  {Bennett}}\ and\ \bibinfo {author} {\bibfnamefont {G.}~\bibnamefont
  {Brassard}},\ }\href {\doibase 10.1016/j.tcs.2014.05.025} {\bibfield
  {journal} {\bibinfo  {journal} {Theor. Comput. Sci.}\ }\textbf {\bibinfo
  {volume} {560}},\ \bibinfo {pages} {7} (\bibinfo {year} {2014})}\BibitemShut
  {NoStop}%
\bibitem [{\citenamefont {Chabaud}\ \emph {et~al.}(2019)\citenamefont
  {Chabaud}, \citenamefont {Douce}, \citenamefont {Grosshans}, \citenamefont
  {Kashefi},\ and\ \citenamefont {Markham}}]{Chabaud2019}%
  \BibitemOpen
  \bibfield  {author} {\bibinfo {author} {\bibfnamefont {U.}~\bibnamefont
  {Chabaud}}, \bibinfo {author} {\bibfnamefont {T.}~\bibnamefont {Douce}},
  \bibinfo {author} {\bibfnamefont {F.}~\bibnamefont {Grosshans}}, \bibinfo
  {author} {\bibfnamefont {E.}~\bibnamefont {Kashefi}}, \ and\ \bibinfo
  {author} {\bibfnamefont {D.}~\bibnamefont {Markham}},\ }\href
  {https://arxiv.org/pdf/1905.12700.pdf http://arxiv.org/abs/1905.12700} {\
  (\bibinfo {year} {2019})},\ \Eprint {http://arxiv.org/abs/1905.12700}
  {arXiv:1905.12700} \BibitemShut {NoStop}%
\bibitem [{\citenamefont {Pirandola}\ \emph {et~al.}(2019)\citenamefont
  {Pirandola}, \citenamefont {Andersen}, \citenamefont {Banchi}, \citenamefont
  {Berta}, \citenamefont {Bunandar}, \citenamefont {Colbeck}, \citenamefont
  {Englund}, \citenamefont {Gehring}, \citenamefont {Lupo}, \citenamefont
  {Ottaviani}, \citenamefont {Pereira}, \citenamefont {Razavi}, \citenamefont
  {Shaari}, \citenamefont {Tomamichel}, \citenamefont {Usenko}, \citenamefont
  {Vallone}, \citenamefont {Villoresi},\ and\ \citenamefont
  {Wallden}}]{Pirandola2019}%
  \BibitemOpen
  \bibfield  {author} {\bibinfo {author} {\bibfnamefont {S.}~\bibnamefont
  {Pirandola}}, \bibinfo {author} {\bibfnamefont {U.~L.}\ \bibnamefont
  {Andersen}}, \bibinfo {author} {\bibfnamefont {L.}~\bibnamefont {Banchi}},
  \bibinfo {author} {\bibfnamefont {M.}~\bibnamefont {Berta}}, \bibinfo
  {author} {\bibfnamefont {D.}~\bibnamefont {Bunandar}}, \bibinfo {author}
  {\bibfnamefont {R.}~\bibnamefont {Colbeck}}, \bibinfo {author} {\bibfnamefont
  {D.}~\bibnamefont {Englund}}, \bibinfo {author} {\bibfnamefont
  {T.}~\bibnamefont {Gehring}}, \bibinfo {author} {\bibfnamefont
  {C.}~\bibnamefont {Lupo}}, \bibinfo {author} {\bibfnamefont {C.}~\bibnamefont
  {Ottaviani}}, \bibinfo {author} {\bibfnamefont {J.}~\bibnamefont {Pereira}},
  \bibinfo {author} {\bibfnamefont {M.}~\bibnamefont {Razavi}}, \bibinfo
  {author} {\bibfnamefont {J.~S.}\ \bibnamefont {Shaari}}, \bibinfo {author}
  {\bibfnamefont {M.}~\bibnamefont {Tomamichel}}, \bibinfo {author}
  {\bibfnamefont {V.~C.}\ \bibnamefont {Usenko}}, \bibinfo {author}
  {\bibfnamefont {G.}~\bibnamefont {Vallone}}, \bibinfo {author} {\bibfnamefont
  {P.}~\bibnamefont {Villoresi}}, \ and\ \bibinfo {author} {\bibfnamefont
  {P.}~\bibnamefont {Wallden}},\ }\href {http://arxiv.org/abs/1906.01645} {\
  (\bibinfo {year} {2019})},\ \Eprint {http://arxiv.org/abs/1906.01645}
  {arXiv:1906.01645} \BibitemShut {NoStop}%
\bibitem [{\citenamefont {Schuld}\ and\ \citenamefont
  {Petruccione}(2018)}]{SCHULD2019}%
  \BibitemOpen
  \bibfield  {author} {\bibinfo {author} {\bibfnamefont {M.}~\bibnamefont
  {Schuld}}\ and\ \bibinfo {author} {\bibfnamefont {F.}~\bibnamefont
  {Petruccione}},\ }\href {\doibase 10.1007/978-3-319-96424-9} {\emph {\bibinfo
  {title} {{Supervised Learning with Quantum Computers}}}}\ (\bibinfo
  {publisher} {SPRINGER},\ \bibinfo {year} {2018})\BibitemShut {NoStop}%
\bibitem [{\citenamefont {Sent{\'{i}}s}\ \emph {et~al.}(2010)\citenamefont
  {Sent{\'{i}}s}, \citenamefont {Bagan}, \citenamefont {Calsamiglia},\ and\
  \citenamefont {Mu{\~{n}}oz-Tapia}}]{Sentis2010}%
  \BibitemOpen
  \bibfield  {author} {\bibinfo {author} {\bibfnamefont {G.}~\bibnamefont
  {Sent{\'{i}}s}}, \bibinfo {author} {\bibfnamefont {E.}~\bibnamefont {Bagan}},
  \bibinfo {author} {\bibfnamefont {J.}~\bibnamefont {Calsamiglia}}, \ and\
  \bibinfo {author} {\bibfnamefont {R.}~\bibnamefont {Mu{\~{n}}oz-Tapia}},\
  }\href {\doibase 10.1103/PhysRevA.82.042312} {\bibfield  {journal} {\bibinfo
  {journal} {Phys. Rev. A - At. Mol. Opt. Phys.}\ }\textbf {\bibinfo {volume}
  {82}} (\bibinfo {year} {2010}),\ 10.1103/PhysRevA.82.042312}\BibitemShut
  {NoStop}%
\bibitem [{\citenamefont {Sent{\'{i}}s}\ \emph {et~al.}(2015)\citenamefont
  {Sent{\'{i}}s}, \citenamefont {Gu{\c t}{\u a}},\ and\ \citenamefont
  {Adesso}}]{Sentis2014}%
  \BibitemOpen
  \bibfield  {author} {\bibinfo {author} {\bibfnamefont {G.}~\bibnamefont
  {Sent{\'{i}}s}}, \bibinfo {author} {\bibfnamefont {M.}~\bibnamefont {Gu{\c
  t}{\u a}}}, \ and\ \bibinfo {author} {\bibfnamefont {G.}~\bibnamefont
  {Adesso}},\ }\href {\doibase 10.1140/epjqt/s40507-015-0030-4} {\bibfield
  {journal} {\bibinfo  {journal} {EPJ Quantum Technol.}\ }\textbf {\bibinfo
  {volume} {2}} (\bibinfo {year} {2015}),\ 10.1140/epjqt/s40507-015-0030-4},\
  \Eprint {http://arxiv.org/abs/1410.8700} {arXiv:1410.8700} \BibitemShut
  {NoStop}%
\bibitem [{\citenamefont {Gu{\c t}{\v a}}\ and\ \citenamefont
  {Kot{\l}owski}(2010)}]{Guta2010a}%
  \BibitemOpen
  \bibfield  {author} {\bibinfo {author} {\bibfnamefont {M.}~\bibnamefont
  {Gu{\c t}{\v a}}}\ and\ \bibinfo {author} {\bibfnamefont {W.}~\bibnamefont
  {Kot{\l}owski}},\ }\href {\doibase 10.1088/1367-2630/12/12/123032} {\bibfield
   {journal} {\bibinfo  {journal} {New J. Phys.}\ }\textbf {\bibinfo {volume}
  {12}},\ \bibinfo {pages} {123032} (\bibinfo {year} {2010})},\ \Eprint
  {http://arxiv.org/abs/1004.2468} {arXiv:1004.2468} \BibitemShut {NoStop}%
\bibitem [{\citenamefont {Lloyd}\ and\ \citenamefont
  {Weedbrook}(2018)}]{Lloyd2018}%
  \BibitemOpen
  \bibfield  {author} {\bibinfo {author} {\bibfnamefont {S.}~\bibnamefont
  {Lloyd}}\ and\ \bibinfo {author} {\bibfnamefont {C.}~\bibnamefont
  {Weedbrook}},\ }\href {\doibase 10.1103/PhysRevLett.121.040502} {\bibfield
  {journal} {\bibinfo  {journal} {Phys. Rev. Lett.}\ }\textbf {\bibinfo
  {volume} {121}} (\bibinfo {year} {2018}),\ 10.1103/PhysRevLett.121.040502},\
  \Eprint {http://arxiv.org/abs/1804.09139} {arXiv:1804.09139} \BibitemShut
  {NoStop}%
\bibitem [{\citenamefont {Fanizza}\ \emph {et~al.}(2019)\citenamefont
  {Fanizza}, \citenamefont {Mari},\ and\ \citenamefont
  {Giovannetti}}]{Fanizza2019}%
  \BibitemOpen
  \bibfield  {author} {\bibinfo {author} {\bibfnamefont {M.}~\bibnamefont
  {Fanizza}}, \bibinfo {author} {\bibfnamefont {A.}~\bibnamefont {Mari}}, \
  and\ \bibinfo {author} {\bibfnamefont {V.}~\bibnamefont {Giovannetti}},\
  }\href {\doibase 10.1109/tit.2019.2916646} {\bibfield  {journal} {\bibinfo
  {journal} {IEEE Trans. Inf. Theory}\ }\textbf {\bibinfo {volume} {65}},\
  \bibinfo {pages} {5931} (\bibinfo {year} {2019})},\ \Eprint
  {http://arxiv.org/abs/1805.03477} {arXiv:1805.03477} \BibitemShut {NoStop}%
\bibitem [{\citenamefont {Blank}\ \emph {et~al.}(2019)\citenamefont {Blank},
  \citenamefont {Park}, \citenamefont {Rhee},\ and\ \citenamefont
  {Petruccione}}]{Blank19}%
  \BibitemOpen
  \bibfield  {author} {\bibinfo {author} {\bibfnamefont {C.}~\bibnamefont
  {Blank}}, \bibinfo {author} {\bibfnamefont {D.~K.}\ \bibnamefont {Park}},
  \bibinfo {author} {\bibfnamefont {J.-K.~K.}\ \bibnamefont {Rhee}}, \ and\
  \bibinfo {author} {\bibfnamefont {F.}~\bibnamefont {Petruccione}},\ }\href
  {http://arxiv.org/abs/1909.02611} {\  (\bibinfo {year} {2019})},\ \Eprint
  {http://arxiv.org/abs/1909.02611} {arXiv:1909.02611} \BibitemShut {NoStop}%
\bibitem [{\citenamefont {Sergioli}\ \emph {et~al.}(2019)\citenamefont
  {Sergioli}, \citenamefont {Giuntini},\ and\ \citenamefont
  {Freytes}}]{Sergioli2019}%
  \BibitemOpen
  \bibfield  {author} {\bibinfo {author} {\bibfnamefont {G.}~\bibnamefont
  {Sergioli}}, \bibinfo {author} {\bibfnamefont {R.}~\bibnamefont {Giuntini}},
  \ and\ \bibinfo {author} {\bibfnamefont {H.}~\bibnamefont {Freytes}},\ }\href
  {\doibase 10.1371/journal.pone.0216224} {\bibfield  {journal} {\bibinfo
  {journal} {PLoS One}\ }\textbf {\bibinfo {volume} {14}} (\bibinfo {year}
  {2019}),\ 10.1371/journal.pone.0216224}\BibitemShut {NoStop}%
\bibitem [{\citenamefont {Benedetti}\ \emph {et~al.}(2019)\citenamefont
  {Benedetti}, \citenamefont {Grant}, \citenamefont {Wossnig},\ and\
  \citenamefont {Severini}}]{Benedetti2019}%
  \BibitemOpen
  \bibfield  {author} {\bibinfo {author} {\bibfnamefont {M.}~\bibnamefont
  {Benedetti}}, \bibinfo {author} {\bibfnamefont {E.}~\bibnamefont {Grant}},
  \bibinfo {author} {\bibfnamefont {L.}~\bibnamefont {Wossnig}}, \ and\
  \bibinfo {author} {\bibfnamefont {S.}~\bibnamefont {Severini}},\ }\href
  {\doibase 10.1088/1367-2630/ab14b5} {\bibfield  {journal} {\bibinfo
  {journal} {New J. Phys.}\ }\textbf {\bibinfo {volume} {21}} (\bibinfo {year}
  {2019}),\ 10.1088/1367-2630/ab14b5},\ \Eprint
  {http://arxiv.org/abs/1806.00463} {arXiv:1806.00463} \BibitemShut {NoStop}%
\bibitem [{\citenamefont {Carleo}\ and\ \citenamefont
  {Troyer}(2016)}]{Carleo2016}%
  \BibitemOpen
  \bibfield  {author} {\bibinfo {author} {\bibfnamefont {G.}~\bibnamefont
  {Carleo}}\ and\ \bibinfo {author} {\bibfnamefont {M.}~\bibnamefont
  {Troyer}},\ }\href {\doibase 10.1126/science.aag2302} {\bibfield  {journal}
  {\bibinfo  {journal} {Science}\ }\textbf {\bibinfo {volume} {355}},\ \bibinfo
  {pages} {602} (\bibinfo {year} {2016})},\ \Eprint
  {http://arxiv.org/abs/1606.02318} {arXiv:1606.02318} \BibitemShut {NoStop}%
\bibitem [{\citenamefont {Torlai}\ and\ \citenamefont
  {Melko}(2017)}]{Torlai2016}%
  \BibitemOpen
  \bibfield  {author} {\bibinfo {author} {\bibfnamefont {G.}~\bibnamefont
  {Torlai}}\ and\ \bibinfo {author} {\bibfnamefont {R.~G.}\ \bibnamefont
  {Melko}},\ }\href {\doibase 10.1103/PhysRevLett.119.030501} {\bibfield
  {journal} {\bibinfo  {journal} {Phys. Rev. Lett.}\ }\textbf {\bibinfo
  {volume} {119}} (\bibinfo {year} {2017}),\ 10.1103/PhysRevLett.119.030501},\
  \Eprint {http://arxiv.org/abs/1610.04238} {arXiv:1610.04238} \BibitemShut
  {NoStop}%
\bibitem [{\citenamefont {{Van Nieuwenburg}}\ \emph {et~al.}(2017)\citenamefont
  {{Van Nieuwenburg}}, \citenamefont {Liu},\ and\ \citenamefont
  {Huber}}]{VanNieuwenburg2017}%
  \BibitemOpen
  \bibfield  {author} {\bibinfo {author} {\bibfnamefont {E.~P.}\ \bibnamefont
  {{Van Nieuwenburg}}}, \bibinfo {author} {\bibfnamefont {Y.~H.}\ \bibnamefont
  {Liu}}, \ and\ \bibinfo {author} {\bibfnamefont {S.~D.}\ \bibnamefont
  {Huber}},\ }\href {\doibase 10.1038/nphys4037} {\bibfield  {journal}
  {\bibinfo  {journal} {Nat. Phys.}\ }\textbf {\bibinfo {volume} {13}},\
  \bibinfo {pages} {435} (\bibinfo {year} {2017})},\ \Eprint
  {http://arxiv.org/abs/1610.02048} {arXiv:1610.02048} \BibitemShut {NoStop}%
\bibitem [{\citenamefont {Carrasquilla}\ and\ \citenamefont
  {Melko}(2017)}]{Carrasquilla2017}%
  \BibitemOpen
  \bibfield  {author} {\bibinfo {author} {\bibfnamefont {J.}~\bibnamefont
  {Carrasquilla}}\ and\ \bibinfo {author} {\bibfnamefont {R.~G.}\ \bibnamefont
  {Melko}},\ }\href {\doibase 10.1038/nphys4035} {\bibfield  {journal}
  {\bibinfo  {journal} {Nat. Phys.}\ }\textbf {\bibinfo {volume} {13}},\
  \bibinfo {pages} {431} (\bibinfo {year} {2017})},\ \Eprint
  {http://arxiv.org/abs/1605.01735} {arXiv:1605.01735} \BibitemShut {NoStop}%
\bibitem [{\citenamefont {Torlai}\ \emph {et~al.}(2018)\citenamefont {Torlai},
  \citenamefont {Mazzola}, \citenamefont {Carrasquilla}, \citenamefont
  {Troyer}, \citenamefont {Melko},\ and\ \citenamefont {Carleo}}]{Torlai2018}%
  \BibitemOpen
  \bibfield  {author} {\bibinfo {author} {\bibfnamefont {G.}~\bibnamefont
  {Torlai}}, \bibinfo {author} {\bibfnamefont {G.}~\bibnamefont {Mazzola}},
  \bibinfo {author} {\bibfnamefont {J.}~\bibnamefont {Carrasquilla}}, \bibinfo
  {author} {\bibfnamefont {M.}~\bibnamefont {Troyer}}, \bibinfo {author}
  {\bibfnamefont {R.}~\bibnamefont {Melko}}, \ and\ \bibinfo {author}
  {\bibfnamefont {G.}~\bibnamefont {Carleo}},\ }\href {\doibase
  10.1038/s41567-018-0048-5} {\bibfield  {journal} {\bibinfo  {journal} {Nat.
  Phys.}\ }\textbf {\bibinfo {volume} {14}},\ \bibinfo {pages} {447} (\bibinfo
  {year} {2018})},\ \Eprint {http://arxiv.org/abs/1703.05334}
  {arXiv:1703.05334} \BibitemShut {NoStop}%
\bibitem [{\citenamefont {Melnikov}\ \emph {et~al.}(2018)\citenamefont
  {Melnikov}, \citenamefont {{Poulsen Nautrup}}, \citenamefont {Krenn},
  \citenamefont {Dunjko}, \citenamefont {Tiersch}, \citenamefont {Zeilinger},\
  and\ \citenamefont {Briegel}}]{Melnikov2018}%
  \BibitemOpen
  \bibfield  {author} {\bibinfo {author} {\bibfnamefont {A.~A.}\ \bibnamefont
  {Melnikov}}, \bibinfo {author} {\bibfnamefont {H.}~\bibnamefont {{Poulsen
  Nautrup}}}, \bibinfo {author} {\bibfnamefont {M.}~\bibnamefont {Krenn}},
  \bibinfo {author} {\bibfnamefont {V.}~\bibnamefont {Dunjko}}, \bibinfo
  {author} {\bibfnamefont {M.}~\bibnamefont {Tiersch}}, \bibinfo {author}
  {\bibfnamefont {A.}~\bibnamefont {Zeilinger}}, \ and\ \bibinfo {author}
  {\bibfnamefont {H.~J.}\ \bibnamefont {Briegel}},\ }\href {\doibase
  10.1073/pnas.1714936115} {\bibfield  {journal} {\bibinfo  {journal} {Proc.
  Natl. Acad. Sci. U. S. A.}\ }\textbf {\bibinfo {volume} {115}},\ \bibinfo
  {pages} {1221} (\bibinfo {year} {2018})},\ \Eprint
  {http://arxiv.org/abs/1706.00868} {arXiv:1706.00868} \BibitemShut {NoStop}%
\bibitem [{\citenamefont {F{\"{o}}sel}\ \emph {et~al.}(2018)\citenamefont
  {F{\"{o}}sel}, \citenamefont {Tighineanu}, \citenamefont {Weiss},\ and\
  \citenamefont {Marquardt}}]{Fosel2018}%
  \BibitemOpen
  \bibfield  {author} {\bibinfo {author} {\bibfnamefont {T.}~\bibnamefont
  {F{\"{o}}sel}}, \bibinfo {author} {\bibfnamefont {P.}~\bibnamefont
  {Tighineanu}}, \bibinfo {author} {\bibfnamefont {T.}~\bibnamefont {Weiss}}, \
  and\ \bibinfo {author} {\bibfnamefont {F.}~\bibnamefont {Marquardt}},\ }\href
  {\doibase 10.1103/PhysRevX.8.031084} {\bibfield  {journal} {\bibinfo
  {journal} {Phys. Rev. X}\ }\textbf {\bibinfo {volume} {8}} (\bibinfo {year}
  {2018}),\ 10.1103/PhysRevX.8.031084},\ \Eprint
  {http://arxiv.org/abs/1802.05267} {arXiv:1802.05267} \BibitemShut {NoStop}%
\bibitem [{\citenamefont {Walln{\"{o}}fer}\ \emph {et~al.}(2019)\citenamefont
  {Walln{\"{o}}fer}, \citenamefont {Melnikov}, \citenamefont {D{\"{u}}r},\ and\
  \citenamefont {Briegel}}]{Wallnofer2019}%
  \BibitemOpen
  \bibfield  {author} {\bibinfo {author} {\bibfnamefont {J.}~\bibnamefont
  {Walln{\"{o}}fer}}, \bibinfo {author} {\bibfnamefont {A.~A.}\ \bibnamefont
  {Melnikov}}, \bibinfo {author} {\bibfnamefont {W.}~\bibnamefont {D{\"{u}}r}},
  \ and\ \bibinfo {author} {\bibfnamefont {H.~J.}\ \bibnamefont {Briegel}},\
  }\href {http://arxiv.org/abs/1904.10797} {\  (\bibinfo {year} {2019})},\
  \Eprint {http://arxiv.org/abs/1904.10797} {arXiv:1904.10797} \BibitemShut
  {NoStop}%
\bibitem [{\citenamefont {Bukov}\ \emph {et~al.}(2018)\citenamefont {Bukov},
  \citenamefont {Day}, \citenamefont {Sels}, \citenamefont {Weinberg},
  \citenamefont {Polkovnikov},\ and\ \citenamefont {Mehta}}]{Bukov2017}%
  \BibitemOpen
  \bibfield  {author} {\bibinfo {author} {\bibfnamefont {M.}~\bibnamefont
  {Bukov}}, \bibinfo {author} {\bibfnamefont {A.~G.}\ \bibnamefont {Day}},
  \bibinfo {author} {\bibfnamefont {D.}~\bibnamefont {Sels}}, \bibinfo {author}
  {\bibfnamefont {P.}~\bibnamefont {Weinberg}}, \bibinfo {author}
  {\bibfnamefont {A.}~\bibnamefont {Polkovnikov}}, \ and\ \bibinfo {author}
  {\bibfnamefont {P.}~\bibnamefont {Mehta}},\ }\href {\doibase
  10.1103/PhysRevX.8.031086} {\bibfield  {journal} {\bibinfo  {journal} {Phys.
  Rev. X}\ }\textbf {\bibinfo {volume} {8}} (\bibinfo {year} {2018}),\
  10.1103/PhysRevX.8.031086},\ \Eprint {http://arxiv.org/abs/1705.00565}
  {arXiv:1705.00565} \BibitemShut {NoStop}%
\bibitem [{\citenamefont {Niu}\ \emph {et~al.}(2019)\citenamefont {Niu},
  \citenamefont {Boixo}, \citenamefont {Smelyanskiy},\ and\ \citenamefont
  {Neven}}]{Niu2019}%
  \BibitemOpen
  \bibfield  {author} {\bibinfo {author} {\bibfnamefont {M.~Y.}\ \bibnamefont
  {Niu}}, \bibinfo {author} {\bibfnamefont {S.}~\bibnamefont {Boixo}}, \bibinfo
  {author} {\bibfnamefont {V.}~\bibnamefont {Smelyanskiy}}, \ and\ \bibinfo
  {author} {\bibfnamefont {H.}~\bibnamefont {Neven}},\ }in\ \href {\doibase
  10.2514/6.2019-0954} {\emph {\bibinfo {booktitle} {AIAA Scitech 2019
  Forum}}},\ Vol.~\bibinfo {volume} {5}\ (\bibinfo  {publisher} {Nature
  Publishing Group},\ \bibinfo {year} {2019})\ p.~\bibinfo {pages} {33},\
  \Eprint {http://arxiv.org/abs/1803.01857} {arXiv:1803.01857} \BibitemShut
  {NoStop}%
\bibitem [{\citenamefont {Silver}\ \emph {et~al.}(2016)\citenamefont {Silver},
  \citenamefont {Huang}, \citenamefont {Maddison}, \citenamefont {Guez},
  \citenamefont {Sifre}, \citenamefont {van~den Driessche}, \citenamefont
  {Schrittwieser}, \citenamefont {Antonoglou}, \citenamefont {Panneershelvam},
  \citenamefont {Lanctot}, \citenamefont {Dieleman}, \citenamefont {Grewe},
  \citenamefont {Nham}, \citenamefont {Kalchbrenner}, \citenamefont
  {Sutskever}, \citenamefont {Lillicrap}, \citenamefont {Leach}, \citenamefont
  {Kavukcuoglu}, \citenamefont {Graepel},\ and\ \citenamefont
  {Hassabis}}]{Silver2016}%
  \BibitemOpen
  \bibfield  {author} {\bibinfo {author} {\bibfnamefont {D.}~\bibnamefont
  {Silver}}, \bibinfo {author} {\bibfnamefont {A.}~\bibnamefont {Huang}},
  \bibinfo {author} {\bibfnamefont {C.~J.}\ \bibnamefont {Maddison}}, \bibinfo
  {author} {\bibfnamefont {A.}~\bibnamefont {Guez}}, \bibinfo {author}
  {\bibfnamefont {L.}~\bibnamefont {Sifre}}, \bibinfo {author} {\bibfnamefont
  {G.}~\bibnamefont {van~den Driessche}}, \bibinfo {author} {\bibfnamefont
  {J.}~\bibnamefont {Schrittwieser}}, \bibinfo {author} {\bibfnamefont
  {I.}~\bibnamefont {Antonoglou}}, \bibinfo {author} {\bibfnamefont
  {V.}~\bibnamefont {Panneershelvam}}, \bibinfo {author} {\bibfnamefont
  {M.}~\bibnamefont {Lanctot}}, \bibinfo {author} {\bibfnamefont
  {S.}~\bibnamefont {Dieleman}}, \bibinfo {author} {\bibfnamefont
  {D.}~\bibnamefont {Grewe}}, \bibinfo {author} {\bibfnamefont
  {J.}~\bibnamefont {Nham}}, \bibinfo {author} {\bibfnamefont {N.}~\bibnamefont
  {Kalchbrenner}}, \bibinfo {author} {\bibfnamefont {I.}~\bibnamefont
  {Sutskever}}, \bibinfo {author} {\bibfnamefont {T.}~\bibnamefont
  {Lillicrap}}, \bibinfo {author} {\bibfnamefont {M.}~\bibnamefont {Leach}},
  \bibinfo {author} {\bibfnamefont {K.}~\bibnamefont {Kavukcuoglu}}, \bibinfo
  {author} {\bibfnamefont {T.}~\bibnamefont {Graepel}}, \ and\ \bibinfo
  {author} {\bibfnamefont {D.}~\bibnamefont {Hassabis}},\ }\href {\doibase
  10.1038/nature16961} {\bibfield  {journal} {\bibinfo  {journal} {Nature}\
  }\textbf {\bibinfo {volume} {529}},\ \bibinfo {pages} {484} (\bibinfo {year}
  {2016})}\BibitemShut {NoStop}%
\bibitem [{\citenamefont {Osaki}\ \emph {et~al.}(1996)\citenamefont {Osaki},
  \citenamefont {Ban},\ and\ \citenamefont {Hirota}}]{Osaki1996}%
  \BibitemOpen
  \bibfield  {author} {\bibinfo {author} {\bibfnamefont {M.}~\bibnamefont
  {Osaki}}, \bibinfo {author} {\bibfnamefont {M.}~\bibnamefont {Ban}}, \ and\
  \bibinfo {author} {\bibfnamefont {O.}~\bibnamefont {Hirota}},\ }\href
  {\doibase 10.1103/PhysRevA.54.1691} {\bibfield  {journal} {\bibinfo
  {journal} {Phys. Rev. A}\ }\textbf {\bibinfo {volume} {54}},\ \bibinfo
  {pages} {1691} (\bibinfo {year} {1996})}\BibitemShut {NoStop}%
\bibitem [{\citenamefont {Weedbrook}\ \emph {et~al.}(2012)\citenamefont
  {Weedbrook}, \citenamefont {Pirandola}, \citenamefont
  {Garc{\'{i}}a-Patr{\'{o}}n}, \citenamefont {Cerf}, \citenamefont {Ralph},
  \citenamefont {Shapiro},\ and\ \citenamefont {Lloyd}}]{RevGauss}%
  \BibitemOpen
  \bibfield  {author} {\bibinfo {author} {\bibfnamefont {C.}~\bibnamefont
  {Weedbrook}}, \bibinfo {author} {\bibfnamefont {S.}~\bibnamefont
  {Pirandola}}, \bibinfo {author} {\bibfnamefont {R.}~\bibnamefont
  {Garc{\'{i}}a-Patr{\'{o}}n}}, \bibinfo {author} {\bibfnamefont {N.~J.}\
  \bibnamefont {Cerf}}, \bibinfo {author} {\bibfnamefont {T.~C.}\ \bibnamefont
  {Ralph}}, \bibinfo {author} {\bibfnamefont {J.~H.}\ \bibnamefont {Shapiro}},
  \ and\ \bibinfo {author} {\bibfnamefont {S.}~\bibnamefont {Lloyd}},\ }\href
  {\doibase 10.1103/RevModPhys.84.621} {\bibfield  {journal} {\bibinfo
  {journal} {Rev. Mod. Phys.}\ }\textbf {\bibinfo {volume} {84}},\ \bibinfo
  {pages} {621} (\bibinfo {year} {2012})},\ \Eprint
  {http://arxiv.org/abs/1110.3234} {arXiv:1110.3234} \BibitemShut {NoStop}%
\bibitem [{\citenamefont {Takeoka}\ and\ \citenamefont
  {Sasaki}(2008)}]{Takeoka2008}%
  \BibitemOpen
  \bibfield  {author} {\bibinfo {author} {\bibfnamefont {M.}~\bibnamefont
  {Takeoka}}\ and\ \bibinfo {author} {\bibfnamefont {M.}~\bibnamefont
  {Sasaki}},\ }\href {\doibase 10.1103/PhysRevA.78.022320} {\bibfield
  {journal} {\bibinfo  {journal} {Phys. Rev. A - At. Mol. Opt. Phys.}\ }\textbf
  {\bibinfo {volume} {78}},\ \bibinfo {pages} {1} (\bibinfo {year} {2008})},\
  \Eprint {http://arxiv.org/abs/0706.1038} {arXiv:0706.1038} \BibitemShut
  {NoStop}%
\bibitem [{\citenamefont {Dolinar}(1973)}]{Dolinar1973}%
  \BibitemOpen
  \bibfield  {author} {\bibinfo {author} {\bibfnamefont {S.~J.}\ \bibnamefont
  {Dolinar}},\ }\href@noop {} {\bibfield  {journal} {\bibinfo  {journal} {Q.
  Prog. Rep. (Research Lab. Electron.}\ }\textbf {\bibinfo {volume} {111}},\
  \bibinfo {pages} {115} (\bibinfo {year} {1973})}\BibitemShut {NoStop}%
\bibitem [{\citenamefont {Kennedy}(1973)}]{Kennedy1973a}%
  \BibitemOpen
  \bibfield  {author} {\bibinfo {author} {\bibfnamefont {R.~S.}\ \bibnamefont
  {Kennedy}},\ }\href {https://dspace.mit.edu/handle/1721.1/56346} {\bibfield
  {journal} {\bibinfo  {journal} {MIT Res. Lab. Electron. Q. Prog. Rep.}\
  }\textbf {\bibinfo {volume} {108}},\ \bibinfo {pages} {219} (\bibinfo {year}
  {1973})}\BibitemShut {NoStop}%
\bibitem [{\citenamefont {Geremia}(2004)}]{Geremia2004}%
  \BibitemOpen
  \bibfield  {author} {\bibinfo {author} {\bibfnamefont {J.}~\bibnamefont
  {Geremia}},\ }\href {\doibase 10.1103/PhysRevA.70.062303} {\bibfield
  {journal} {\bibinfo  {journal} {Phys. Rev. A}\ }\textbf {\bibinfo {volume}
  {70}},\ \bibinfo {pages} {062303} (\bibinfo {year} {2004})},\ \Eprint
  {http://arxiv.org/abs/0407205} {arXiv:0407205 [quant-ph]} \BibitemShut
  {NoStop}%
\bibitem [{\citenamefont {Takeoka}\ \emph {et~al.}(2005)\citenamefont
  {Takeoka}, \citenamefont {Sasaki}, \citenamefont {{Van Loock}},\ and\
  \citenamefont {L{\"{u}}tkenhaus}}]{Takeoka2005}%
  \BibitemOpen
  \bibfield  {author} {\bibinfo {author} {\bibfnamefont {M.}~\bibnamefont
  {Takeoka}}, \bibinfo {author} {\bibfnamefont {M.}~\bibnamefont {Sasaki}},
  \bibinfo {author} {\bibfnamefont {P.}~\bibnamefont {{Van Loock}}}, \ and\
  \bibinfo {author} {\bibfnamefont {N.}~\bibnamefont {L{\"{u}}tkenhaus}},\
  }\href {\doibase 10.1103/PhysRevA.71.022318} {\bibfield  {journal} {\bibinfo
  {journal} {Phys. Rev. A - At. Mol. Opt. Phys.}\ }\textbf {\bibinfo {volume}
  {71}},\ \bibinfo {pages} {1} (\bibinfo {year} {2005})},\ \Eprint
  {http://arxiv.org/abs/0410133} {arXiv:0410133 [quant-ph]} \BibitemShut
  {NoStop}%
\bibitem [{\citenamefont {Takeoka}\ \emph {et~al.}(2006)\citenamefont
  {Takeoka}, \citenamefont {Sasaki},\ and\ \citenamefont
  {L{\"{u}}tkenhaus}}]{Takeoka2006}%
  \BibitemOpen
  \bibfield  {author} {\bibinfo {author} {\bibfnamefont {M.}~\bibnamefont
  {Takeoka}}, \bibinfo {author} {\bibfnamefont {M.}~\bibnamefont {Sasaki}}, \
  and\ \bibinfo {author} {\bibfnamefont {N.}~\bibnamefont {L{\"{u}}tkenhaus}},\
  }\href {\doibase 10.1103/PhysRevLett.97.040502} {\bibfield  {journal}
  {\bibinfo  {journal} {Phys. Rev. Lett.}\ }\textbf {\bibinfo {volume} {97}},\
  \bibinfo {pages} {040502} (\bibinfo {year} {2006})}\BibitemShut {NoStop}%
\bibitem [{\citenamefont {Cook}\ \emph {et~al.}(2007)\citenamefont {Cook},
  \citenamefont {Martin}, \citenamefont {Geremia}, \citenamefont {Chase},\ and\
  \citenamefont {Geremia}}]{Cook2007}%
  \BibitemOpen
  \bibfield  {author} {\bibinfo {author} {\bibfnamefont {R.~L.}\ \bibnamefont
  {Cook}}, \bibinfo {author} {\bibfnamefont {P.~J.}\ \bibnamefont {Martin}},
  \bibinfo {author} {\bibfnamefont {J.~M.}\ \bibnamefont {Geremia}}, \bibinfo
  {author} {\bibfnamefont {B.~A.}\ \bibnamefont {Chase}}, \ and\ \bibinfo
  {author} {\bibfnamefont {J.~M.}\ \bibnamefont {Geremia}},\ }\href {\doibase
  10.1038/nature05655} {\bibfield  {journal} {\bibinfo  {journal} {Nature}\
  }\textbf {\bibinfo {volume} {446}},\ \bibinfo {pages} {774} (\bibinfo {year}
  {2007})}\BibitemShut {NoStop}%
\bibitem [{\citenamefont {{Da Silva}}\ \emph {et~al.}(2013)\citenamefont {{Da
  Silva}}, \citenamefont {Guha},\ and\ \citenamefont {Dutton}}]{DaSilva2013}%
  \BibitemOpen
  \bibfield  {author} {\bibinfo {author} {\bibfnamefont {M.~P.}\ \bibnamefont
  {{Da Silva}}}, \bibinfo {author} {\bibfnamefont {S.}~\bibnamefont {Guha}}, \
  and\ \bibinfo {author} {\bibfnamefont {Z.}~\bibnamefont {Dutton}},\ }\href
  {\doibase 10.1103/PhysRevA.87.052320} {\bibfield  {journal} {\bibinfo
  {journal} {Phys. Rev. A - At. Mol. Opt. Phys.}\ }\textbf {\bibinfo {volume}
  {87}} (\bibinfo {year} {2013}),\ 10.1103/PhysRevA.87.052320},\ \Eprint
  {http://arxiv.org/abs/arXiv:1201.6625} {arXiv:arXiv:1201.6625} \BibitemShut
  {NoStop}%
\bibitem [{\citenamefont {Nair}\ \emph {et~al.}(2014)\citenamefont {Nair},
  \citenamefont {Guha},\ and\ \citenamefont {Tan}}]{Nair2014}%
  \BibitemOpen
  \bibfield  {author} {\bibinfo {author} {\bibfnamefont {R.}~\bibnamefont
  {Nair}}, \bibinfo {author} {\bibfnamefont {S.}~\bibnamefont {Guha}}, \ and\
  \bibinfo {author} {\bibfnamefont {S.~H.}\ \bibnamefont {Tan}},\ }\href
  {\doibase 10.1103/PhysRevA.89.032318} {\bibfield  {journal} {\bibinfo
  {journal} {Phys. Rev. A - At. Mol. Opt. Phys.}\ }\textbf {\bibinfo {volume}
  {89}},\ \bibinfo {pages} {1} (\bibinfo {year} {2014})},\ \Eprint
  {http://arxiv.org/abs/1212.2048} {arXiv:1212.2048} \BibitemShut {NoStop}%
\bibitem [{\citenamefont {Rosati}\ \emph
  {et~al.}(2016{\natexlab{b}})\citenamefont {Rosati}, \citenamefont {Mari},\
  and\ \citenamefont {Giovannetti}}]{Rosati16a}%
  \BibitemOpen
  \bibfield  {author} {\bibinfo {author} {\bibfnamefont {M.}~\bibnamefont
  {Rosati}}, \bibinfo {author} {\bibfnamefont {A.}~\bibnamefont {Mari}}, \ and\
  \bibinfo {author} {\bibfnamefont {V.}~\bibnamefont {Giovannetti}},\ }\href
  {\doibase 10.1103/PhysRevA.93.062315} {\bibfield  {journal} {\bibinfo
  {journal} {Phys. Rev. A}\ }\textbf {\bibinfo {volume} {93}},\ \bibinfo
  {pages} {062315} (\bibinfo {year} {2016}{\natexlab{b}})},\ \Eprint
  {http://arxiv.org/abs/1602.03989} {arXiv:1602.03989} \BibitemShut {NoStop}%
\bibitem [{\citenamefont {Dimario}\ and\ \citenamefont
  {Becerra}(2018)}]{DiMario2018a}%
  \BibitemOpen
  \bibfield  {author} {\bibinfo {author} {\bibfnamefont {M.~T.}\ \bibnamefont
  {Dimario}}\ and\ \bibinfo {author} {\bibfnamefont {F.~E.}\ \bibnamefont
  {Becerra}},\ }\href {\doibase 10.1103/PhysRevLett.121.023603} {\bibfield
  {journal} {\bibinfo  {journal} {Phys. Rev. Lett.}\ }\textbf {\bibinfo
  {volume} {121}},\ \bibinfo {pages} {023603} (\bibinfo {year}
  {2018})}\BibitemShut {NoStop}%
\bibitem [{\citenamefont {Becerra}\ \emph {et~al.}(2013)\citenamefont
  {Becerra}, \citenamefont {Fan}, \citenamefont {Baumgartner}, \citenamefont
  {Goldhar}, \citenamefont {Kosloski},\ and\ \citenamefont
  {Migdall}}]{Becerra2013a}%
  \BibitemOpen
  \bibfield  {author} {\bibinfo {author} {\bibfnamefont {F.~E.}\ \bibnamefont
  {Becerra}}, \bibinfo {author} {\bibfnamefont {J.}~\bibnamefont {Fan}},
  \bibinfo {author} {\bibfnamefont {G.}~\bibnamefont {Baumgartner}}, \bibinfo
  {author} {\bibfnamefont {J.}~\bibnamefont {Goldhar}}, \bibinfo {author}
  {\bibfnamefont {J.~T.}\ \bibnamefont {Kosloski}}, \ and\ \bibinfo {author}
  {\bibfnamefont {a.}~\bibnamefont {Migdall}},\ }\href {\doibase
  10.1038/nphoton.2012.316} {\bibfield  {journal} {\bibinfo  {journal} {Nat.
  Photonics}\ }\textbf {\bibinfo {volume} {7}},\ \bibinfo {pages} {147}
  (\bibinfo {year} {2013})}\BibitemShut {NoStop}%
\bibitem [{\citenamefont {M{\"{u}}ller}\ and\ \citenamefont
  {Marquardt}(2015)}]{Muller2015}%
  \BibitemOpen
  \bibfield  {author} {\bibinfo {author} {\bibfnamefont {C.~R.}\ \bibnamefont
  {M{\"{u}}ller}}\ and\ \bibinfo {author} {\bibfnamefont {C.}~\bibnamefont
  {Marquardt}},\ }\href {\doibase 10.1088/1367-2630/17/3/032003} {\bibfield
  {journal} {\bibinfo  {journal} {New J. Phys.}\ }\textbf {\bibinfo {volume}
  {17}},\ \bibinfo {pages} {1} (\bibinfo {year} {2015})},\ \Eprint
  {http://arxiv.org/abs/1412.6242} {arXiv:1412.6242} \BibitemShut {NoStop}%
\bibitem [{\citenamefont {Guha}\ and\ \citenamefont {Wilde}(2012)}]{Guha2012}%
  \BibitemOpen
  \bibfield  {author} {\bibinfo {author} {\bibfnamefont {S.}~\bibnamefont
  {Guha}}\ and\ \bibinfo {author} {\bibfnamefont {M.~M.}\ \bibnamefont
  {Wilde}},\ }\href {\doibase 10.1109/ISIT.2012.6284250} {\bibfield  {journal}
  {\bibinfo  {journal} {IEEE Int. Symp. Inf. Theory - Proc.}\ ,\ \bibinfo
  {pages} {546}} (\bibinfo {year} {2012})},\ \Eprint
  {http://arxiv.org/abs/1202.0533} {arXiv:1202.0533} \BibitemShut {NoStop}%
\bibitem [{\citenamefont {Wilde}\ and\ \citenamefont {Guha}(2013)}]{Wilde2013}%
  \BibitemOpen
  \bibfield  {author} {\bibinfo {author} {\bibfnamefont {M.~M.}\ \bibnamefont
  {Wilde}}\ and\ \bibinfo {author} {\bibfnamefont {S.}~\bibnamefont {Guha}},\
  }\href {\doibase 10.1109/TIT.2012.2218792} {\bibfield  {journal} {\bibinfo
  {journal} {IEEE Trans. Inf. Theory}\ }\textbf {\bibinfo {volume} {59}},\
  \bibinfo {pages} {1175} (\bibinfo {year} {2013})},\ \Eprint
  {http://arxiv.org/abs/1109.2591} {arXiv:1109.2591} \BibitemShut {NoStop}%
\bibitem [{\citenamefont {Rosati}\ and\ \citenamefont
  {Giovannetti}(2016)}]{Rosati16b}%
  \BibitemOpen
  \bibfield  {author} {\bibinfo {author} {\bibfnamefont {M.}~\bibnamefont
  {Rosati}}\ and\ \bibinfo {author} {\bibfnamefont {V.}~\bibnamefont
  {Giovannetti}},\ }\href {\doibase 10.1063/1.4953690} {\bibfield  {journal}
  {\bibinfo  {journal} {J. Math. Phys.}\ }\textbf {\bibinfo {volume} {57}},\
  \bibinfo {pages} {062204} (\bibinfo {year} {2016})},\ \Eprint
  {http://arxiv.org/abs/1506.04999} {arXiv:1506.04999} \BibitemShut {NoStop}%
\bibitem [{\citenamefont {Rosati}\ \emph {et~al.}(2017)\citenamefont {Rosati},
  \citenamefont {Mari},\ and\ \citenamefont {Giovannetti}}]{Rosati2017}%
  \BibitemOpen
  \bibfield  {author} {\bibinfo {author} {\bibfnamefont {M.}~\bibnamefont
  {Rosati}}, \bibinfo {author} {\bibfnamefont {A.}~\bibnamefont {Mari}}, \ and\
  \bibinfo {author} {\bibfnamefont {V.}~\bibnamefont {Giovannetti}},\ }\href
  {\doibase 10.1103/PhysRevA.96.012317} {\bibfield  {journal} {\bibinfo
  {journal} {Phys. Rev. A}\ }\textbf {\bibinfo {volume} {96}},\ \bibinfo
  {pages} {012317} (\bibinfo {year} {2017})},\ \Eprint
  {http://arxiv.org/abs/1703.05701} {arXiv:1703.05701} \BibitemShut {NoStop}%
\bibitem [{\citenamefont {Xiang}\ \emph {et~al.}(2017)\citenamefont {Xiang},
  \citenamefont {Zhang}, \citenamefont {Jiang},\ and\ \citenamefont
  {Rabl}}]{Xiang2017}%
  \BibitemOpen
  \bibfield  {author} {\bibinfo {author} {\bibfnamefont {Z.~L.}\ \bibnamefont
  {Xiang}}, \bibinfo {author} {\bibfnamefont {M.}~\bibnamefont {Zhang}},
  \bibinfo {author} {\bibfnamefont {L.}~\bibnamefont {Jiang}}, \ and\ \bibinfo
  {author} {\bibfnamefont {P.}~\bibnamefont {Rabl}},\ }\href {\doibase
  10.1103/PhysRevX.7.011035} {\bibfield  {journal} {\bibinfo  {journal} {Phys.
  Rev. X}\ }\textbf {\bibinfo {volume} {7}},\ \bibinfo {pages} {011035}
  (\bibinfo {year} {2017})},\ \Eprint {http://arxiv.org/abs/1611.10241}
  {arXiv:1611.10241} \BibitemShut {NoStop}%
\bibitem [{\citenamefont {DiMario}\ \emph {et~al.}(2019)\citenamefont
  {DiMario}, \citenamefont {Kunz}, \citenamefont {Banaszek},\ and\
  \citenamefont {Becerra}}]{DiMario2019}%
  \BibitemOpen
  \bibfield  {author} {\bibinfo {author} {\bibfnamefont {M.~T.}\ \bibnamefont
  {DiMario}}, \bibinfo {author} {\bibfnamefont {L.}~\bibnamefont {Kunz}},
  \bibinfo {author} {\bibfnamefont {K.}~\bibnamefont {Banaszek}}, \ and\
  \bibinfo {author} {\bibfnamefont {F.~E.}\ \bibnamefont {Becerra}},\ }\href
  {\doibase 10.1038/s41534-019-0177-4} {\bibfield  {journal} {\bibinfo
  {journal} {npj Quantum Inf.}\ }\textbf {\bibinfo {volume} {5}} (\bibinfo
  {year} {2019}),\ 10.1038/s41534-019-0177-4},\ \Eprint
  {http://arxiv.org/abs/1907.12515} {arXiv:1907.12515} \BibitemShut {NoStop}%
\bibitem [{\citenamefont {Bellman}(2003)}]{Bellman2003}%
  \BibitemOpen
  \bibfield  {author} {\bibinfo {author} {\bibfnamefont {R.}~\bibnamefont
  {Bellman}},\ }\href@noop {} {\emph {\bibinfo {title} {{Dynamic Programming
  (Reprinted version)}}}}\ (\bibinfo  {publisher} {Dover Publications},\
  \bibinfo {year} {2003})\ p.\ \bibinfo {pages} {384}\BibitemShut {NoStop}%
\bibitem [{\citenamefont {Watkins}(1989)}]{Watkins1989}%
  \BibitemOpen
  \bibfield  {author} {\bibinfo {author} {\bibfnamefont {C.}~\bibnamefont
  {Watkins}},\ }\emph {\bibinfo {title} {{``Learning from delayed rewards''.
  PhD thesis}}},\ \href {http://www.cs.rhul.ac.uk/{~}chrisw/thesis.html} {Ph.D.
  thesis},\ \bibinfo  {school} {Cambridge} (\bibinfo {year} {1989})\BibitemShut
  {NoStop}%
\bibitem [{\citenamefont {Sutton}\ and\ \citenamefont {{G.
  Barto}}(2018)}]{Sutton2018}%
  \BibitemOpen
  \bibfield  {author} {\bibinfo {author} {\bibfnamefont {R.}~\bibnamefont
  {Sutton}}\ and\ \bibinfo {author} {\bibfnamefont {A.}~\bibnamefont {{G.
  Barto}}},\ }\href@noop {} {\emph {\bibinfo {title} {{Reinforcement Learning
  Sutton}}}}\ (\bibinfo  {publisher} {MIT Press},\ \bibinfo {year}
  {2018})\BibitemShut {NoStop}%
\bibitem [{\citenamefont {Lattimore}\ and\ \citenamefont
  {Szepesvari}(2018)}]{banditbook}%
  \BibitemOpen
  \bibfield  {author} {\bibinfo {author} {\bibfnamefont {T.}~\bibnamefont
  {Lattimore}}\ and\ \bibinfo {author} {\bibfnamefont {C.}~\bibnamefont
  {Szepesvari}},\ }\href@noop {} {\bibfield  {journal} {\bibinfo  {journal}
  {Cambridge Univ. Press}\ ,\ \bibinfo {pages} {542}} (\bibinfo {year}
  {2018})}\BibitemShut {NoStop}%
\bibitem [{\citenamefont {Auer}\ \emph {et~al.}(2002)\citenamefont {Auer},
  \citenamefont {Cesa-Bianchi},\ and\ \citenamefont {Fischer}}]{Auer2002}%
  \BibitemOpen
  \bibfield  {author} {\bibinfo {author} {\bibfnamefont {P.}~\bibnamefont
  {Auer}}, \bibinfo {author} {\bibfnamefont {N.}~\bibnamefont {Cesa-Bianchi}},
  \ and\ \bibinfo {author} {\bibfnamefont {P.}~\bibnamefont {Fischer}},\ }\href
  {\doibase 10.1023/A:1013689704352} {\bibfield  {journal} {\bibinfo  {journal}
  {Mach. Learn.}\ }\textbf {\bibinfo {volume} {47}},\ \bibinfo {pages} {235}
  (\bibinfo {year} {2002})}\BibitemShut {NoStop}%
\bibitem [{\citenamefont {Russo}\ \emph {et~al.}(2018)\citenamefont {Russo},
  \citenamefont {{Van Roy}}, \citenamefont {Kazerouni}, \citenamefont
  {Osband},\ and\ \citenamefont {Wen}}]{Russo2018}%
  \BibitemOpen
  \bibfield  {author} {\bibinfo {author} {\bibfnamefont {D.~J.}\ \bibnamefont
  {Russo}}, \bibinfo {author} {\bibfnamefont {B.}~\bibnamefont {{Van Roy}}},
  \bibinfo {author} {\bibfnamefont {A.}~\bibnamefont {Kazerouni}}, \bibinfo
  {author} {\bibfnamefont {I.}~\bibnamefont {Osband}}, \ and\ \bibinfo {author}
  {\bibfnamefont {Z.}~\bibnamefont {Wen}},\ }\href {\doibase
  10.1561/2200000070} {\enquote {\bibinfo {title} {{A tutorial on Thompson
  sampling}},}\ } (\bibinfo {year} {2018}),\ \Eprint
  {http://arxiv.org/abs/1707.02038} {arXiv:1707.02038} \BibitemShut {NoStop}%
\bibitem [{\citenamefont {Jin}\ \emph {et~al.}(2018)\citenamefont {Jin},
  \citenamefont {Allen-Zhu}, \citenamefont {Bubeck},\ and\ \citenamefont
  {Jordan}}]{qlprovably}%
  \BibitemOpen
  \bibfield  {author} {\bibinfo {author} {\bibfnamefont {C.}~\bibnamefont
  {Jin}}, \bibinfo {author} {\bibfnamefont {Z.}~\bibnamefont {Allen-Zhu}},
  \bibinfo {author} {\bibfnamefont {S.}~\bibnamefont {Bubeck}}, \ and\ \bibinfo
  {author} {\bibfnamefont {M.~I.}\ \bibnamefont {Jordan}},\ }\bibfield
  {booktitle} {\emph {\bibinfo {booktitle} {Advances in Neural Information
  Processing Systems 31}},\ }\href
  {http://papers.nips.cc/paper/7735-is-q-learning-provably-efficient.pdf} {\ ,\
  \bibinfo {pages} {4863} (\bibinfo {year} {2018})}\BibitemShut {NoStop}%
\bibitem [{\citenamefont {Singh}\ \emph {et~al.}(1994)\citenamefont {Singh},
  \citenamefont {Jaakkola},\ and\ \citenamefont {Jordan}}]{Singh1994}%
  \BibitemOpen
  \bibfield  {author} {\bibinfo {author} {\bibfnamefont {S.~P.}\ \bibnamefont
  {Singh}}, \bibinfo {author} {\bibfnamefont {T.}~\bibnamefont {Jaakkola}}, \
  and\ \bibinfo {author} {\bibfnamefont {M.~I.}\ \bibnamefont {Jordan}},\ }in\
  \href {\doibase 10.1016/b978-1-55860-335-6.50042-8} {\emph {\bibinfo
  {booktitle} {Mach. Learn. Proc. 1994}}}\ (\bibinfo {year} {1994})\ pp.\
  \bibinfo {pages} {284--292}\BibitemShut {NoStop}%
\bibitem [{\citenamefont {Mnih}\ \emph {et~al.}(2013)\citenamefont {Mnih},
  \citenamefont {Kavukcuoglu}, \citenamefont {Silver}, \citenamefont {Graves},
  \citenamefont {Antonoglou}, \citenamefont {Wierstra},\ and\ \citenamefont
  {Riedmiller}}]{Mnih2013}%
  \BibitemOpen
  \bibfield  {author} {\bibinfo {author} {\bibfnamefont {V.}~\bibnamefont
  {Mnih}}, \bibinfo {author} {\bibfnamefont {K.}~\bibnamefont {Kavukcuoglu}},
  \bibinfo {author} {\bibfnamefont {D.}~\bibnamefont {Silver}}, \bibinfo
  {author} {\bibfnamefont {A.}~\bibnamefont {Graves}}, \bibinfo {author}
  {\bibfnamefont {I.}~\bibnamefont {Antonoglou}}, \bibinfo {author}
  {\bibfnamefont {D.}~\bibnamefont {Wierstra}}, \ and\ \bibinfo {author}
  {\bibfnamefont {M.}~\bibnamefont {Riedmiller}},\ }\href
  {http://arxiv.org/abs/1312.5602} {\  (\bibinfo {year} {2013})},\ \Eprint
  {http://arxiv.org/abs/1312.5602} {arXiv:1312.5602} \BibitemShut {NoStop}%
\bibitem [{\citenamefont {Shani}\ \emph {et~al.}(2013)\citenamefont {Shani},
  \citenamefont {Pineau},\ and\ \citenamefont {Kaplow}}]{Shani2013}%
  \BibitemOpen
  \bibfield  {author} {\bibinfo {author} {\bibfnamefont {G.}~\bibnamefont
  {Shani}}, \bibinfo {author} {\bibfnamefont {J.}~\bibnamefont {Pineau}}, \
  and\ \bibinfo {author} {\bibfnamefont {R.}~\bibnamefont {Kaplow}},\ }\href
  {\doibase 10.1007/s10458-012-9200-2} {\bibfield  {journal} {\bibinfo
  {journal} {Auton. Agent. Multi. Agent. Syst.}\ }\textbf {\bibinfo {volume}
  {27}},\ \bibinfo {pages} {1} (\bibinfo {year} {2013})}\BibitemShut {NoStop}%
\bibitem [{\citenamefont {Egorov}(2015)}]{Egorov2015}%
  \BibitemOpen
  \bibfield  {author} {\bibinfo {author} {\bibfnamefont {M.}~\bibnamefont
  {Egorov}},\ }\href {http://arxiv.org/abs/1903.07765} {\emph {\bibinfo {title}
  {{Deep reinforcement learning with pomdps}}}},\ \bibinfo {type} {Tech. Rep.}\
  (\bibinfo  {institution} {Technical Report, Stanford University},\ \bibinfo
  {year} {2015})\ \Eprint {http://arxiv.org/abs/1903.07765} {arXiv:1903.07765}
  \BibitemShut {NoStop}%
\bibitem [{\citenamefont {Zhu}\ \emph {et~al.}(2018)\citenamefont {Zhu},
  \citenamefont {Li}, \citenamefont {Poupart},\ and\ \citenamefont
  {Miao}}]{Zhu2018}%
  \BibitemOpen
  \bibfield  {author} {\bibinfo {author} {\bibfnamefont {P.}~\bibnamefont
  {Zhu}}, \bibinfo {author} {\bibfnamefont {X.}~\bibnamefont {Li}}, \bibinfo
  {author} {\bibfnamefont {P.}~\bibnamefont {Poupart}}, \ and\ \bibinfo
  {author} {\bibfnamefont {G.}~\bibnamefont {Miao}},\ }\href
  {http://arxiv.org/abs/1704.07978 http://arxiv.org/abs/1804.06309} {\
  (\bibinfo {year} {2018})},\ \Eprint {http://arxiv.org/abs/1804.06309}
  {arXiv:1804.06309} \BibitemShut {NoStop}%
\bibitem [{\citenamefont {Silver}\ \emph {et~al.}(2014)\citenamefont {Silver},
  \citenamefont {Lever}, \citenamefont {Heess}, \citenamefont {Degris},
  \citenamefont {Wierstra},\ and\ \citenamefont {Riedmiller}}]{ddpg}%
  \BibitemOpen
  \bibfield  {author} {\bibinfo {author} {\bibfnamefont {D.}~\bibnamefont
  {Silver}}, \bibinfo {author} {\bibfnamefont {G.}~\bibnamefont {Lever}},
  \bibinfo {author} {\bibfnamefont {N.}~\bibnamefont {Heess}}, \bibinfo
  {author} {\bibfnamefont {T.}~\bibnamefont {Degris}}, \bibinfo {author}
  {\bibfnamefont {D.}~\bibnamefont {Wierstra}}, \ and\ \bibinfo {author}
  {\bibfnamefont {M.}~\bibnamefont {Riedmiller}},\ }in\ \href@noop {} {\emph
  {\bibinfo {booktitle} {Proceedings of the 31st International Conference on
  International Conference on Machine Learning - Volume 32}}},\ \bibinfo
  {series and number} {ICML’14}\ (\bibinfo  {publisher} {JMLR.org},\ \bibinfo
  {year} {2014})\ p.\ \bibinfo {pages} {I–387–I–395}\BibitemShut
  {NoStop}%
\bibitem [{\citenamefont {Szepesvári}(2010)}]{algsrl}%
  \BibitemOpen
  \bibfield  {author} {\bibinfo {author} {\bibfnamefont {C.}~\bibnamefont
  {Szepesvári}},\ }\bibfield  {booktitle} {\emph {\bibinfo {booktitle}
  {Algorithms for Reinforcement Learning}},\ }\href
  {http://dx.doi.org/10.2200/S00268ED1V01Y201005AIM009} {\ \bibinfo {series}
  {Synthesis Lectures on Artificial Intelligence and Machine Learning}
  (\bibinfo {year} {2010})}\BibitemShut {NoStop}%
\bibitem [{\citenamefont {Thompson}(1933)}]{Thompson1933}%
  \BibitemOpen
  \bibfield  {author} {\bibinfo {author} {\bibfnamefont {W.~R.}\ \bibnamefont
  {Thompson}},\ }\href {\doibase 10.2307/2332286} {\bibfield  {journal}
  {\bibinfo  {journal} {Biometrika}\ }\textbf {\bibinfo {volume} {25}},\
  \bibinfo {pages} {285} (\bibinfo {year} {1933})}\BibitemShut {NoStop}%
\bibitem [{\citenamefont {Lai}\ and\ \citenamefont {Robbins}(1985)}]{Lai1985}%
  \BibitemOpen
  \bibfield  {author} {\bibinfo {author} {\bibfnamefont {T.~L.}\ \bibnamefont
  {Lai}}\ and\ \bibinfo {author} {\bibfnamefont {H.}~\bibnamefont {Robbins}},\
  }\href {\doibase 10.1016/0196-8858(85)90002-8} {\bibfield  {journal}
  {\bibinfo  {journal} {Adv. Appl. Math.}\ }\textbf {\bibinfo {volume} {6}},\
  \bibinfo {pages} {4} (\bibinfo {year} {1985})}\BibitemShut {NoStop}%
\bibitem [{\citenamefont {Agrawal}(1995)}]{Agrawal1995}%
  \BibitemOpen
  \bibfield  {author} {\bibinfo {author} {\bibfnamefont {R.}~\bibnamefont
  {Agrawal}},\ }\href {\doibase 10.2307/1427934} {\bibfield  {journal}
  {\bibinfo  {journal} {Adv. Appl. Probab.}\ }\textbf {\bibinfo {volume}
  {27}},\ \bibinfo {pages} {1054} (\bibinfo {year} {1995})}\BibitemShut
  {NoStop}%
\bibitem [{\citenamefont {Thompson}(1935)}]{Thompson1935}%
  \BibitemOpen
  \bibfield  {author} {\bibinfo {author} {\bibfnamefont {W.~R.}\ \bibnamefont
  {Thompson}},\ }\href {\doibase 10.2307/2371219} {\bibfield  {journal}
  {\bibinfo  {journal} {Am. J. Math.}\ }\textbf {\bibinfo {volume} {57}},\
  \bibinfo {pages} {450} (\bibinfo {year} {1935})}\BibitemShut {NoStop}%
\bibitem [{\citenamefont {Scott}(2010)}]{Scott2010}%
  \BibitemOpen
  \bibfield  {author} {\bibinfo {author} {\bibfnamefont {S.~L.}\ \bibnamefont
  {Scott}},\ }\href {\doibase 10.1002/asmb.874} {\bibfield  {journal} {\bibinfo
   {journal} {Appl. Stoch. Model. Bus. Ind.}\ }\textbf {\bibinfo {volume}
  {26}},\ \bibinfo {pages} {639} (\bibinfo {year} {2010})}\BibitemShut
  {NoStop}%
\bibitem [{\citenamefont {Kaufmann}\ \emph {et~al.}(2012)\citenamefont
  {Kaufmann}, \citenamefont {Korda},\ and\ \citenamefont {Munos}}]{TSoptimal}%
  \BibitemOpen
  \bibfield  {author} {\bibinfo {author} {\bibfnamefont {E.}~\bibnamefont
  {Kaufmann}}, \bibinfo {author} {\bibfnamefont {N.}~\bibnamefont {Korda}}, \
  and\ \bibinfo {author} {\bibfnamefont {R.}~\bibnamefont {Munos}},\ }in\
  \href@noop {} {\emph {\bibinfo {booktitle} {Algorithmic Learning Theory}}},\
  \bibinfo {editor} {edited by\ \bibinfo {editor} {\bibfnamefont {N.~H.}\
  \bibnamefont {Bshouty}}, \bibinfo {editor} {\bibfnamefont {G.}~\bibnamefont
  {Stoltz}}, \bibinfo {editor} {\bibfnamefont {N.}~\bibnamefont {Vayatis}}, \
  and\ \bibinfo {editor} {\bibfnamefont {T.}~\bibnamefont {Zeugmann}}}\
  (\bibinfo  {publisher} {Springer Berlin Heidelberg},\ \bibinfo {address}
  {Berlin, Heidelberg},\ \bibinfo {year} {2012})\ pp.\ \bibinfo {pages}
  {199--213}\BibitemShut {NoStop}%
\bibitem [{\citenamefont {Garivier}\ \emph {et~al.}(2019)\citenamefont
  {Garivier}, \citenamefont {M{\'{e}}nard},\ and\ \citenamefont
  {Stoltz}}]{workTSFOLK}%
  \BibitemOpen
  \bibfield  {author} {\bibinfo {author} {\bibfnamefont {A.}~\bibnamefont
  {Garivier}}, \bibinfo {author} {\bibfnamefont {P.}~\bibnamefont
  {M{\'{e}}nard}}, \ and\ \bibinfo {author} {\bibfnamefont {G.}~\bibnamefont
  {Stoltz}},\ }\href {\doibase 10.1287/moor.2017.0928} {\bibfield  {journal}
  {\bibinfo  {journal} {Math. Oper. Res.}\ }\textbf {\bibinfo {volume} {44}},\
  \bibinfo {pages} {377} (\bibinfo {year} {2019})},\ \Eprint
  {http://arxiv.org/abs/1602.07182} {arXiv:1602.07182} \BibitemShut {NoStop}%
\bibitem [{\citenamefont {Bubeck}\ \emph {et~al.}(2011)\citenamefont {Bubeck},
  \citenamefont {Munos},\ and\ \citenamefont {Stoltz}}]{simpleRegretMunoz}%
  \BibitemOpen
  \bibfield  {author} {\bibinfo {author} {\bibfnamefont {S.}~\bibnamefont
  {Bubeck}}, \bibinfo {author} {\bibfnamefont {R.}~\bibnamefont {Munos}}, \
  and\ \bibinfo {author} {\bibfnamefont {G.}~\bibnamefont {Stoltz}},\ }\href
  {\doibase 10.1016/j.tcs.2010.12.059} {\bibfield  {journal} {\bibinfo
  {journal} {Theor. Comput. Sci.}\ }\textbf {\bibinfo {volume} {412}},\
  \bibinfo {pages} {1832} (\bibinfo {year} {2011})}\BibitemShut {NoStop}%
\end{thebibliography}%

\end{document}